\documentclass[aps,pra,nofootinbib,twocolumn,superscriptaddress,10pt]{revtex4-1}
\pdfoutput=1
\def\Title {Bogoliubov Excitations of Disordered Bose-Einstein Condensates}
\usepackage[utf8]{inputenc}

\usepackage[pdftex]{graphicx}
\usepackage{amsmath,amsfonts,amssymb,dsfont,cases}

\usepackage[pdftex, a4paper, bookmarks, bookmarksopen,
bookmarksopenlevel=1, bookmarksnumbered=true,
pdftitle={\Title},colorlinks, linkcolor=blue, urlcolor=blue]{hyperref}

\usepackage[american]{babel}

\renewcommand{\Re}{{\rm Re}}
\renewcommand{\Im}{{\rm Im}}

\DeclareMathOperator{\PV}{P}	% Cauchy principle value
\newcommand{\vc}[1]{\boldsymbol{#1}}
\renewcommand{\r}{{\vc r}}
\renewcommand{\k}{{\vc k}}
\newcommand{\e}{{\vc e}}
\newcommand{\q}{{\vc q}}
\newcommand{\p}{{\vc p}}
\newcommand{\K}{{\vc K}}
\newcommand{\G}{\mathcal{G}}
\newcommand{\V}{\mathcal{V}}
\newcommand{\matr}[4]{\left(\begin{array}{cc}#1&#2\\#3&#4\end{array}\right)} % 2x2 Matrix
\newcommand{\cvect}[2]{\left(\begin{array}{c}#1\\#2\end{array}\right)}         % Column vector
\newcommand{\vnab}{{\vc\nabla}}
\newcommand{\pder}[2]{\frac{\partial #1}{\partial #2}}		% partial derivative
\newcommand{\tfder}[2]{\delta #1/\delta #2}		%
\newcommand{\rmd}{\mathrm{d}}

\newcommand{\ls}{l_\text{s}}
\newcommand{\ltr}{l_\text{tr}}
\newcommand{\lloc}{l_\text{loc}}

\newcommand{\intdd}[1]{\int \frac{{\rm d}^d #1}{(2\pi)^d}}	% k-space integral
\newcommand{\intddr}{\int {\rm d}^d r}				% realspace integral

\newcommand{\dn}{\delta n}			% density fluctuation
\newcommand{\dph}{\delta \varphi}		% phase fluctuation
\newcommand{\gh}[1]{\hat\gamma_{#1}}		% quantized Bg amplitude
\newcommand{\ghd}[1]{\hat\gamma^\dagger_{#1}}	%
\newcommand{\epn}[1]{\epsilon^0_{#1}}		% free kinetic energy
\newcommand{\ep}[1]{\epsilon_{#1}}		% Bogoliubov energy
\newcommand{\avg}[1]{\overline{{#1}}}				% synonym
\newcommand{\Psitd}[1]{\Phi^{(#1)}} 				%Imprint in the Gross-Pitaevskii wave-function

\newcommand{\bra}[1]{\left\langle #1 \right |}
\newcommand{\ket}[1]{\left| #1 \right\rangle}

\newenvironment{smallpmatrix}{\left(\begin{smallmatrix}}%
{\end{smallmatrix}\right)}

\newcommand{\EKomm}[2]{\bigl\langle \bigl[ #1, #2 \bigr]  \bigr\rangle} 
\newcommand{\matrs}[4]{\begin{smallpmatrix}
#1 & #2 \\
#3 & #4 
\end{smallpmatrix}
} % slim small 2x2 Matrix

\begin{document}
\title{\Title}
\author{Christopher Gaul}
\affiliation{Departamento de F\'isica de Materiales, Universidad Complutense, E-28040 Madrid, Spain}
\author{Cord A.\ M\"uller}
\affiliation{Centre for Quantum Technologies, National University of Singapore, Singapore 117543, Singapore}
\date{\today} 

\begin{abstract}
We describe repulsively interacting Bose-Einstein condensates in spatially correlated disorder potentials of arbitrary dimension.
The first effect of disorder is to deform the mean-field condensate. 
Secondly, the quantum excitation spectrum and condensate population are affected.
By a saddle-point expansion of the many-body Hamiltonian around the
deformed mean-field ground state, we derive the fundamental quadratic Hamiltonian of quantum fluctuations. 
Importantly, a basis is used such that excitations are orthogonal to the deformed condensate.  
Via Bogoliubov-Nambu perturbation theory, we compute the effective excitation dispersion, including mean free paths and localization lengths.
Corrections to the speed of sound and average density of states are calculated, due to correlated disorder in arbitrary dimensions, extending to the
case of weak lattice potentials. 
 \end{abstract}

\maketitle

%------------------------------------------
\section{Introduction} 
\label{chptIntroduction}
%------------------------------------------

The intriguing interplay of Bose
statistics, interaction, and disorder is one of the most prominent
problems in condensed matter physics, known
as the \emph{dirty boson problem} \cite{Giamarchi1987,*Giamarchi1988,Fisher1989}. 
Experimentally, it was first studied with 
superfluid Helium in aerosol glasses (Vycor) \cite{Crooker1983,*Chan1988,*Wong1990}.  
Over the past years, several groups have loaded ultracold
atoms into optical potentials and studied Bose-Einstein condensates (BECs) in the presence of disorder under
very clean laboratory conditions 
\cite{Clement2005,Schulte2005,Lye2005,Chen2008,White2009,Dries2010}.

In this paper, we study the situation where Bose statistics and
interaction are the dominant effects, and the disorder weakly perturbs the homogeneous situation. 
In this regime, the presence of a well-populated condensate makes 
Bogoliubov's theory \cite{Bogoliubov1947} the most economic description, 
because it describes quantum fluctuations around the best mean-field approximation of the condensate. 

What happens to a homogeneous condensate if a weak external potential is
switched on?  How are the quantum fluctuations affected? 
These questions constitute the \emph{inhomogeneous Bogoliubov problem}, a
problem of notorious difficulty, due to the broken translational
symmetry \cite{Nozieres1999}. 
Bogoliubov theories for disordered BEC have been formulated by Lee and
Gunn \cite{Lee1990}, Huang and Meng \cite{Huang1992} and Giorgini,
Pitaevskii and Stringari \cite{Giorgini1994}, complemented by
\cite{Kobayashi2002,Astrakharchik2002,Bilas2006,Lugan2007a,Falco2007,Fontanesi2009,Hu2009}, among
others. Yet, none of the existing theories covers 
spatially correlated disorder and all dimensionalities, which come
into focus after recent experimental advances
\cite{Lewenstein2007,Bloch2008,Sanchez-Palencia2010,Modugno2010,Robert-de-Saint-Vincent2010}.

Moreover, some approaches appear questionable from
a conceptual point of view. 
Indeed, the primary effect of an external potential is to
deform the condensate itself. This is most obvious for cold-atom
BECs in traps, where the condensate forms in a non-uniform spatial
mode that results from the competition between
interaction, kinetic, and potential energy. Therefore,
it is awkward, if not outright inappropriate, to construct a Bogoliubov
theory in terms of fluctuations that are still
defined as deviations from the uniform condensate of the
homogeneous case
\cite{Huang1992,Kobayashi2002,Astrakharchik2002,Falco2007,Hu2009}.
Instead, Bogoliubov's ansatz warrants to first determine the deformed
condensate mode on the mean-field level.  
Accordingly, we discuss the number-conserving deformation of the condensate caused by a weak external
potential in Section \ref{secMeanfield}. 

The deformed condensate is the vacuum of Bogoliubov fluctuations, to whose description we turn in a second step.
Great care must be taken to ensure that the excitations occur in modes that remain orthogonal
to the inhomogeneous ground state---even in a disordered situation where the ground state depends on each realization of the random potential.
We have found it helpful to tackle this formidable problem by a variational saddle-point expansion, a powerful method central to the solution of many problems in statistical and 
quantum mechanics \cite{Kleinert2009,Giamarchi1996}. 
Such an expansion yields all relevant terms in a systematic
manner, without the need for deciding ad hoc which terms are to be 
kept or discarded. Section \ref{secSaddlepoint} contains a
full account of our formulation, leading to the fundamental inhomogeneous Bogoliubov
Hamiltonian that is quadratic in the fluctuations.  

We emphasize that our approach describes both effects, the deformation of the ground state and the scattering of excitations, on the same footing and to the same order in the external potential.
Moreover, our approach involves quantized fluctuations that always remain orthogonal to the inhomogeneous Bogoliubov vacuum.
A fully analytical description is presented up to second order in disorder strength. 

The excitation spectrum of any system provides precious information
about its (thermo-)dynamic properties. For instance, the homogeneous interacting
Bose gas has a gapless excitation spectrum. This is consistent with the fact that the
low-energy excitations are the Goldstone modes \cite{Goldstone1962}
associated with the spontaneous $U(1)$ symmetry breaking in the BEC
phase. These low-energy excitations are collective in character,
with many interacting particles oscillating back and forth as in a sound wave.
And really, the dispersion relation at low energy is linear, its slope being the sound velocity. 
The sound velocity is intimately linked to numerous important quantities like
specific heat and compressibility, and moreover, by a classical argument due to Landau, equal to the critical velocity of superfluidity \cite{Dalfovo1999,Pethick2002,Pitaevskii2003}. 

Since an external potential couples to the particle density and does not interfere with the $U(1)$
symmetry, it is not expected to induce an excitation gap. However, 
inhomogeneity should certainly affect the speed of sound, which is a
nonlinear function of the particle density. 
Thus, it is of particular interest to predict the speed-of-sound correction in
disordered Bose gases. But curiously, the state of affairs for this
key quantity is far from satisfactory. 
The simplest Bogoliubov theories cannot predict a change in excitation
dispersion at all \cite{Huang1992,Kobayashi2002,Astrakharchik2002}. 
More elaborate calculations by Giorgini \textit{et
al.}~\cite{Giorgini1994} predict a certain positive correction for uncorrelated
disorder in three dimensions, a result which has been exactly reproduced by Lopatin
and Vinokur \cite{Lopatin2002} and Falco \textit{et al.}\ \cite{Falco2007}.
This is contradicted by Yukalov and Graham
\cite{Yukalov2007,Yukalov2007a} who report a \emph{decrease} of the sound
velocity in three dimensions, even in the case of uncorrelated
disorder. A negative correction is also found, in all dimensions, for spatially
correlated disorder with a correlation length much
longer than the condensate healing length \cite{Gaul2009a}.  

Clearly, there is a need for a unified theory that describes the dispersion
relation of Bogoliubov excitations in presence of disorder with
spatial correlation. In section \ref{secDispersion}, we provide such a
theory, at least perturbatively for weak external potentials, 
by applying standard diagrammatic Green function techniques to the
inhomogeneous Bogoliubov Hamiltonian derived in
Sec.~\ref{secSaddlepoint}. We compute the
ensemble-averaged disorder correction to the single-excitation
spectrum in general,  including the elastic scattering rate, and corrections to
sound velocity and density of states. 

In Section \ref{secResults}, these general results are discussed in
greater detail, with particular emphasis on the case of correlated disorder.
Numerous analytical results are found in certain limiting regions of the parameter
space, which is spanned by condensate healing length, excitation wave length, and
disorder correlation length.  Specific results pertaining to optical speckle
potentials are collected in Appendix \ref{secSpeckle}. In passing,
we recover the localization
properties of Bogoliubov excitations in one dimension as described earlier by Bilas and Pavloff
\cite{Bilas2006} and Lugan
\textit{et al.}\ \cite{Lugan2007a}. We briefly connect to the case of
weak lattice potentials and confirm predictions by Taylor and Zaremba
\cite{Taylor2003} and Liang \emph{et al.}\
\cite{Liang2008} within our formalism.  We also reconfirm  
the positive correction of the speed of
sound by uncorrelated disorder in three
dimensions \cite{Giorgini1994,Lopatin2002,Falco2007}. 
It turns out, though, that this result is hardly generic, because in
lower dimensions and for correlated disorder in general, one
always finds a negative correction. We confirm our analytical
predictions  in one dimension by numerical simulations on the
mean field level, as well as by exact numerical
diagonalization of the Bogoliubov-de Gennes equations.  

Finally, Section \ref{secConclusions} concludes and closes the paper
on some open questions.

%----------------------------------------------------------------------------
\section{The inhomogeneous Bogoliubov Hamiltonian}
\label{secRelevantHamiltonian}
%----------------------------------------------------------------------------

A weakly interacting Bose gas is described by 
the (grand canonical) Hamiltonian \cite{Dalfovo1999,Pethick2002,Pitaevskii2003}
\begin{align}\label{eqManyParticleHamiltonian}
 \hat E = \int &{\rm d}^d r\,
\hat \Psi^\dagger \left[\frac{-\hbar^2}{2m} \nabla^2 + V(\r) - \mu + \frac{g}{2}\hat \Psi^\dagger
 \hat\Psi
 \right] \hat \Psi,
\end{align}
in terms of particle annihilation and creation operators 
$\hat\Psi=\hat\Psi(\r)$ and $\hat\Psi^\dagger=\hat\Psi^\dagger(\r)$, respectively, which obey
the canonical commutator relations 
\begin{equation}\label{commutators.eq}
\begin{split}
\bigl[\hat \Psi(\r) , \hat \Psi(\r')\bigr] & = \bigl[\hat \Psi^\dagger(\r) , \hat \Psi^\dagger(\r')\bigr] = 0,\\
\bigl[\hat \Psi(\r) , \hat \Psi^\dagger(\r')\bigr] &= \delta(\r-\r')  .
\end{split}
\end{equation}
Atom-atom interaction is taken into account in the form of s-wave
scattering. The s-wave scattering length
$a_{\rm s}$ determines the interaction parameter,  $g={4\pi \hbar^2} a_{\rm s} / {m}$ in three
dimensions, with similar relations in quasi-two and quasi-one
dimensional geometries. We will treat the case of repulsive
interaction with a constant interaction parameter $g>0$.
This interaction potential is a good approximation in the regime of
low energy and dilute gases, where the gas parameter $(n a_{\rm s}^3)^{1/2}$ is small,
i.e., the average particle distance $n^{-1/3}$ is much larger than the
scattering length~$a_{\rm s}$.
For dilute ultracold gases (unlike superfluid helium), this parameter is typically very small 
$(n a_{\rm s}^3)^{1/2} \approx 0.01$ \cite{Pitaevskii1998}. 

We will work in the canonical ensemble with a fixed total number of particles
$N = \intddr \bigl\langle \hat \Psi^\dagger(\r) \hat \Psi(\r)
\bigr\rangle$. The chemical potential $\mu$ serves as the Lagrange parameter that
has to be adjusted accordingly, as function of external control parameters.  
One of these external control fields is the inhomogeneous external potential 
$V(\r)$. In the laboratory, this typically comprises a
global trapping potential as well as, say, optical lattices and/or disorder potentials. 
In the following, we will concentrate on
the situation where the global trapping potential is very smooth, ideally a
very large box, and  $V(\r)$ then describes the local spatial fluctuations around the homogeneous background. 
 
Below a critical temperature, the Bose gas forms a BEC \cite{EinsteinBEC}, where a macroscopically large fraction of particles populates the ground state of the single-particle density matrix.
In the absence of interaction, this is just the ground state of the potential $V(\r)$, but also 
in a dilute interacting Bose gas a well defined condensate mode appears, as proven rigorously in the homogeneous case and three dimensions \cite{Lieb2002,Erdos2007}. Within a mean-field description (or equivalently, Hartree-Fock theory) the condensate spontaneously breaks the U(1) gauge invariance of the Hamiltonian \eqref{eqManyParticleHamiltonian} by settling on a global phase.
In lower dimensions and within confining potentials, quasi-condensates
\cite{Mora2003} exist, whose phase coherence is not of truly infinite
range, but can extend over large enough distances such that the
condensate shows the tell-tale signatures of a phase-coherent matter wave. 

Bogoliubov's theory \cite{Bogoliubov1947} takes advantage of the macroscopically occupied ground state of the BEC and splits the field operator into a mean-field condensate and quantized fluctuations:
\begin{equation}\label{eqBogoliubovPrescription}
\hat \Psi(\r) = \Phi(\r) + \delta\hat\Psi(\r) .
\end{equation}
The small parameter of this expansion is again the gas parameter $(n a_{\rm s}^3)^{1/2}$ \cite{Lee1957}, and for dilute condensed atomic gases, Bogoliubov theory proves to be a very adequate description. 

Following this approach, we will first describe how the external
potential $V(\r)$ affects the condensate mode, strictly within mean
field. In a second step, we determine the relevant Hamiltonian of the quantum
fluctuations around this modified ground state. We emphasize from the
outset that a consistent Bogoliubov theory requires to calculate both steps to the same
order in $V(\r)$; otherwise one runs the risk of describing only half of
the relevant physics. Instead of deciding ad hoc which terms should be
kept and which not, we resort to a well-controlled saddle-point expansion of the many-body
Hamiltonian around the mean-field ground state. 

%---------------------------------------------------------------------------
\subsection{Deformed mean-field ground state}\label{secMeanfield}

The mean-field approach, known as Gross-Pitaevskii (GP) theory \cite{Gross1963,Pitaevskii2003},
neglects the quantum fluctuations and replaces the field operators by a complex field $\Psi=\Psi(\r)$, 
such that the many-body Hamiltonian \eqref{eqManyParticleHamiltonian} reduces to the GP energy functional: 
\begin{align}\label{eqGPEnergyFunctionalPsi}
E = \intddr \biggl\lbrace
\frac{\hbar^2}{2m} |\vc{\nabla} \Psi|^2 + \big[ V(\r)- \mu \big] |\Psi|^2
+\frac{g}{2}|\Psi|^4	\biggr\rbrace .
\end{align}
We wish to determine its ground state as function of the external potential $V(\r)$. 
By definition, the ground state $\Psi_0(\r)=\Phi(\r)$ minimizes the energy functional \eqref{eqGPEnergyFunctionalPsi}. 
It obeys the stationarity condition $\left.\tfder{E}{\Psi^*} \right|_0=0$, also known as the stationary GP equation
\begin{equation} \label{eqStationaryGPE} 
-\frac{\hbar^2}{2m} {\nabla^2\Phi(\r)} + (g |\Phi(\r)|^2 - \mu) {\Phi(\r)} =
- V(\r) {\Phi(\r)}  .
\end{equation}
For a stationary potential $V(\r)$, the condensate's kinetic energy is always minimized by choosing a fixed global phase,
thereby ruling out superfluid flow or vortices, and without loss of generality we may take $\Phi(\r) \in \mathbb{R}$ in the following.

In the homogeneous case $V(\r)=0$, the repulsive interaction spreads
the density over the entire available volume, and $n = |\Phi|^2 = \mu/g$. In the inhomogeneous case, however,
the condensate wave function depends via Eq.~\eqref{eqStationaryGPE}
nonlinearly on $V(\r)$. Numerically, the condensate
$\Psi(\r)$ can be computed very efficiently, for any given potential $V(\r)$, by propagating the GP equation
in imaginary time \cite{Dalfovo1996}. 

What analytical tools are available? 
If the external potential and the condensate wave function vary only very smoothly, the first term in Eq.~\eqref{eqStationaryGPE}, the kinetic energy or quantum pressure, is negligible.
In this so-called Thomas-Fermi (TF) regime, the density profile is then determined by the balance of interaction and external potential, 
\begin{equation}\label{eqThomasFermi}
n_{\rm TF}(\r) = \frac{1}{g} [ \mu - V(\r) ]
\end{equation}
where $V(\r)<\mu$ and $n_{\rm TF}(\r)=0$ else. 
But in a potential that varies on short length scales, the kinetic energy term becomes relevant, 
and the TF result no longer suffices.

If the external potential is small, $V\ll  g n \approx \mu$, the imprint on the condensate amplitude can be computed 
perturbatively \cite{Sanchez-Palencia2006}. We expand  
\begin{align}\label{eqSmoothingExpansionGPGs}
 \Phi(\r) =  \Psitd0 + \Psitd1(\r) + \Psitd2(\r) + \ldots,
\end{align}
around the homogeneous solution $\Psitd0 =\sqrt{n}$ in powers of
the small parameter $V/\mu \ll 1$.
In order to maintain a fixed average  particle density $L^{-d}\intddr
|\Phi(\r)|^2 = n$, also 
the chemical potential is adjusted at each order, 
\begin{align}
\mu &= \mu^{(0)} + \mu^{(1)}+ \mu^{(2)} + \ldots, & \mu^{(0)} &= g n . 
\end{align}
We insert these expansions into Eq.~\eqref{eqStationaryGPE} and
collect orders up to $V^2/\mu^2$. 
Because the kinetic energy $\epn{k} = \hbar^2k^2/2m$ is diagonal in $k$-space, 
   solving for the $\Psitd{i}$ and
$\mu^{(i)}$ is best done in momentum representation, 
$\Phi_\k= \bra{\k}\Phi\rangle= L^{-d/2}\intddr e^{- i \k\cdot\r} \Phi(\r)$ and  
$V_\k = \bra{\k +\k'}V\ket{\k'} = L^{-d} \intddr
e^{-i\k\cdot\r}V(\r)$.  

The first-order imprint of the potential in the condensate amplitude reads
\begin{equation}\label{eqSmoothingPsi1}
\Psitd1_\k  = -\frac{ (1-\delta_{\k 0})V_\k}{ \epn{k}+2 g n} N^{1/2} . 
\end{equation} 
As expected for linear response, the shift is directly proportional to the potential's matrix element
$V_\k$. The Kronecker delta stems from the first-order shift
$\mu^{(1)}=V_0$ that compensates the potential average, 
ensuring $\Psitd1_0 = 0$ as required by conservation of average
particle density.  

In the denominator, the comparison between interaction $g n$ and kinetic energy
defines a characteristic length scale of the BEC, the \emph{healing length} $\xi = \hbar/\sqrt{2 m g n}$.
Factoring out $g n$, one is left with $2+\epn{k}/g n= 2+k^2\xi^2$ in the denominator. This term becomes constant for
long-range potential variations with $k\xi \to 0$, and
one recovers the
TF imprint  \eqref{eqThomasFermi} for the
density $n(\r) = |\Phi(\r)|^2 = n + n^{(1)}(\r)$, in Fourier components $n^{(1)}_{\k\text{TF}} =-V_\k/g$. 
In the contrary case $k\xi\gg1$, this denominator suppresses short-scale potential
variations. Indeed, the condensate avoids rapid
variations, which cost too much kinetic energy, and responds only to
a smoothed component of the external potential \cite{Sanchez-Palencia2006}. 

\begin{figure}[tbp]
 \centerline{\includegraphics[angle=270,width=\linewidth]{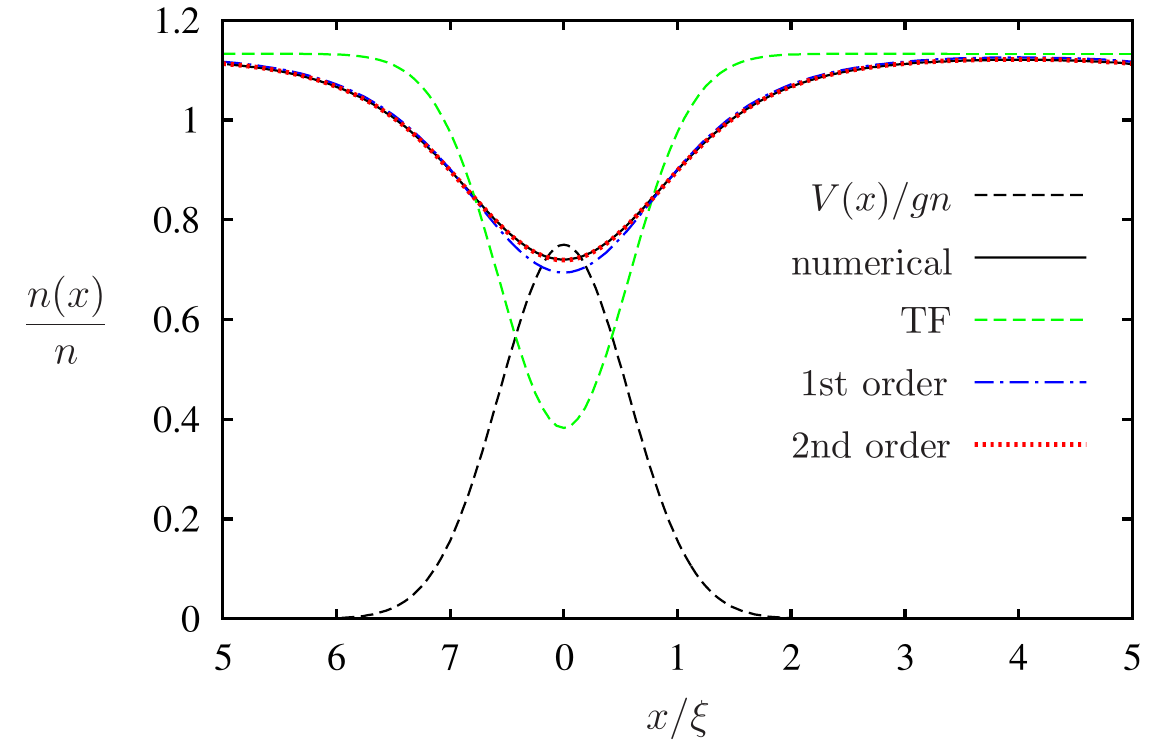}}
\caption[Condensate density profile in presence of an impurity]
{(Color online)  Condensate density $n(x)$ deformed by an impurity potential 
 $V(x) = V\exp(-x^2/\sigma^2)$ (dashed black) 
 with $V=0.75 g n$ and $\sigma= 0.8\xi$.
The numerical solution of the GP equation \eqref{eqStationaryGPE} 
[solid black, under periodic boundary conditions within the shown
interval] 
differs significantly from the TF result \eqref{eqThomasFermi}
 [dashed green]. Including first-order \eqref{n1k} and second-order smoothing
  \eqref{n2k} terms improves
 the agreement.\label{fig_smoothing}} 
\end{figure}

\autoref{fig_smoothing} shows a 1D real-space plot of the condensate
density deformed by a rather strong Gaussian impurity potential
of width $\sigma = 0.8\xi$. For such a small impurity, the full numerical
solution differs greatly from the simple TF formula
\eqref{eqThomasFermi}.
The first-order smoothing result \eqref{eqSmoothingPsi1} already gives much better agreement.
However, we need to push the expansion even further, in order to obtain consistent second-order results later on.

%-------------------------------
 %---------------------------

Solving for the second order imprint brings about terms of two different types. Viewing the GP equation \eqref{eqStationaryGPE} as a scattering equation for the field $\Phi$  \cite{Wellens2009, GaulPhD2010}, one finds first a contribution from double scattering by the external potential $V(\r)$ with free propagation in between. 
Secondly, there is a contribution from the interaction of two single-scattered amplitudes $\Phi^{(1)}$.
Altogether, including the chemical potential shift, the second-order condensate deformation reads
\begin{align}\label{eqSmoothingPsi2}
 \Phi^{(2)}_\k 
   &= \frac{1}{N^{1/2}} \sum_\p \Phi^{(1)}_{\k-\p} \Phi^{(1)}_{\p}  \frac{ (1-\delta_{\k0})\epn{p} -g n}{2 g n + \epn{k}} .
\end{align}

For future reference, we also write down the leading-order terms for the Fourier components of the condensate density $n(\r) = |\Phi(\r)|^2$, 
the inverse field $\check \Phi(\r) = n/\Phi(\r)$, as well as the inverse density $\check
n(\r) = n^2/n(\r)$.
To linear order, one has of course $\check\Phi^{(1)}_\k = -  \Phi^{(1)}_\k$ as well as 
\begin{equation}\label{n1k}
n^{(1)}_\k =  -\check
n^{(1)}_\k = -2 n \frac{ (1-\delta_{\k 0})V_\k}{2 g n + \epn{k}},
\end{equation} 
using the Fourier convention $n_\k = L^{-d} \intddr e^{-i \k\cdot\r} n(\r)$ for $n$ and $\check n$, in the same way as for $V$.
Eq.~\eqref{n1k} is the linear response of the condensate
density to the external potential \cite{Giorgini1994}. 

To second order, one finds  
\begin{align}
\check \Phi^{(2)}_\k
   &= \frac{1}{N^{1/2}} \sum_\p \Phi^{(1)}_{\k-\p} \Phi^{(1)}_{\p} \frac{\epn{k} -(1-\delta_{\k0}) \epn{p} + 3 g n}{2 g n + \epn{k}} ,\\
n^{(2)}_\k 
   &= \frac{1-\delta_{\k0}}{L^d}\sum_\p \Phi^{(1)}_{\k-\p} \Phi^{(1)}_{\p} \frac{ 2\epn{p}+\epn{k}}{2 g n + \epn{k}}, \label{n2k}\\
\check n^{(2)}_\k 
   &= \frac{1}{L^d}\sum_\p \Phi^{(1)}_{\k-\p} \Phi^{(1)}_{\p} \frac{3\epn{k} - 2(1-\delta_{\k0})\epn{p} + 8 g n}{2 g n + \epn{k}}.
\end{align}
In \autoref{fig_smoothing}, the results of second-order smoothing are
practically indistinguishable from the full solution of the GP equation.

We note at last that even an external potential with zero mean causes
a negative shift of the chemical potential, 
\begin{equation}\label{eqMu2}
\mu^{(2)} 
= - \frac{1}{N} \sum_{\q} \epn{q}|\Psitd{1}_\q|^2 
= -\sum_\q \epn{q} \frac{ |V_\q|^2 (1-\delta_{\q 0})}{(2g n+ \epn{q})^2}.
\end{equation}
A negative chemical potential shift must occur non-perturbatively, as can be seen by spatially integrating  
the GP equation \eqref{eqStationaryGPE}  after dividing
by $\Phi(\r)$: the positivity of the kinetic energy entails that the
chemical potential shift, at fixed average particle density, must be
negative  \cite{Lee1990}. 

This concludes our calculation of the inhomogeneous GP ground state,
and we turn to the fluctuations around this condensate.

%------------------------------------------------------------
\subsection{Bogoliubov Excitations}\label{secSaddlepoint}

Using the Bogoliubov ansatz \eqref{eqBogoliubovPrescription}, we
expand the Hamiltonian \eqref{eqManyParticleHamiltonian} in powers of $\delta \hat\Psi$ and $\delta \hat\Psi^\dagger$. 
To zeroth order, we find the GP ground-state energy $E_0 = E[\Phi(\r)]$.
The linear term vanishes, because $\Phi(\r)$ minimizes the energy functional \eqref{eqGPEnergyFunctionalPsi}. 
The relevant contribution is then the quadratic part, $\hat E=E_0+\hat H$. 
Third-order and forth-order terms in the fluctuations are neglected. They describe interaction between the excitations and become only relevant for larger densities or higher temperatures \cite{Katz2002}. 

For reasons that will become clear in Sec.~\ref{secBasis} below, the inhomogeneous Bogoliubov Hamiltonian is best expressed in density-phase variables.
From
$\hat\Psi = \exp\{i\delta\hat\varphi \} \sqrt{n+\delta \hat n}
= \Phi+ \delta\hat n/2\Phi +i\Phi \delta\hat \varphi+\dots $ follows 
\begin{subequations}\label{eq_dPsi_dn_dphi}
\begin{align}
 \delta \hat n (\r)       &= \Phi(\r)             \left\lbrace \delta \hat \Psi^\dagger(\r) + \delta \hat \Psi(\r) \right\rbrace , \\
 \delta \hat \varphi (\r) &= \frac{i}{2\Phi(\r)}  \left\lbrace \delta \hat \Psi^\dagger(\r) - \delta \hat \Psi(\r) \right\rbrace ,
\end{align}
\end{subequations}
up to higher orders in $\delta \hat\Psi$. The commutators
\eqref{commutators.eq} imply that density and phase (fluctuation) operators are conjugate, 
$\bigl[\delta \hat n(\r), \delta \hat \varphi(\r') \bigr] = i 
\delta(\r-\r')$. 
By expanding the many-body Hamiltonian to second order in the
fluctuations around the mean-field solution, the relevant Hamiltonian
for the excitations is found as
\cite{Gaul2008}
\begin{align}
\hat H  
= \int {\rm d}^d r \biggl\lbrace 
\frac{\hbar^2}{2m} \biggl[  &\left(\vnab \frac{\delta \hat n}{2\Phi(\r)} \right)^2
+ \frac{\left[\nabla^2 \Phi(\r)\right]}{4 \Phi^{3}(\r)} \delta \hat n^2 \nonumber \\
& + \Phi^2(\r) (\vnab \delta \hat \varphi)^2\biggr] + \frac{g}{2} \,\delta \hat n^2 \biggr\rbrace .\label{eqBogoliubovHamiltonian}
\end{align}
With Eq.~\eqref{eqBogoliubovHamiltonian}, the problem is reduced to a Hamiltonian that is quadratic in the excitations. 
To this order, there are no mixed terms of $\delta \hat n$ and $\delta \hat \varphi$.
The perturbing potential $V(\r)$ does not appear directly.
Instead, it enters nonlinearly via the condensate function
$\Phi(\r)$, which can be predetermined 
by solving the GP equation \eqref{eqStationaryGPE} or calculated
perturbatively, as explained in the previous Sec.~\ref{secMeanfield}. 

Before further discussing the impact of the external potential, we briefly consider the excitations of the 
homogeneous system.

%------------------------------------------------------------
\subsubsection{Homogeneous Bogoliubov Hamiltonian}\label{secFreeBogoliubov}
\begin{figure}[tb]%
\centerline{\includegraphics[angle=-90,width=\linewidth]{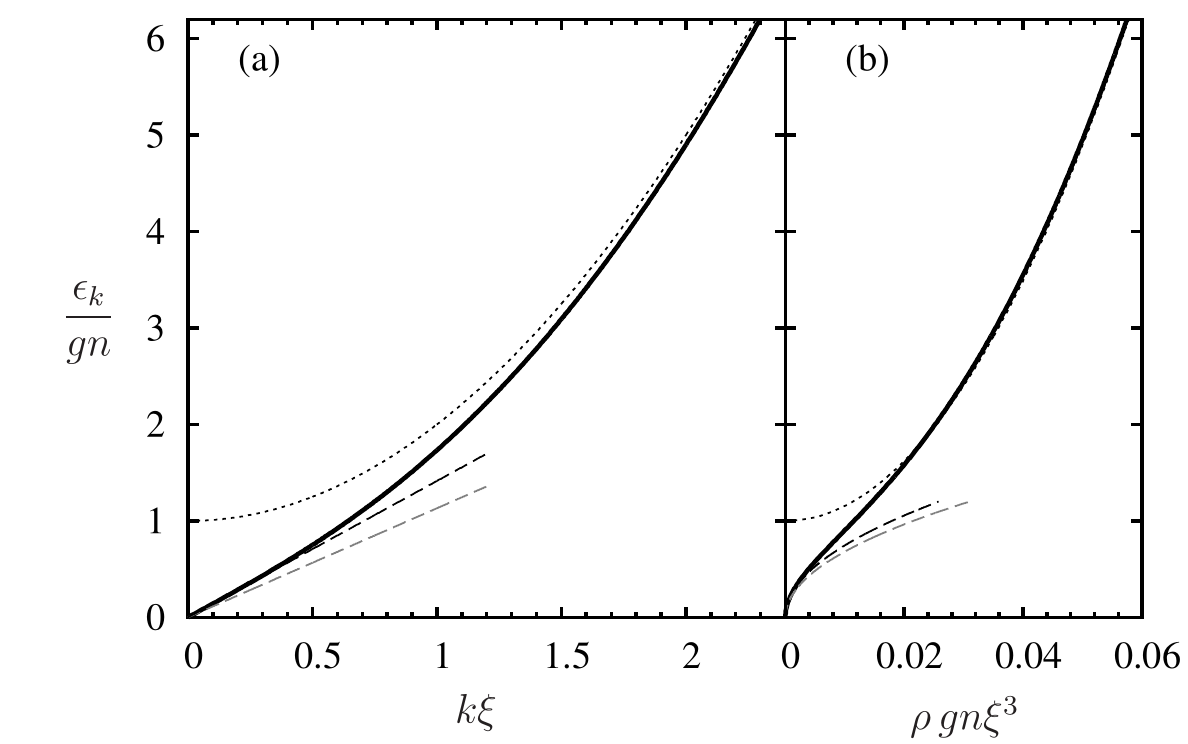}}
\caption{Bogoliubov excitations in the homogeneous condensate. 
(a) Bogoliubov dispersion relation \eqref{eqBgDispersion} (solid).
(b) Density of states \eqref{eqDOS_Bg} in three dimensions (solid).
In both panels, dashed and dotted lines show the low-energy and high-energy asymptotics, respectively.
The gray dashed lines foreshadow the disorder corrections provided in
Sec.~\ref{secResults}.} 
\label{figCleanBogoliubov}
\end{figure}

In the homogeneous case  $V(\r) = 0$, the Bogoliubov Hamiltonian \eqref{eqBogoliubovHamiltonian} becomes translation invariant and thus diagonal in the momentum representation  
$\delta \hat n_\k = L^{-d/2}\intddr e^{-i\k\cdot\r} \delta\hat n(\r)$ 
and $\delta \hat \varphi_\k=L^{-d/2}\intddr e^{-i\k\cdot\r} \delta\hat \varphi(\r)$: 
\begin{align}
\hat H^{(0)}
&= \sum_{\k} 
\left[
n \epsilon^0_{k} \, {\delta\hat \varphi_{\k}^\dagger}{\delta\hat \varphi_\k}
+ \frac{2g n+\epsilon^0_{k}}{4n}\,\delta \hat n_\k^\dagger\delta \hat n_\k 
\right]  \label{eqFreeGP}
\end{align}
with $\epn{k} = \hbar^2 k^2/2m$.
This Hamiltonian looks diagonal, but the Heisenberg equations of motion for 
$\delta \hat n_\k=\delta\hat n_{-\k}^\dagger$ and $\delta \hat \varphi_\k=\delta\hat\varphi_{-\k}^\dagger$, 
which obey
\begin{equation}\label{commutation_deln_delphi_k.eq} 
[\delta\hat n_\k,\delta\hat\varphi_{\k'}^\dagger]=i\delta_{\k\k'}, 
\end{equation}
are still coupled. 
This is resolved by a Bogoliubov transformation \cite{Bogoliubov1947}, coupling density and phase fluctuations to quasiparticle creation and annihilation operators $\hat \gamma^\dagger_\k$  and $\hat \gamma_\k$:
\begin{equation}\label{eqBgTrafo1}
\cvect{\gh{\k}}{\ghd{-\k}}
 = A_k 
\cvect{i\sqrt{n}\,\delta \hat \varphi_{\k}}{\delta \hat
  n_\k/2\sqrt{n}}, \quad A_k  =
\matr{a_k}{a_k^{-1}}{-a_k}{a_k^{-1}}. 
\end{equation}
A transformation of this kind, with the free parameter $a_k$, guarantees that the quasiparticles obey bosonic commutation relations 
\begin{align}\label{eqCommutatorGamma}
 \bigl[\gh{\k},\ghd{\k'}\bigr] &= \delta_{\k \k'}, &
 \bigl[\gh{\k},\gh{\k'}\bigr] =  \bigl[\ghd{\k},\ghd{\k'}\bigr] &= 0  .
\end{align}
The Hamiltonian \eqref{eqFreeGP} becomes diagonal,  
\begin{equation}\label{eqBgHamiltonianDiagonal}
\hat H^{(0)} = \sum_{\k} \, \ep{k} \ghd{\k}\gh{\k} ,
\end{equation}
by choosing 
\begin{equation}\label{eq_ak}
a_k = \left(\frac{\epsilon_k}{\epsilon^0_k}\right)^{1/2} = \left(\frac{k^2\xi^2}{2+k^2\xi^2}\right)^{\frac{1}{4}} .
\end{equation}
The excitations are found to have the famous Bogoliubov dispersion relation \cite{Bogoliubov1947}
\begin{equation}\label{eqBgDispersion}
\ep{k} = \sqrt{\epn{k}(2g n+\epn{k})} = g n\, k \xi \sqrt{2+k^2\xi^2},
\end{equation}
plotted in \autoref{figCleanBogoliubov}(a) for reference. 

%------------------------------
%------------------------------

In the high-energy or large-momentum regime $k\xi\gg1$, 
the excitations are essentially free particles with dispersion $\epn{k}$, 
shifted by the condensate background energy, 
\begin{equation}\label{cleanBgPartcl}
\ep{k} \approx \epn{k} + g n . 
\end{equation} 
In the low-energy regime, the interaction dominates over the bare
kinetic energy. A single excitation involves many individual
particles, comparable to a classical sound wave. Indeed, 
the dispersion relation is linear, $\ep{k} = \hbar c k $, with the bare sound velocity $c=\sqrt{g n/m}$.
According to a classical argument due to Landau \cite[chapter 10.1]{Pethick2002}, 
this linear dispersion at low energies implies superfluidity with critical velocity 
$v_c = \min_k(\ep{k}/{\hbar k}) = c$.

The transition from sound-wave to single-particle excitations also reflects in the \emph{density of states} (DOS)
\begin{align}\label{eqDOS_Bg}
 \rho(\epsilon) &= \frac{S_d k_\epsilon^{d-1}}{(2\pi)^d} 
 \left|\pder{k}{\epsilon}\right|_{k_\epsilon} 
\Theta(\epsilon)  
\end{align}
with  $S_d = 2,2\pi,4\pi$ the surface of the $d$-dimensional sphere in
$d=1,2,3$, respectively. The $k$-vector at energy $\epsilon$ is given by $k^2_\epsilon \xi^2 =
[1+(\epsilon/g n)^2]^{1/2}-1$. 
Eq.~\eqref{eqDOS_Bg} shows the transition from a sound-wave DOS
$\rho_{\rm sw}(\epsilon) \propto \epsilon^{d-1}$ to particle DOS 
$\rho_{\rm pt}(\epsilon+g n) \propto \epsilon^{\frac{d}{2}-1}$, as
illustrated in \autoref{figCleanBogoliubov}(b) for $d=3$.

%%%%%%%%%%%%%%%%%%%%%%%%%%%%%%%%%%%%%%%%%%%%%%%%%%%%%%%%%%%%%%%%%%%%%%%%%%%%%%
\subsubsection{Inhomogeneous Bogoliubov  Hamiltonian}\label{secInhomBgHamiltonian}

Let us come back to the Hamiltonian
\eqref{eqBogoliubovHamiltonian} including 
the full imprint of $V(\r)$ in the condensate $\Phi(\r)$. 
This inhomogeneity breaks translation invariance, so the Hamiltonian cannot be diagonal
in momentum representation. However, it is still quadratic  
in the fluctuations without any term mixing $\delta \hat n$ and $\delta \hat
\varphi$, with the general structure 
\begin{align}\label{eqEnergyFunctionalStructure}
\hat H 
&= \frac{1}{2}\sum_{\k,\k'}
\biggl\lbrace
 n\, \delta \hat \varphi_\k^\dagger S_{\k\k'} \delta \hat\varphi_{\k'}
+ \frac{1}{4n}  \delta \hat n_\k^\dagger  R_{\k \k'} {\delta \hat n_{\k'}}
 \biggr\rbrace.
\end{align}
The coupling matrices $S_{\k\k'}$ and $R_{\k \k'}$ contain the
relevant information about the Fourier components of the condensate
field $\Phi(\r)$ and its inverse $\check\Phi(\r) = n/\Phi(\r)$, as
well as of their gradients.

There is only a single term involving the phase gradients in the
Hamiltonian \eqref{eqBogoliubovHamiltonian}, proportional to the
density. Upon Fourier transformation, the coupling matrix reads   
\begin{align}\label{eqScattSgeneral}
 S_{\k\k'} 
=\frac{\hbar^2}{mn} \k\cdot\k'  n_{\k-\k'} .  
\end{align}
Its diagonal elements are $S_{\k\k}=2 \epn{k}$, to all orders of
$V(\r)$, by conservation of average density. In the homogeneous case,
it reduces to $S^{(0)}_{\k\k'} = 2\epn{k}\delta_{\k\k'}$. As a function
of the condensate field components, it can be rewritten as 
\begin{align}\label{eqScattSgeneral.fields}
 S_{\k\k'} 
= \frac{2g}{L^d} \sum_{\p} \k\cdot\k' \, \xi^2 \, \Phi_{\k-\p}\Phi_{\p-\k'}.
\end{align}

In the Hamiltonian \eqref{eqBogoliubovHamiltonian},  the density
fluctuation $\delta\hat n$
appears in several, complicated looking terms. But the corresponding
coupling matrix $R_{\k\k'}$ can be
brought in a form very similar to \eqref{eqScattSgeneral.fields} by
using the Fourier components of the inverse field: 
\begin{equation}\label{eqScattRgeneral}
 R_{\k \k'} =  \frac{2g}{L^d} \sum_{\p} \tilde r_{\k \p \k'} \,
 \check\Phi_{\k-\p}\check\Phi_{\p-\k'} + 4g n \delta_{\k\k'},
\end{equation}
with 
$\tilde r_{\k\p\k'} 
= \bigl[ p^2 + 2(\k'-\p)\cdot(\k-\p) + \frac{1}{2}(\k'-\p)^2+\frac{1}{2}(\k-\p)^2\bigr] \xi^2$.
In the homogeneous case, it reduces to $R^{(0)}_{\k\k'} =
2(2gn+\epn{k})\delta_{\k\k'}$. In the inhomogeneous case, 
the background-mediated coupling between fluctuations, as expressed by Eqs.\ \eqref{eqScattSgeneral} and
\eqref{eqScattRgeneral},  
is non-perturbative in the potential strength [as long as the
Bogoliubov ansatz \eqref{eqBogoliubovPrescription} is valid]. These expressions hold
for arbitrary potentials, if only the condensate $\Phi$ and its
inverse $\check\Phi$ are correctly determined. 
Notably, the commutation relations
\eqref{commutation_deln_delphi_k.eq} for the fluctuation operators, as 
defined in Eq.\ \eqref{eq_dPsi_dn_dphi}, remain valid in the inhomogeneous setting. 

\begin{figure}
 \includegraphics[width=0.8\linewidth]{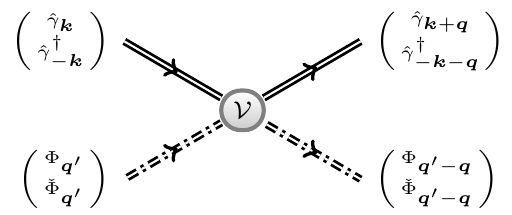}
\caption[Bogoliubov scattering vertex]{Universal Bogoliubov scattering
  vertex \eqref{eqDefWY}. Bogoliubov excitations $(\gh{\k},\ghd{-\k})$
  are scattered by an effective vertex, nonperturbatively determined by the
  deformed condensate and its inverse, 
$(\Phi_\q, \check \Phi_\q$).}\label{figBgScattUniversal} 
\end{figure}

For  further analysis in terms of Bogoliubov quasiparticles, we
transform to the Bogoliubov basis \eqref{eqBgTrafo1} of the
homogeneous case. 
We separate the homogeneous contribution from the inhomogeneous 
contribution in the coupling matrices, $S_{\k\k'} = S_{\k\k'}^{(0)} +
S_{\k\k'}^{(V)}$ and $R_{\k\k'} = R_{\k\k'}^{(0)} +
R_{\k\k'}^{(V)}$ and define the effective Bogoliubov excitation
scattering vertex
\begin{equation}\label{eqDefWY}
\frac{(A_k^{-1})^\text{t}}{2}
\matr{S_{\k\k'}^{(V)}}{0}{0}{R_{\k\k'}^{(V)}} 
\frac{A_{k'}^{-1}}{2}
 = \matr{W_{\k\k'}}{Y_{\k\k'}}{Y_{\k\k'}}{W_{\k\k'}} 
=:\V_{\k\k'} ,
\end{equation}
depicted in Fig.\ \ref{figBgScattUniversal}.
This brings the inhomogeneous Bogoliubov Hamiltonian
\eqref{eqBogoliubovHamiltonian} in the form
$\hat H= \hat H^{(0)} +\hat H^{(V)}$ or  
\begin{equation}\label{eqInhomBgHamiltonian_gamma}
\hat H = \sum_{\k} \, \ep{k} \ghd{\k}\gh{\k} +
\frac{1}{2} \sum_{\k,\k'} 
(\ghd{\k},\gh{-\k})  \begin{pmatrix}W& Y\\ Y& W\end{pmatrix}_{\k\k'} \cvect{\gh{\k'}}{\ghd{-\k'}}.
\end{equation}

%-------------------------
%------------------------

Let us reflect on what has been achieved at this point. By a saddle-point expansion of the general many-body Hamiltonian
\eqref{eqManyParticleHamiltonian}, we have derived the Hamiltonian
describing the dynamics of Bogoliubov excitations in inhomogeneous
external potentials. These excitations
are defined as in the homogeneous case, with now an inhomogeneous
contribution to the Hamiltonian, 
the coupling matrix $\V_{\k\k'} $, that provokes
scattering between different $\k$-modes. This coupling has
a much richer structure than a simple potential scattering term 
$V_\q \hat a^\dagger_{\k+\q} \hat a_\k$ for single particles. It is
both nonlinear in the potential and contains off-diagonal
contributions, because the underlying
condensate background depends nonlinearly on the potential and
mediates anomalous scattering between quasiparticle excitations.  
Nevertheless, we have managed to identify the relevant scattering
vertex $\mathcal V$, which allows us to
set up a systematic perturbation theory. As a first step for fully
analytical calculations, we have to expand the scattering vertex to
lowest orders in $V$.

%----------------
\subsubsection{Perturbative expansion of the Bogoliubov scattering
  vertex}
\label{ssecExpansionV}
We expand the scattering matrix elements in powers of the
inhomogeneous potential $V_\k$, 
using the smoothing theory exposed in
Sec.~\ref{secMeanfield}: 
\begin{align}
  \includegraphics{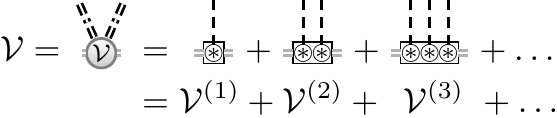}\label{eqExpansionV}
\end{align}
Each dashed dangling line here represents the bare external
potential $V_\q$ [not to be confounded with the dash-dotted double
lines representing the background condensate fields $(\Phi_\q,
\check \Phi_\q)$ in Fig.\ \ref{figBgScattUniversal}]. 
The scattering vertices $\V^{(j)}$ of order $j$
can be derived systematically by 
\begin{enumerate}
\item[(i)]   computing the ground state \eqref{eqSmoothingExpansionGPGs} to the desired order;
\item[(ii)]  computing the inverse field $\check\Phi_\q  = (n/\Phi)_\q$;
\item[(iii)] collecting all terms of order $j$ in Eqs.~\eqref{eqScattSgeneral} and \eqref{eqScattRgeneral} in order to obtain $S^{(j)}$ and $R^{(j)}$,
and 
\item[(iv)] applying the transformation \eqref{eqDefWY} in order to obtain $W^{(j)}$ and $Y^{(j)}$.
\end{enumerate}
Let us make this procedure explicit for the first orders $j=1,2$.

The first-order scattering amplitudes  
\begin{equation} \label{W1Y1}
W^{(1)}_{\k \k'} = w^{(1)}_{\k \k'} V_{\k-\k'}, \quad Y^{(1)}_{\k \k'} = y^{(1)}_{\k \k'} V_{\k-\k'}, 
\end{equation} 
are both directly proportional to the potential's matrix element
$V_{\k-\k'}$, as required by conservation of momentum. All information
about the interaction and the background condensate is factorized into
the envelope functions 
\begin{subequations}\label{w1y1}
\begin{align}
w^{(1)}_{\k \k'} = 
\frac{ (1-\delta_{\k \k'}) a_k a_{k'}\xi^2} {2+\xi^2(\k'-\k)^2}
\left[ k^2 + k'^2 -\k\cdot\k'
- \frac{\k\cdot\k'}{a_k^{2} a_{k'}^{2}} \right]  \label{eqW1}, 
\end{align}
\begin{align}
y^{(1)}_{\k \k'} = 
\frac{ (1-\delta_{\k \k'}) a_k a_{k'}\xi^2} {2+\xi^2(\k'-\k)^2}
\left[ k^2 + k'^2 -\k\cdot\k'
+ \frac{ \k\cdot\k'}{a_k^{2} a_{k'}^{2}} \right]
 \label{eqY1},
\end{align}
\end{subequations}
with $a_k$ from Eq.~\eqref{eq_ak}.
We have previously studied the scattering of Bogoliubov excitations
by an isolated impurity, as described by these matrix
elements \cite{Gaul2008}. An interesting feature of the transition from sound-like to
particle-like excitations is that the amplitude for elastic scattering
$|\k|=|\k'|$ by an angle $\theta$ is proportional to the remarkably simple envelope
function  
\begin{equation}\label{eqElastScattAmpl}
\left.w^{(1)}_{\k\k'}\right|_{k'=k} =  \frac{\epn{k}}{\ep{k}}
\underbrace{\frac{k^2\xi^2(1-\cos\theta) - \cos\theta } {k^2\xi^2(1-\cos\theta) + 1}}_{\displaystyle{=:A(k\xi,\theta)}}  (1-\delta_{\k \k'})
. 
\end{equation} 
This envelope is responsible for a node in the scattering amplitude at
$\cos\theta_0=k^2\xi^2/(1+k^2\xi^2)$, thus interpolating between the p-wave
scattering of a sound wave with $\theta_0=\pi/2$ and the s-wave
scattering of a single particle \cite{Gaul2008}.

The second-order couplings 
\begin{align}
  S^{(2)}_{\k\k'} &= \frac{1}{g n}\sum_{\p} s^{(2)}_{\k \p
    \k'}V_{\k-\p} V_{\p-\k'}, \\
  R^{(2)}_{\k\k'} &=  \frac{1}{g n}  \sum_{\p} r^{(2)}_{\k \p \k'} V_{\k-\p} V_{\p-\k'}
\end{align}
feature the kernels 
\begin{widetext}
\begin{align}
s^{(2)}_{\k \p \k'} 
 &= 2\xi^2 \k \cdot \k'
    \frac{ [(\k-\k')^2+(\k-\p)^2+(\p-\k')^2]\xi^2}
         {[2+(\k-\k')^2\xi^2][2+(\k-\p)^2\xi^2][2+(\p-\k')^2\xi^2]} (1-\delta_{\k\k'})(1-\delta_{\k\p})(1-\delta_{\p\k'}) , \\
r^{(2)}_{\k \p \k'}
 &= 2\xi^2 \biggl\lbrace 
p^2 + 2(\k-\p)\cdot(\k'-\p) + (\k-\p)^2  \nonumber \\ & \qquad 
+ 2(k^2 +k'^2 -\k\cdot\k') 
\frac{3+ (\k-\k')^2\xi^2 - \xi^2 (\k-\p)^2 (1-\delta_{\k\k'})}{2+(\k-\k')^2\xi^2}
\biggr\rbrace \,
\frac{1-\delta_{\k\p}}{2+(\k-\p)^2\xi^2} \ \frac{1-\delta_{\p\k'}}{2+(\p-\k')^2\xi^2} .
\end{align}
\end{widetext}
Later, the ensemble average over the disorder 
will restore translation invariance. The relevant 
diagonal elements are $s^{(2)}_{\k \p \k} = 0 $ (as required by Eq.\ \eqref{eqScattSgeneral} and conservation of particle number) and 
\begin{align}\label{eqEnv2Diag}
 r^{(2)}_{\k \p \k} = 2 \xi^2 \frac{p^2 + 3 (\k-\p)^2 + 3 k^2}{[2+(\k-\p)^2\xi^2]^2} (1-\delta_{\k\p}).
\end{align}
Finally, the matrix elements are transformed according to Eq.~\eqref{eqDefWY},
which yields the second-order diagonal scattering amplitudes 
\begin{align}\label{eq_W2Y2}
  W^{(2)}_{\k \k} = Y^{(2)}_{\k \k} 
 &= \sum_{\p}   w^{(2)}_{\k \p \k}  V_{\k-\p} V_{\p-\k} 
\end{align}
where $w^{(2)}_{\k\p\k}=a_k^2  r^{(2)}_{\k \p \k}/4g n$ in terms of Eqs.\ \eqref{eq_ak} and \eqref{eqEnv2Diag}.

\subsubsection{One-dimensional setting}\label{ssec1DTransmission}

Before proceeding with the general theory, we briefly digress into one
dimension to discuss the link
to previous works. 
 Using the general formalism developed so
far, we can easily compute
the reflection of a Bogoliubov excitation
by a delta-like impurity $V(x) =
V \sigma_0 \delta(x)$ to lowest order in $V \sigma_0$.  In one dimension only
exact backward scattering occurs, and the problem involves the coupling element
$W_{-k,k}$. In the Born approximation, to second order in $V$, we
find the transmission 
\begin{align}\label{eqScattering1DBorn}
T = 1 - \frac{V^2}{4g^2n^2 }\frac{k^2\sigma_0^2}{(k^2\xi^2+1)^2} .
\end{align}
The impurity becomes perfectly transparent at long wave lengths
$k\sigma_0\to0$. At the
crossover $k\xi\approx1$ from sound waves to particles, strong backscattering leads
to a transmission minimum, whereas at high energies the transmission increases again. 
The result \eqref{eqScattering1DBorn} is in quantitative
agreement with the results of Bilas and Pavloff \cite{Bilas2006},
taken in the limit of a weak impurity. It may be useful to note that within our formalism, we do not require the
explicit knowledge of Bogoliubov eigenstates, nor distinguish between propagating
and evanescent modes; all the physics is built into the effective
scattering vertex $\V$. 

Also Kagan, Kovrizhin, and Maksimov \cite{Kagan2003} have considered tunneling
across an impurity, which in their case suppressed the condensate density very
strongly. We agree in the aspect of perfect transmission at low
energies. At high energies, however, Kagan \textit{et al.}\ do not find a revival of transmission.
This is reasonable because their strong impurity deeply depresses the
condensate on the spatial scale of $\xi$, and for wave lengths shorter
than $\xi$, transmission remains suppressed.

%%%%%%%%%%%%%%%%%%%%%%%%%%%%%%%%%%%%%%%%%%%%%%%%%%%%%%%%%%%%%%%%%%
\subsubsection{Appropriate basis for the inhomogeneous Bogoliubov problem}\label{secBasis}

\newcommand{\dPsi}{\delta\hat\Psi(\r)}
\newcommand{\dPsid}{\delta\hat\Psi^\dagger(\r)}

Before deriving physical quantities from the effective Hamiltonian
\eqref{eqInhomBgHamiltonian_gamma}, which will be the subject of the
following section \ref{secDispersion}, it remains to justify our choice of basis for 
inhomogeneous Bogoliubov excitations. 

The Hamiltonian \eqref{eqBogoliubovHamiltonian}
is quadratic in the fluctuations, no matter whether written in
terms of the single-particle basis $\dPsi,\dPsid$ or the hydrodynamic
basis $\delta\hat n(\r),\delta\hat\varphi(\r)$, and thus 
can always be diagonalized:
$\hat H = \sum_\nu \hbar\omega_\nu \hat\beta_\nu^\dagger \hat \beta_\nu$.
Here, the eigenmodes $\nu$  are populated by
bosonic quasiparticles,  for which 
\begin{align}\label{eqBgDiadonalInverted}
 \hat\beta_\nu &= \intddr\left[ u_\nu^*(\r) \dPsi + v_\nu^*(\r) \dPsid \right] ,\\
\label{eqBosonicCommutator}
 [\hat \beta_\mu,\hat \beta_\nu^\dagger] &= \delta_{\mu\nu} , \qquad
 [\hat \beta_\mu^\dagger,\hat \beta_\nu^\dagger] = [\hat
 \beta_\mu,\hat \beta_\nu] = 0. 
\end{align}
The eigenfunctions $u_\nu(\r)$  and  $v_\nu(\r)$ are solutions of the Bogoliubov-de~Gennes equation, a non-Hermitian eigen\-value problem:
\begin{align}
\left[ \matr{H(\r)}{g n(\r)}{g n(\r)}{H(\r)}\sigma_3 - \hbar \omega_\nu \right] \cvect{u_\nu(\r) }{v_\nu(\r)} = 0,\label{eqBg-de-Gn}
\end{align}
with $H(\r) = -\frac{\hbar^2}{2m}\nabla^2+V(\r)-\mu  + 2 g n(\r)$ and the Pauli matrix $\sigma_3 =
\matrs{1}{0}{0}{-1}$.
In the case of broken translation symmetry, these modes are not
indexed by a wave vector $\k$. 
They do, however, fulfill the bi-orthogonality relation 
\begin{align}\label{eqBiOrt}
\int {\rm d}^d r \left[ u^*_\nu(\r) u_\mu(\r) - v^*_\nu(\r) v_\mu(\r) \right] = \delta_{\mu\nu} 
\end{align}
and the orthogonality with respect to the condensate \cite{Fetter1972}
\begin{align}\label{eqBiOrtGS}
\intddr \, \Phi^*(\r) \left[ u_\nu(\r) - v_\nu(\r) \right] = 0 \, .
\end{align}
This latter relation expresses the bi-orthogonality
\eqref{eqBiOrt} with respect to the  
zero-frequency Goldstone mode related to the spontaneously broken
$U(1)$ symmetry of the BEC \cite{Lewenstein1996,Goldstone1962}.
The orthogonality relations allow the inversion of Eq.~\eqref{eqBgDiadonalInverted}:
\begin{align}\label{eqBgDiagFormula}
\dPsi = \sum_\nu \bigl[u_\nu(\r) \hat\beta_\nu - v_\nu^*(\r) \hat\beta_\nu^\dagger \bigr].
\end{align}

In presence of external inhomogeneities, and in particular for a disorder potential to which we will turn shortly [Sec.~\ref{secDispersion}], this eigenbasis explicitly depends on each
potential realization, which renders it useless for analytical
calculations. 
Instead, we construct a basis starting from the plane waves that diagonalize the
clean Hamiltonian, while satisfying the 
orthogonality relations \eqref{eqBiOrt} and \eqref{eqBiOrtGS}, \emph{even in the inhomogeneous case}.  
The price to pay for using a plane-wave index is of course 
that the disorder leads to scattering between these eigenstates. 
But this is always the case in disordered systems, and standard perturbation theory applies. 

Still, one has essentially two choices. (i) One can define 
Bogoliubov operators by an expansion over single-particle plane-wave modes:
\begin{equation}
\dPsi =  \sum_{\k} \left( u^{(0)}_\k(\r) \gh{\k}^{(0)} -
  v_{\k}^{(0)}(\r)^*\gh{\k}^{(0)\dagger}\right) \label{eqBasedOnSingleParticle}  
\end{equation} 
with 
\begin{equation}\label{uv0.eq} 
u_\k^{(0)}(\r) = u_k L^{-d/2} e^{i\k\cdot\r}, \quad  v_\k^{(0)}(\r) = v_k
L^{-d/2} 
e^{i\k\cdot\r}.
\end{equation} 
 The coefficients $u_k =
\frac{1}{2}(a_k^{-1}+a_k)$ and $v_k =
\frac{1}{2}(a_k^{-1}-a_k)$ are custom-tailored to
satisfy the bi-orthogonality \eqref{eqBiOrt}  for all $\k \neq 0$, because $u_k^2-v_k^2 = 1$.
But if now the disorder potential is switched on, the condensate $\Phi(\r)$ is
deformed and not orthogonal to the plane waves anymore. Testing the condition \eqref{eqBiOrtGS}, we find
\begin{equation}
\intddr \, \Phi(\r) \bigl[ u_\k^{(0)}(\r) - v_\k^{(0)}(\r) \bigr] = (u_k - v_k) \Phi_{-\k} \neq 0 .
\end{equation}
This overlap with the ground state has disastrous consequences for the theory. 
If one tries to work with these operators, the coupling $W_{\k\k'}$ diverges for $k \rightarrow
0$, and perturbation theory will break down, no matter how small the
external potential. 

(ii) One can define the excitations via the
hydrodynamic fluctuations \eqref{eq_dPsi_dn_dphi}, 
\begin{equation}
 \dPsi =  \frac{\delta \hat n(\r)}{2 \Phi(\r)} 
         +  i \Phi(\r) \delta\hat\varphi(\r) . 
\end{equation}
Contrary to case (i), now the disorder is present from the outset,
such that the fluctuations originate from the disorder-shifted
reference point $\Phi(\r)$, which corresponds to the Bogoliubov vacuum of the
true excitations  \eqref{eqBgDiadonalInverted}. 
The inverse Bogoliubov transformation \eqref{eqBgTrafo1},  
\begin{align}
 \delta\hat n_\k & = a_k\sqrt{n}\left(\gh{\k} + \ghd{-\k}\right) ,
 \label{deltan_gk} \\   
\delta\hat \varphi_\k & = \frac{1}{2i a_k\sqrt{n}}\left(\gh{\k} -
  \ghd{-\k}\right),  \label{deltaphi_gk}
\end{align}
then leads to a decomposition of the form \eqref{eqBgDiagFormula},
\begin{equation}  \label{eqBasedOnHydro} 
\dPsi =  \sum_{\k} \left( u_\k(\r) \gh{\k} -
  v_{\k}(\r)^*\ghd{\k}\right), 
\end{equation}
over mode functions 
\begin{subequations}
\begin{align}
u_\k(\r) &= \frac{1}{2}\left(\frac{\Phi(\r)}{a_k\sqrt{n}}
+ \frac{a_k\sqrt{n}}{\Phi(\r)}\right) \frac{e^{i\k\cdot\r}}{L^{{d}/{2}}}, \\
v_\k(\r) &= \frac{1}{2}\left(\frac{\Phi(\r)}{ a_k\sqrt{n}}
- \frac{a_k\sqrt{n}}{\Phi(\r)}\right) \frac{e^{i\k\cdot\r}}{L^{{d}/{2}}}.
\end{align}
\end{subequations}
In the homogeneous case $\Phi=\sqrt{n}$, these functions reduce
exactly to the plane-wave amplitudes \eqref{uv0.eq}. 
In the inhomogeneous case, the plane waves are found to be modified in such a way that the modes still satisfy the bi-orthogonality \eqref{eqBiOrt}. 
Moreover, they also respect the orthogonality to the deformed ground state \eqref{eqBiOrtGS}, because
$\Phi(\r) \left[u_\k(\r) - v_\k(r) \right]$ is a plane wave with zero spatial average for all $\k \neq 0$.

In conclusion, the Bogoliubov quasiparticles defined in terms of density and phase via \eqref{deltan_gk} and
\eqref{deltaphi_gk}, or equivalently by \eqref{eqBgTrafo1}, fulfill all requirements for the study of the disordered Bogoliubov problem.
They can be labeled by a wave vector $\k$, which is independent of the disorder realization $V(\r)$,
they fulfill the required bi-orthogonality relation \eqref{eqBiOrt},
and most importantly, they decouple from the inhomogeneous condensate
ground state. 

%-------------------------------------------------------------------------------------------
\section{Modified excitation dispersion}
\label{secDispersion} 
%--------------------------------------------------------------------------------------------

In the previous section, we have set up the general formalism for
describing Bogoliubov excitations in a weak external potential, by
deriving the relevant Hamiltonian \eqref{eqInhomBgHamiltonian_gamma}
in the form $\hat H = \hat H^{(0)} + \hat H^{(V)}$, where $\hat H^{(0)}$ describes the
clean system, and $\hat H^{(V)}$ the disorder.
This structure permits using the machinery of perturbation theory \cite{Bruus2004,Mahan2000,Akkermans2007}.
Presently, we explore the consequences for the excitation dispersion relation and the corresponding
density of states. These quantities can be computed via 
zero-temperature single-excitation Green functions, for which we 
calculate the self-energy to order $V^2$. From the self-energy, we
determine physical quantities like mean free paths and corrections to
the speed of sound. Since we will mainly focus on the case
where $V(\r)$ is a disorder potential, we use a notation adapted to that scenario in the
following. But the general theory applies to
arbitrary potentials and notably covers the case of weak lattice
potentials, to which we devote a brief discussion in Sec.~\ref{lattice.sec} below.

%-------------------------------------------------------------------
\subsection{Green functions}\label{secGreenFunctions}

The matrix structure of the scattering vertex $\V$ defined in
Eq.~\eqref{eqDefWY} suggests introducing the Bogoliubov-Nambu (BN) pseudo spinors $\hat\Gamma_{\k}=
(\gh{\k},\ghd{-\k})^\text{t}$ in terms of which the Hamiltonian \eqref{eqInhomBgHamiltonian_gamma}
takes a more compact appearance: 
\begin{equation}\label{eqInhomBgHamiltonian_Gamma}
\hat H = \frac{1}{2} \sum_{\k} \ep{k} \hat\Gamma_{\k}^\dagger\hat\Gamma_{\k} +
\frac{1}{2} \sum_{\k,\k'} 
\hat\Gamma^\dagger_{\k}
\V_{\k \k'}
\hat\Gamma_{\k'}.
\end{equation}
The Heisenberg equation of motion for the BN spinor reads 
\begin{equation}\label{eqBgEOM}
 i \hbar \pder{}{t} \hat \Gamma_{\k} 
= \sigma_3
    \sum_{\k'} 
      \bigl[ \ep{\k} \delta_{\k \k'} + \V_{\k \k'}  \bigr] \hat\Gamma_{\k'} .
 \end{equation}
The multiplication by the Pauli matrix $\sigma_3$ is characteristic for the dynamics within the Bogoliubov-de~Gennes symmetry class that describes bosonic excitations of interacting systems \cite{Gurarie2003}.
We see that the Bogoliubov excitation in mode $\k$ is scattered to mode $\k'$ by the potential $\V_{\k\k'}$, 
the momentum transfer being provided by the underlying 
condensate $\Phi$ and its inverse profile $\check\Phi$, as represented
by the vertex in \autoref{figBgScattUniversal}. 
The effective excitation spectrum belonging to the equation of motion \eqref{eqBgEOM}
can be derived by studying the corresponding Green function.

Many-body Green functions contain all the information about how a quasiparticle created
in  state $\k'$  at time 0 propagates to  state $\k$ where it is destroyed at time
$t$. For the present, we need only the retarded Green functions at
temperature $T=0$.
Taking advantage of the Nambu structure, one defines a 
matrix-valued 
Nambu-Green function \cite{Bruus2004}
\begin{equation}\label{eqGeneralizedGreenFunction}
 \G_{\k\k'}(t) = \frac{\Theta(t)}{i\hbar}  \EKomm{\hat\Gamma_
   {\k}(t)}{\hat\Gamma^\dagger_{\k'}(0)} = \matr{G_{\k\k'}(t)}{F^\dagger_{\k\k'}(t)}{F_{\k\k'}(t)}{G^\dagger_{\k\k'}(t)}
\end{equation}
from the single-(quasi)particle retarded Green function  
\begin{equation}\label{GreenG}
G_{\k\k'}(t) = \frac{1}{i\hbar} \Theta(t) \EKomm{\gh { \k}(t)}{\ghd{\k'}(0)}
\end{equation}
and the anomalous Green function 
\begin{equation}\label{GreenF}
F_{\k\k'}(t) =  \frac{1}{i\hbar}\Theta(t) \EKomm{\ghd{-\k}(t)}{\ghd{\k'}(0)}.
\end{equation}
Here, $\langle \cdot \rangle$ stands for the 
expectation value in the Bogoliubov vacuum $\ket{0}$ defined by
$\hat\gamma_\k\ket{0}=0$ for all $\k$.  
The equation of motion of $\G$ under the Hamiltonian \eqref{eqInhomBgHamiltonian_Gamma} reads
\begin{align}\label{eqEOMmathcalG}
 i\hbar \frac{\rm d}{{\rm d}t}  \G 
=  \sigma_3  \delta(t) +  \sigma_3 
\left[
{\epsilon} +\mathcal{V} \right]
\G.
\end{align}
In this compact notation, 
$\epsilon_{\k\k'} = \epsilon_\k \delta_{\k\k'} \mathds{1} $.

%------------------------------------------------------------
\subsection{Perturbation theory}\label{secDiagrammatic}

In absence of disorder $V=0$, the equation of motion
\eqref{eqEOMmathcalG} is readily solved in frequency domain.
The anomalous Green function $F^{(0)}_{\k\k'}(\omega)=0$ vanishes, 
and the conventional retarded Green function is found as 
$G^{(0)}_{\k\k'}(\omega) = \delta_{\k\k'} G_{0\k}(\omega)$ with 
\begin{align}\label{eqG0}
 G_{0\k}(\omega) 
= \lim_{\eta\rightarrow 0} \frac{1}{\hbar\omega -\epsilon_k + i \eta}
=: \frac{1}{\hbar\omega -\epsilon_k + i 0} \, .
\end{align}
The infinitesimal shift $+i0$ stems from the causality factor
$\Theta(t)$ that is characteristic for the retarded Green function \eqref{GreenG}.
The Nambu-Green function for the clean system thus reads
\begin{align}\label{eqG0matrix}
\G_{0\k}(\omega) 
= \matr{G_{0\k}(\omega)}     {0}
       {0}  {G_{0\k}^*(-\omega)}.
\end{align}
With this, equation \eqref{eqEOMmathcalG} can be written in the form
\begin{align}\label{eqStartingPointPerturbation}
 \G  = \left[ \mathcal{G}_0^{-1} - \mathcal{V} \right]^{-1}
\end{align}
which is a suitable starting point for diagrammatic perturbation theory.

Equation \eqref{eqStartingPointPerturbation} permits a series
expansion in powers of $\mathcal{V}$ for the full Green function,
ensemble-averaged over the disorder: 
\begin{align}
\avg{\G} &= \G_0 + \G_0 \overline{\V} \G_0 + \G_0 \overline{\mathcal{V} \G_0 \mathcal{V}} \G_0 +
\ldots \label{eqPerturbationSeries}
\end{align}
Without loss of generality, we assume in the following that the
disorder potential is centered, i.e.\ $\avg{V}=0$. Then, second-order
and higher moments of the disorder potential have to be computed,
$\avg{V_{\k_1}V_{\k_2}}$, $\avg{V_{\k_1}V_{\k_2}V_{\k_3}}$, etc. Depending
on the disorder distribution, these moments may factorize into
independent terms. E.g., the moments of a Gaussian random process 
factorize completely into products of pair correlations. Thus, the
series \eqref{eqPerturbationSeries}  
contains  \emph{reducible} contributions from
products of disorder correlations that can 
be separated into independent factors by removing a single Green function $\mathcal{G}_0$.
This redundancy can be avoided by defining the self-energy $\Sigma$
via the Dyson equation
\begin{align}\label{eqDyson}
\avg{\G} = \G_0 + \G_0 \Sigma \avg{\G}. 
\end{align}
The self-energy contains precisely 
all irreducible contributions of the disorder-averaged right hand side
of Eq.~\eqref{eqPerturbationSeries}. Moreover, it directly describes the disorder-induced corrections to
the spectrum, as becomes evident from the formal
solution $\avg{\G}^{-1} = \G_0^{-1} - \Sigma$. 

In principle, any desired order in the disorder potential $V$ of the series $\Sigma = \Sigma^{(1)}+ \Sigma^{(2)} +
\Sigma^{(3)} + \ldots$ can be determined by first expanding  $\Sigma=\avg{\V}
+ \avg{\V\G_0\V}+\dots$ into
powers of the nonlinear scattering vertex $\V$ and then 
using the perturbative
expansion \eqref{eqExpansionV}, while retaining only the irreducible
contributions. 
In practice, of course, the
number of diagrams grows very rapidly with the order. Since all
first-order terms vanish by virtue of $\avg{V}=0$,  we 
concentrate on terms of order $V^2$.  This so-called Born
truncation of the full series is valid for weak potentials. 
We find two contributions:
\begin{equation}
 \includegraphics{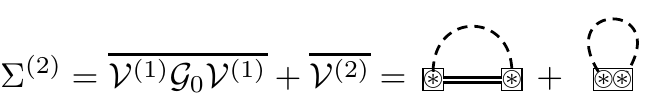} 
  \label{eqSigmaBorn} 
\end{equation}
 The general structure of the first contribution
$ \avg{\V^{(1)}\G_0 \V^{(1)}}$ 
is well-known from single particles in disorder
\cite{Rammer1998,Vollhardt1980,Kuhn2005,Kuhn2007a}: the particle is scattered once by the
bare disorder into a different mode and then scattered once more, back
into the original mode. The second contribution $\avg{\V^{(2)}}$ is specific to the Bogoliubov
problem and the nonlinear background of the GP equation
\eqref{eqStationaryGPE}: it describes the single scattering of the
excitation by a background fluctuation that is itself second order in
the disorder amplitude. 

%------------------------------------------------
\subsection{Self-energy}\label{secSelfE}

We focus now on the upper left-hand block $\Sigma^{(2)}_{11}$ of the Nambu 
self-energy matrix, which relates to the normal retarded Green function and
describes the change in the quasiparticle dispersion. 

Spelling out the two contributions \eqref{eqSigmaBorn} in terms of the
first- and second-order
scattering matrix elements \eqref{W1Y1} and \eqref{eq_W2Y2}, we find
\begin{equation} 
\Sigma^{(2)}_{11\k\k'}(\omega) = \sum_\p Z_{\k\p\k'}(\omega)
\avg{V_{\k-\p}V_{\p-\k'}}. 
\end{equation} 
Under the sum, we distinguish two essential factors, the potential
correlator and a kernel function. Concerning the potential, we assume
for notational convenience that the
disorder is homogeneous and isotropic under the ensemble average, with
a $k$-space pair correlator
\begin{equation}\label{eqDisorderCorrelator}
 \avg{V_\q V_{-\q'}} = L^{-d}\delta_{\q \q'} V^2 \sigma^d  C_d(q\sigma).
\end{equation}
The dimensionless function $C_d(q\sigma)$
characterizes the potential correlations persisting on the length
$\sigma$. Our formulation allows for a straightforward extension
to anisotropic disorder \cite{Robert-de-Saint-Vincent2010} or lattice
potentials, see Sec.\ \ref{lattice.sec} below. 

Because this disorder average restores homogeneity, we only need the  
kernel function for $\k'=\k$:  
\begin{equation}\label{SelfEKernel}
Z_{\k\p\k}(\omega)=  \frac{[w^{(1)}_{\k\p}]^2}{\hbar\omega - \epsilon_{\p}+i0}
 - \frac{[y^{(1)}_{\k\p}]^2}{\hbar\omega + \epsilon_{\p}+i0} +
 w^{(2)}_{\k\p\k},   
\end{equation}
with the envelope functions defined in Eqs.~\eqref{eqW1},
\eqref{eqY1}, and \eqref{eq_W2Y2}. This kernel depends solely on the healing length $\xi$.
Finally, the retarded normal self-energy $\Sigma^{(2)}_{11\k\k'}(\omega) =
\delta_{\k\k'} \Sigma(k,\omega)$ takes the functional form 
\begin{equation}  \label{eqSigma2}
\Sigma(k,\omega) = V^2\sigma^d \intdd{q} Z_{\k(\k+\q)\k}(\omega)
C_d(q\sigma). 
\end{equation} 
This expression is the second main achievement of the present
work. All physical results presented below 
follow by straightforward calculations from here.

%--------------------
\subsection{Disorder-modified dispersion relation}\label{secDisorderModifiedDispersion}

In order to grasp the significance of the self-energy, it is useful to
define the spectral function $S(k,\omega) = -2 \Im \avg{G}(k,\omega)$
\cite{Bruus2004}, which contains all information about the frequency
and lifetime of the excitations. 
In the clean system with dispersion $\epsilon_k=\hbar\omega$, the
spectral function is 
\begin{equation} 
S_0(k,\omega) = -2 \Im G_0(k,\omega) = 2\pi \delta(\hbar\omega-\ep{k}) .
\end{equation}
In presence of disorder, this function gets modified, while retaining its
normalization $ \int ({\rm d}\hbar\omega/2\pi) {S(k,\omega)} = 1$; 
this allows interpreting the spectral function as the energy
distribution of a quasiparticle with wave vector~$\k$. To leading
order in $V^2/(g n)^2$,   
\begin{align}\label{eqSpectralFunctionBorn}
S(k,\omega) =
\frac{-2 \Im \Sigma(k,\omega)}
{\left[\hbar\omega-(\ep{k} + \Re\Sigma(k,\omega))\right]^2 + \left[\Im\Sigma(k,\omega)\right]^2}.
\end{align}
The self-energy's real and imaginary part enter in characteristic ways.
First, the Bogoliubov modes are expressed, according to the reasoning of Sec.~\ref{secBasis}, in the plane-wave basis, which is not the eigenbasis of the
Bogoliubov Hamiltonian in presence of disorder. 
Thus, $\k$ are not ``good quantum numbers'', and the Bogoliubov modes
suffer scattering. This broadens their
dispersion relation, or
equivalently implies the existence of an elastic scattering rate (inverse lifetime) 
$\gamma_k=\tau_k^{-1}= - 2 \Im\Sigma(k)/\hbar$. 
Here, the notation $\Sigma(k)=\Sigma(k,\ep{k})$ indicates that one can take the self-energy on shell
to the lowest order $V^2$ considered. 
In terms of length scales, the
scattering rate defines the elastic scattering mean free path $\ls$
via 
\begin{equation}\label{els}
\ls^{-1} = \frac{\gamma_k}{v_\text{g}} = \frac{-2\Im\Sigma(k)}{\partial_k\ep{k}}    
\end{equation}
with the usual definition of the group velocity, $\hbar v_\text{g}=\partial_k\ep{k} = 2g n\xi(1+k^2\xi^2)[2+k^2\xi^2]^{-1/2}$. 

Secondly, the peak of the spectral density is shifted to  
\begin{align}\label{ModDispersion}
 \avg{\epsilon}_{k} = \ep{k} + \Re\Sigma(k), 
\end{align}
which defines the disorder-modified dispersion relation, again to 
order $V^2$. Notably, Eq.\ \eqref{ModDispersion} 
describes the impact of disorder on the speed of sound in the
low-energy regime. Indeed, a short calculation shows that 
the kernel function \eqref{SelfEKernel}
behaves  like $Z_{\k(\k+\q)\k}\sim k$ as $k\to0$, such that there
is always a finite correction to the speed of sound. Within our
theory, the disordered
potential conserves the linear character of the dispersion relation at
low energy, as required by the existence of the zero-frequency Goldstone
mode due to the spontaneously broken U(1) symmetry.

%--------------------
\subsection{Average density of states}\label{secAVDOS}

The quasiparticle dispersion enters practically all thermodynamic
quantities that determine how the disordered condensate responds to
external excitations, both at zero and finite temperature. Often, one
only needs to know the density of states. In a disordered system, 
the spectral function \eqref{eqSpectralFunctionBorn}---remember its
r\^ole as the probability density for a Bogoliubov quasiparticle $\k$ 
to have energy $\hbar\omega$---determines the average density of
states (AVDOS) per unit volume as \cite{Bruus2004,Akkermans2007}
\begin{equation}\label{eqDOS} 
 \avg{\rho}(\hbar \omega) = \intdd{k} \frac{S(k,\omega)}{2\pi} \, .
\end{equation}
As function of frequency and for weak disorder, the spectral function
\eqref{eqSpectralFunctionBorn} is very well approximated by a Lorentzian centered at
$\avg{\epsilon}_k/\hbar$ with small width $\gamma_k \ll \ep{k} $. 
The following Sec.~\ref{secResults}
will show that the relative scattering rate $\gamma_k/\ep{k}$ of low-energy, sound-wave
excitations tends to zero, and that the main effect of disorder is the
dispersion shift \eqref{ModDispersion} (in contrast to the case of
single particles in disorder, where the scattering rate is the
dominant quantity \cite{Kuhn2007a}). In the sound-wave regime, 
we can therefore approximate 
$S(k,\omega) = 2\pi \delta(\hbar\omega- \avg{\epsilon}_k)$  in Eq.\ \eqref{eqDOS}. 
With this, the
shift $\avg{\rho}(\epsilon) =\rho(\epsilon) +
\Delta\avg{\rho}(\epsilon)$ from the clean DOS in $d$ dimensions [Eq.~\eqref{eqDOS_Bg}]  
reads 
\begin{align}\label{eqAVDOSsound}
 \frac{\Delta \avg{\rho}(\epsilon)}{\rho(\epsilon)} 
  = - \left.\left[d + k \pder{}{k}\right] 
 \frac{\Re\Sigma(k)}{k \partial_k \ep{k}}\right|_{k=k_\epsilon} 
\end{align}
and is thus found to be a function of the dispersion 
shift \eqref{ModDispersion}. 

%--------------------
\subsection{Parameter space of the disordered Bogoliubov problem}

Because the self-energy is evaluated to order $V^2$, all  
the corrections it implies will be of this same
order. For notational brevity, we henceforth denote the small parameter of this
expansion by
\begin{equation}
\label{def.v2} 
v^2:= \frac{V^2}{(g n)^2}\ll1. 
\end{equation} 

Furthermore, we observe that the self-energy \eqref{eqSigma2}
depends on three different length scales: the excitation wave length
$\lambda = 2\pi/k$, the healing length $\xi$, and the disorder
correlation length $\sigma$. 
The resulting physics can only depend on the value of these lengths relative to each other: 
\begin{itemize}
 \item The correlation parameter $\zeta=\sigma/\xi$ indicates whether the disordered
   condensate background is in the Thomas-Fermi regime
   ($\zeta\gg1 $) or in the smoothing regime ($\zeta \ll 1$), see Sec.~\ref{secMeanfield}.
\item The  reduced wavenumber $k\xi$ indicates whether the excitations are sound waves ($k\xi \ll 1$) or
particles ($k\xi \gg 1$), see Sec.~\ref{secFreeBogoliubov}. 
\item The parameter $k\sigma$ 
discriminates the effectively $\delta$-correlated regime ($k\sigma \ll 1$) from a
very smooth scattering potential ($k\sigma \gg 1$). 
\end{itemize}
%
%-------------
\begin{figure}
\includegraphics[width=0.7\linewidth]{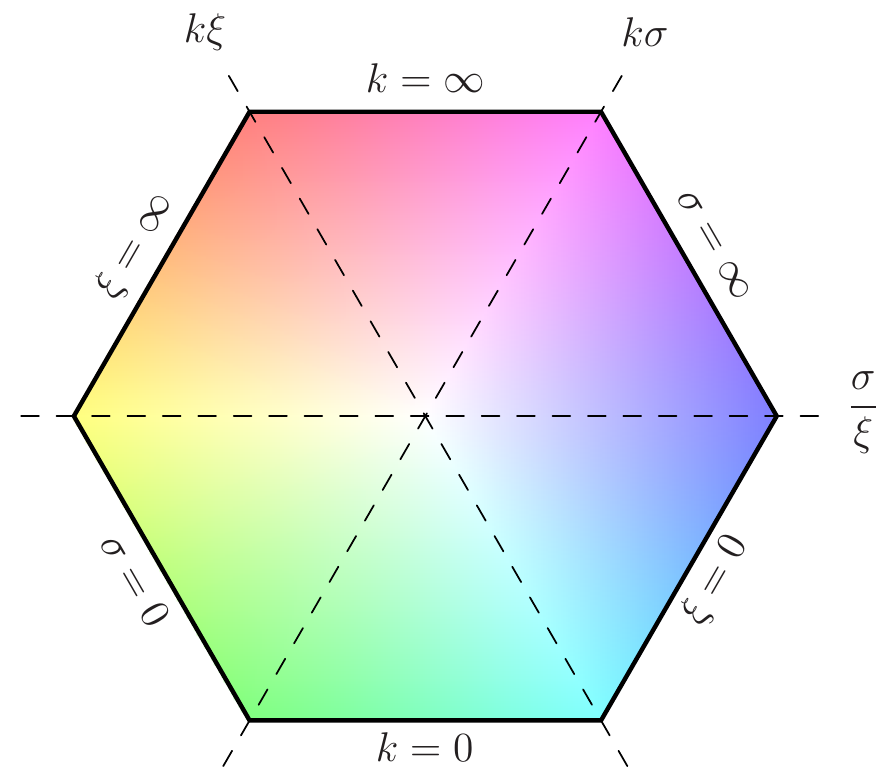}
\caption{(Color online) Parameter space of the disordered Bogoliubov
  problem, spanned by three length scales: healing length $\xi$,
  excitation wave vector $k$, and disorder correlation length $\sigma$.
  At opposing vertices, the three dimensionless parameters
  $k\xi$, $k\sigma$, and $\sigma/\xi$ take their extreme values $0$
  or  $\infty$. On the six edges, one of the length scales itself is 
  either $0$ or $\infty$.}
\label{fig_parahexhex}
\end{figure}
%-----------
%
These parameters are not independent, for any
given two of them determine the third, e.g.\
$(k\sigma)/(k\xi)=\sigma/\xi$. The parameter space thus is a
two-dimensional manifold. Nonetheless, it is useful to keep all three
parameters to describe the various physical regimes. We have therefore
found it convenient to map the entire parameter space to a hexagon,
see Fig.~\ref{fig_parahexhex}. The three symmetry axes connecting the
vertices carry the three dimensionless parameters, such that the
extreme values $0$ and $\infty$ occur at opposite vertices. 
The symmetry axes perpendicular to the edges then represent the values
of the length scales themselves, with their extreme values
$0,\infty$ taken on the entire edges. 
This construction is analogous to the representation of RGB color
space by hue and saturation at fixed lightness. Each of the three
dimensionless parameters describes one of the channels, e.g. $k\xi$ 
 the red channel with $k\xi=0$ mapped to cyan and $k\xi=\infty$
 mapped to red. Similarly, $k\sigma$ and $\sigma/\xi$ define the green and
 blue channel, respectively, which completes the color space, as shown in 
Fig.~\ref{fig_parahexhex}.

%------------------------------------------
\section{Results}
\label{secResults}
%------------------------------------------

All of the physical quantities that we compute in the following depend
crucially on the correlation parameter $\zeta=\sigma/\xi$ measuring
the disorder potential correlation in units of the condensate healing
length. In general, the results will even depend on the specific pair
correlation function $C_d(k\sigma)$ defined in Eq.~\eqref{eqDisorderCorrelator}.
For concreteness and direct applicability to cold-atom experiments, we will study in detail the case 
of optical speckle patterns, some properties of which are
summarized in Sec.~\ref{secSpeckle}. However, many analytical results
in limiting cases are independent of the specific pair correlation
and are thus universally applicable. 
 
%------------------------------
\subsection{Mean free paths}
\label{secMeanFreePath}

%--------------
\subsubsection{Elastic scattering mean free path} 
\label{secEls} 

%-------------
\begin{figure}
\includegraphics[width=0.95\linewidth]{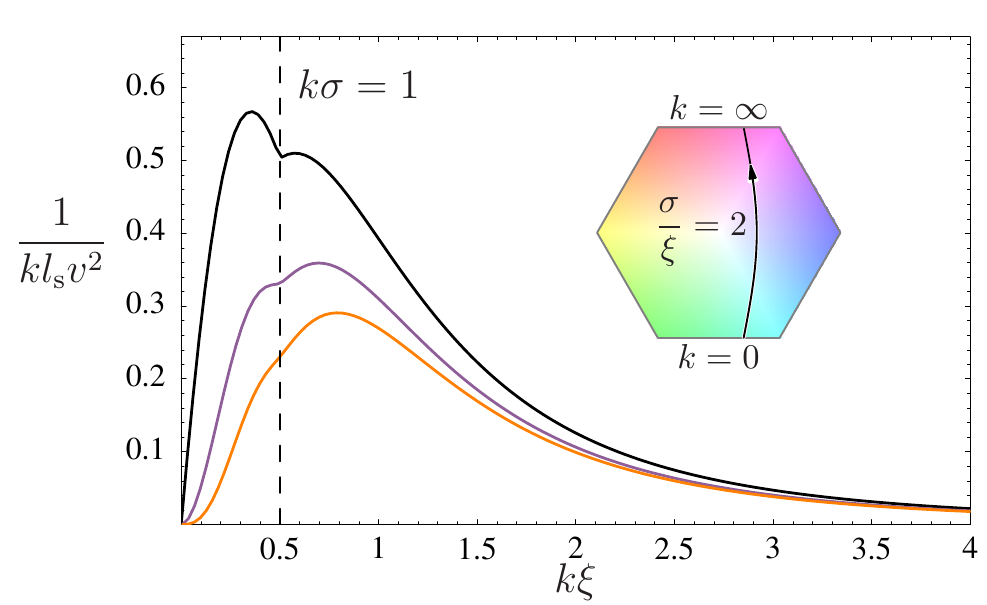}
\caption{(Color online) Inverse elastic scattering mean free path
  \eqref{eqkls} as function of Bogoliubov wave number $k\xi$ for  
disorder with fixed correlation parameter $\zeta=\sigma/\xi=2$, in $d=1,2,3$ dimensions (top to bottom).
For small or large $k\xi$, universal limiting behavior is found, see text. The
feature at $k\sigma=1$ is particular to the speckle disorder
[Eq.~\eqref{eqSpeckleCorr}]. 
Inset: the plot's path in the parameter space, \autoref{fig_parahexhex}.
}
\label{fig_ls}
\end{figure}
%-----------

First, we  evaluate the elastic scattering mean free path, 
in the dimensionless form $1/(k\ls) =
-2\Im\Sigma(k)/(\hbar k v_\text{g})$, which is  the small
parameter of weak-disorder expansions in standard quantum transport
theories \cite{Rammer1998}.  
The only possibility for an imaginary part to
occur in the on-shell self-energy 
\eqref{eqSigma2} is by the imaginary part of the
Green function, $\Im
(\epsilon_k-\epsilon_{\k+\q}+i0)^{-1} = -\pi
\delta(\epsilon_k-\epsilon_{\k+\q})$, multiplying the normal
scattering amplitude. This restricts the integral over
the intermediate state to the energy shell. There,  
the scattering element simplifies according to
Eq.~\eqref{eqElastScattAmpl}, and we find  
\begin{align}\label{eqkls}
 \frac{1}{k l_{\rm s}} 
= \frac{\pi v^2}{2} \frac{k^d\sigma^d}{(1+k^2\xi^2)^2}
\int \frac{{\rm d}\Omega_d}{(2\pi)^d}  A(k\xi,\theta)^2 C_d\bigl(2k\sigma\sin\textstyle{\frac{\theta}{2}}\bigr).
\end{align}

Fig.~\ref{fig_ls} shows this inverse scattering mean free path plotted as function of
$k\xi$ for a speckle potential [Eq.~\eqref{eqSpeckleCorr}] with a fixed correlation ratio $\zeta=\sigma/\xi=2$
in dimensions $d=1,2,3$. 
The $k$-dependent fraction in front of the integral in \eqref{eqkls} ensures that the mean free path
diverges both for very low and high momenta, where one recovers essentially a clean
system. In-between, around $k\xi= 1$, 
there is a minimum mean free path. Fig.~\ref{fig_ls} also shows a feature at $k\sigma = 1$, which is
specific to the speckle correlation function used here, due to the non-analyticity
at the boundary of its support. Indeed, in
$d=1$, at this point the contribution of the backscattering process
$k\to{-}k$ becomes impossible at the level of the Born approximation,
which explains the kink at $k\sigma=1$. In higher dimensions, the
angular integration lifts the singularity to higher derivatives, so
that it becomes less conspicuous.  

Obviously, the mean free path \eqref{eqkls} depends on the two dimensionless
parameters $k\xi$ and $k\sigma$. Let us discuss some interesting
limiting cases, which are easily identified in our parameter space
representation.  

(i) First, the hydrodynamic limit $\xi\to0$ is 
found on the lower right-hand edge of  Fig.~\ref{fig_parahexhex}.
Here, we recover exactly the
elastic mean free path for scattering of sound waves  
\cite{Gaul2009a}, which only depends on the remaining parameter $k\sigma$. 

(ii) Secondly, the limit $\xi\to\infty$ is found on the
upper left-hand edge of Fig.~\ref{fig_parahexhex}.
There, $\xi$ drops out together with the
interaction energy $g n$ from \eqref{eqkls}, which reduces 
exactly to the elastic mean free path for single-particle scattering,
as calculated in \cite{Kuhn2005,Kuhn2007a}. This reduced mean free
path $k\ls$ can again only depend on $k\sigma$.  

More generally, let us consider a fixed correlation ratio $\sigma/\xi$ of order
unity, and look at the asymptotics as function of $k\xi$, as plotted in
Fig.~\ref{fig_ls}. 

(iii) As $k\xi\to0$, also $k\sigma\to0$. Then, the potential correlator becomes
isotropic, $C_d(2k\sigma\sin\frac{\theta}{2})\to C_d(0)$, and 
pulls out of the integral over $A(0,\theta)^2=\cos^2\theta$. This
integral yields $\int \rmd\Omega_d \cos^2\theta = S_d/d$ (a result showing the
``concentration of measure'' of the hypersphere's surface around its
equator \cite{Milman1986}). Thus, we obtain 
\begin{equation} \label{kels_hydro} 
\frac{1}{k\ls} = \frac{\pi S_d  v_\delta^2}{2d(2\pi)^d} (k\xi)^d [1+O(k\xi,k\sigma)]
\end{equation}  
where $S_d$ is the unit sphere's surface
($S_1=2$, $S_2=2\pi$, $S_3=4\pi$), and we define the effective $\delta$-correlation disorder
strength   
\begin{equation} \label{def.vdelta} 
v_\delta^2 = C_d(0)\frac{\sigma^d V^2}{\xi^d (g n)^2}.   
\end{equation} 
The low-$k$ behavior $1/k\ls \propto (k\xi)^d$ indeed appears clearly
in Fig.~\ref{fig_ls}.
In our parameter space Fig.~\ref{fig_parahexhex}, this is the asymptotic behavior of curves
starting from the lower edge $k=0$, for intermediate
values of $\sigma/\xi$, i.e.\ rather in the center of the edge. 
Note that $l_{\rm s}^{-1} \propto k^{d-1}$ is proportional
to the surface of the energy shell, i.e.\ the number of states available for
elastic scattering. In the limit $k \rightarrow 0$, the elastic energy shell shrinks and
the scattering mean free path diverges, even when measured in units of
$k^{-1}$.  

(iv) Conversely, as $k\xi\to\infty$, also $k\sigma\to\infty$. But as soon
as $k\sigma\gg1$,  the disorder potential allows
practically only forward scattering, and we can make a small-angle
approximation to all functions of $\theta$, such as  
$2k\sigma\sin{\frac{\theta}{2}}\to k\sigma\theta$. Then, the final
result can be cast into the form 
\begin{equation} 
\frac{1}{k\ls} = \frac{V^2}{E_\sigma^2}\frac{f_d(\sigma/\xi)}{(k\sigma)^3}  [1+O(1/k\xi,1/k\sigma)]
\end{equation} 
where $E_\sigma=\hbar^2/(m\sigma^2)$ is the characteristic correlation
energy \cite{Kuhn2005,Kuhn2007a}, and $f_1(\zeta)=C_1(0)$ 
as well as 
\begin{equation} 
f_d(\zeta)= S_{d-1} \int_0^\infty\frac{\rmd u\,u^{d-2}}{(2\pi)^{d-1}}\frac{(2\zeta^2-u^2)^2}{(2\zeta^2+u^2)^2} C_d(u)
\end{equation}
in $d=2,3$.
And indeed, all curves in Fig.~\ref{fig_ls} show this decrease as $k^{-3}$ for large momenta.
In our parameter space
Fig.~\ref{fig_parahexhex}, this is the asymptotic behavior of curves
arriving at the upper edge $k=\infty$ for intermediate values of
$\sigma/\xi$. 

For extreme values of $\sigma/\xi$, i.e.\ towards the far left or 
right of the parameter space, one can also find interesting 
asymptotics when the wave length $2\pi/k$ lies between $\sigma$ and $\xi$.

(v) Consider first the case of a potential in the deep Thomas-Fermi
regime $\sigma\gg \xi$ and excitations with $k\sigma \gg 1 \gg k\xi$.
This describes hydrodynamic excitations, i.e.\ a set of parameters
approaching the right outermost vertex  of Fig.~\ref{fig_parahexhex} along the edge $\xi=0$.
Then, by the same reasoning as in the previous case (iv), strongly peaked forward scattering leads to 
$1/k\ls \propto k\sigma$. 
This linear increase of the inverse scattering mean free path would na{\"\i}vely predict infinitely strong scattering as $k$ increases.
Within the full description, however, this unphysical behavior stops as soon as $k\xi\approx 1$ is reached, crossing over to case (iv) with an even simpler description since now $\zeta\gg1$ for which
$f_d(\infty)= S_{d-1}(2\pi)^{1-d} \int_0^\infty \rmd u\,u^{d-2} C_d(u)$ in $d=2,3$.

(vi) Conversely, consider finally a potential in the deep smoothing
regime $\sigma/\xi\ll1$ and excitations with   $k\sigma \ll 1 \ll
k\xi$. This 
describes particle excitations, i.e.\ a set of parameters
approaching the
left outermost vertex  of Fig.~\ref{fig_parahexhex} along the edge $\xi=\infty$. 
Then, by the same reasoning as in 
case (iii), isotropic scattering leads to  
$1/k\ls \propto (k\sigma)^{d-4}$. This low-$k$ divergence of the scattering mean
free path na{\"\i}vely predicts infinitely strong scattering as
$k\sigma\ll1$ and severely limits the validity of simple perturbation theory
for the single-particle case \cite{Kuhn2005,Kuhn2007a}. Not so here,
where the divergence is avoided once $k\xi\approx 1$ is
reached, and the interaction energy comes
into play, crossing over to case (iii). 

In summary, our perturbation theory provides valid expressions for the
elastic scattering rate or inverse mean free path in the full space of
parameters.
At a given value of $\zeta = \sigma/\xi$, the scattering rate is always a bounded function of $k$, multiplied with the small parameter $v^2\ll1$, which vindicates the use of the momentum basis as a starting point for the perturbation theory.

%--------------
\subsubsection{Transport mean free path} 
\label{secEltr} 

%-------------
\begin{figure}
\includegraphics[width=0.9\linewidth]{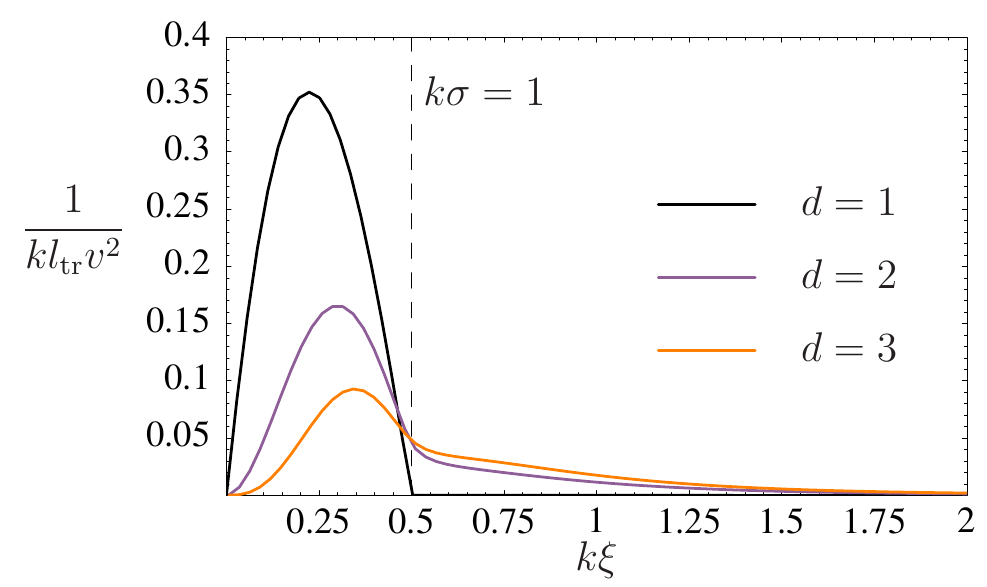}%
\caption{(Color online) Inverse transport mean free path, \eqref{eqkls} including the
  vertex correction factor $(1-\cos\theta)$ under the integral, for a speckle disorder
\eqref{eqSpeckleCorr} with fixed correlation parameter
$\zeta=\sigma/\xi=2$. }
 \label{figInverseklslB} 
\end{figure}
%------------

If a disordered BEC is brought out of equilibrium, it will respond via
its excitations. Therefore, it is of interest to study the transport
properties of Bogoliubov excitations. In principle, a full-fledged quantum
transport theory requires to calculate particle-hole propagators,
which is certainly doable using the Hamiltonian \eqref{eqInhomBgHamiltonian_Gamma}, but
beyond the scope of the present article. Still, the previous results
on the scattering mean free path can be generalized, with very limited
additional effort, to the Boltzmann
transport mean free path \cite{Kuhn2005,Kuhn2007a} that measures the
diffusive randomization of the direction of motion. 
This transport mean free path $\ltr$ is defined by the same integral expression \eqref{eqkls},
where the integrand is multiplied by a factor $(1-\cos\theta)$.
Fig.~\ref{figInverseklslB} shows a plot of $1/k\ltr$ as function of
$k\xi$ for a speckle disorder
\eqref{eqSpeckleCorr} with fixed correlation parameter
$\zeta=\sigma/\xi=2$ in dimensions $d=1,2,3$. 

In one dimension, the only contribution to the inverse transport mean free path is the backscattering contribution $k\to {-k}$, such that 
\begin{align}\label{eqklB1D}
 \frac{1}{k \ltr} 
= \frac{v^2}{2} \frac{k\sigma\, C_1(2k\sigma) }{(1+k^2\xi^2)^2}.
\end{align}
Due to the finite support of the speckle correlation function
\eqref{eqSpeckleCorr1D} backscattering is impossible for $k\sigma > 1$, and the inverse transport mean free path vanishes
(within the Born approximation, and here we do not consider higher-order
corrections to $\ltr$ \cite{Lugan2009,Gurevich2009}), as clearly apparent from
Fig.~\ref{figInverseklslB}. In
dimensions $d \geq 2$, there are finite contributions from small
scattering angles. Adapting the reasoning of case (iv) from the
previous section, one finds that $1/k\ltr\propto (k\sigma)^{-5}$, 
with a prefactor that can be determined similarly.

%--------------
\subsubsection{Localization length} 
\label{ssLocLength}

Just as phonons and particles, Bogoliubov excitations are
expected to localize in disordered environments. Again, a full
calculation is out of reach within the present article, but we can estimate the
localization lengths of our Bogoliubov excitations in correlated
disorder, based on general
results on localization of particles and phonons. 

In one-dimensional disordered systems, 
the localization length $\lloc = 2\ltr$,
which describes exponential localization, 
is directly proportional to the backscattering length that we just
calculated \cite{Thouless1973}. From \eqref{eqklB1D} we deduce 
\begin{equation}\label{eqLocLength}
 \frac{1}{k\lloc} = \frac{v^2}{4} \frac{ k\sigma \, C_1(2 k\sigma)}{(1+k^2\xi^2)^2}, 
\end{equation}
which agrees perfectly with \cite{Lugan2007a}, and also with
\cite{Bilas2006}, in the limits $\sigma \rightarrow 0$ and $\xi
\rightarrow 0$ investigated there.
Those phase-formalism approaches are particularly suited for 1D systems,
whereas our Green-function theory permits going to higher dimensions without conceptual
difficulties.

\begin{figure}[tb]
 \includegraphics[angle=270,width=\linewidth]{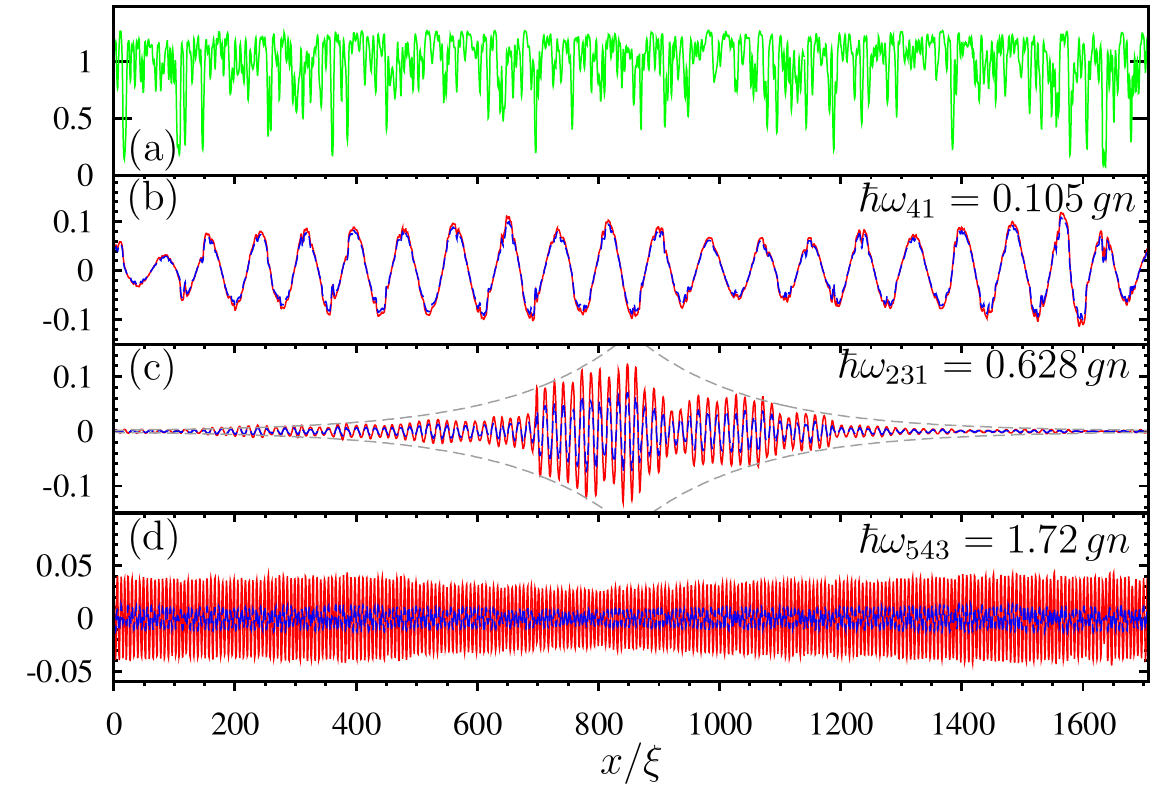}
\caption{(Color online) 
(a) Condensate density $n(x)/n$ of a BEC (system size $L\approx 1700\xi$, periodic boundary conditions)
in a blue-detuned speckle potential with amplitude $V/gn=0.3$ and
correlation length $\sigma =\xi$. 
(b)--(d) Selected Bogoliubov modes $u_\nu(x)$ (solid red) and
$v_\nu(x)$ (dashed blue), obtained by exact diagonalization of the Bogoliubov-de~Gennes
equation \eqref{eqBg-de-Gn}. Low- and high-energy modes [(b) and (d)] are extended,
while localization is most pronounced at intermediate energies [(c)].
The dashed gray line in panel (c) shows the exponential envelope predicted by Eq.~\eqref{eqLocLength}.
}\label{figLocModes}
\end{figure}

In two dimensions, the localization length is related to the
transport mean free path via 
$ \lloc = \ltr \exp\{\frac{\pi}{2} k \ltr \}$. 
This result can be derived using scaling theory arguments that hold
very generally for single-particle excitations, and also
the localization length of phonons has been shown to scale
exponentially with $\ltr$ \cite{John1983}. 

In three dimensions, localized and delocalized states can coexist, as
function of 
energy separated by a mobility edge.
Phonons are localized at high energies and particles are localized at
low energies \cite{John1983}. These opposite characteristics imply
that when the disorder is increased, localized modes will start to
appear at energies close to the point where $\ltr$ is minimum.   

So in all dimensions, localization will be observed most readily,
within finite systems, for modes that have the shortest localization
length. Our results on the transport mean-free path in correlated
potentials show (in agreement with the 1D results of \cite{Lugan2007a}) that 
modes around $k\xi=1$ will be the first to appear localized.

In one dimension, we have quantitatively verified the prediction
\eqref{eqLocLength} by means of an exact diagonalization of the
inhomogeneous Bogoliubov-de~Gennes equation \eqref{eqBg-de-Gn}, after
solving the stationary GP equation \eqref{eqStationaryGPE} for the
condensate. 
Fig.~\ref{figLocModes} shows that indeed only
Bogoliubov modes at intermediate energies $\hbar\omega_\nu \approx
0.6 gn$ appear localized in the finite system.
The observed localization length is compatible with the prediction, Eq.~\eqref{eqLocLength}.

All lengths calculated so far have the property that they diverge in the limit $k\xi \ll 1$. In other words, sound waves can propagate over long distances and for long times in these disordered systems. 
It is therefore meaningful to compute the renormalized speed of sound.  

%------------------------------------------------------------------
\subsection{Speed of sound}\label{secSpeedOfSound}

As shown in Sec.~\ref{secDisorderModifiedDispersion}, the disorder
potential shifts the dispersion relation by $\Delta
\avg{\epsilon}_k=\Re\Sigma(k)$. Using Eqs.\ \eqref{ModDispersion} and
\eqref{eqSigma2}, the relative dispersion shift 
takes the form 
\begin{align}\label{eqDeltaE}
 \frac{\Delta\avg{\epsilon}_k}{\epsilon_k v^2} = \sigma^d \intdd{q}
 z_{\k \q} C_d(q\sigma). 
\end{align}
The kernel $z_{\k \q}$ obtains from the real part of the on-shell
kernel $Z_{\k (\k+\q) \k}(\epsilon_k)$, Eq.~\eqref{SelfEKernel}: 
\begin{align}\label{eqKernelDeltaE}
  z_{\k \q}
 = \frac{(gn)^2}{\epsilon_k} 
  \biggl[ \PV  \frac{[w^{(1)}_{\k (\k+\q)}]^2}{\epsilon_{k} - \epsilon_{\k+\q}}
 - \frac{[y^{(1)}_{\k (\k+\q)}]^2}{\epsilon_k + \epsilon_{\k+\q}} +
 w^{(2)}_{\k (\k+\q) \k} \biggr] ,
\end{align}
with the envelopes defined in Eqs.~\eqref{eqW1},
\eqref{eqY1}, and \eqref{eq_W2Y2}.
$\PV$ denotes the principal value. These equations reduce to much simpler
expressions in different limiting regions of the parameter space,
Fig.~\ref{fig_parahexhex}. We mostly focus on the low-energy, sound
excitations that are of primary interest. Analytical results will be 
confronted with data from a numerical simulation in
Sec.~\ref{secNumericalStudy} below. At last, we show in
Sec.~\ref{lattice.sec} that our theory also covers the case of weak
lattice potentials.

%%%%%%%%%%%%%%%%%%%%%%%%%%%%%%%%%%%%%%%%%%%%%%%%%%%%%%%%%%%%%%%%%%%%%%%%%%%
\subsubsection{Limiting cases}\label{secSpeedOfSoundLimits}
Similar to the proceeding in Sec.~\ref{secMeanFreePath}, we compute
analytical results in the limiting cases located at the edges and corners
of the parameter space, Fig.~\ref{fig_parahexhex}. 

(i) We start with the hydrodynamic limit $\xi \to 0$, the lower right edge
of  Fig.~\ref{fig_parahexhex}. 
The kernel \eqref{eqKernelDeltaE} simplifies to
\begin{equation}\label{eqKernelHydro}
 - \frac{1}{2k^2} \PV \frac{ (k^2+\k \cdot\q)^2}{ q^2+2\k\cdot\q },
\end{equation}
after which Eq.~\eqref{eqDeltaE} reproduces exactly Eq.~(29) of Ref.\ \cite{Gaul2009a}.
Notably, the dispersion shift is negative in all dimensions and for
any value of $\sigma/\xi\gg1$, as anticipated
in the schematic plot of Fig.~\ref{figCleanBogoliubov}(a).
The limiting values are
\begin{subequations}\label{eqHydroLimits}
\begin{numcases}
{\frac{\Delta \avg{\epsilon}_k}{\epsilon_k v^2} = }
	-1/(2 d), 		& $ k\xi \ll k\sigma \ll 1 $ , \label{limits1.eq}\\ 
	-(2+d)/8 , 		& $ k\xi \ll 1 \ll k\sigma $ . \label{limits2.eq}
\end{numcases}
\end{subequations}
These limiting values are expected to hold over an extended range of $k\sigma$. Thus,
they define a shift in the local slope of the dispersion relation. In
other words, $k$-modes in
that particular range have a modified sound velocity. The magnitude of the correction depends significantly on the
excitation's ability to resolve the correlations ($k\sigma\gg1$) or not
($k\sigma\ll1$). As noted in \cite{Gaul2009a}, in the latter case the
correction decreases with dimension, but increases in the former, implying
that the curves for different dimensions must cross around
$k\sigma=1$.  

In passing, we stress that even in the very long-range correlated
limit $\sigma /\xi \to \infty$, these results are not trivial.
Indeed, one could try and use a simple static local density approximation (LDA)
in order to derive the result \eqref{limits2.eq} for  
correlation lengths much longer than the excitation wave length.
In this regime, the background appears locally homogeneous to the wave, and the local
sound velocity $c(\r)=\sqrt{g n(\r)/m}$ is proportional to the condensate field amplitude $\Phi(\r)$.
Thus, LDA expects $\Delta \avg{c}/c$ to be given by $\avg{\Phi}/\Phi$,
which can be easily computed from Eq.~\eqref{eqSmoothingPsi2} to
yield $\Delta \avg{c}_{\rm LDA}/c = -v^2/8$. But this fails to reproduce
Eq.~\eqref{limits2.eq}.  Indeed, static LDA cannot capture the scattering dynamics
(shown by the first diagram of Eq.\ \eqref{eqSigmaBorn}), which is
essential for correctly determining the sound velocity.

(ii) In the regime of particle-like excitations $k\xi\to\infty$
(covering the cases (ii), (iv), and (vi) of
Sec.~\ref{secMeanFreePath}), 
the Hamiltonian \eqref{eqManyParticleHamiltonian} becomes
non-interacting. Consequently, one should expect the entire Bogoliubov problem
to reduce to the problem of single particles in disorder. 
Indeed, Bogoliubov excitations in the particle regime see both, the external potential and the condensate background.
Sampled at high wave numbers $k\xi \gg 1$, the condensate background is smooth and
cannot induce scattering. Fittingly, we found in
Sec.~\ref{secEls} that the elastic scattering mean free path reduces in this
limit to the single-particle expression. In contrast, the deeply inelastic
processes contributing to Eq.~\eqref{eqDeltaE} remain sensitive to the
condensate background, as encoded by the anomalous
and second-order couplings, \eqref{eqY1} and \eqref{eq_W2Y2}, which 
do not simply vanish in the limit $\xi\to\infty$.
We find that the leading-order correction to the dispersion relation (for $k\xi \gg 1$ and $k\sigma$ not too small),
\begin{align}\label{eqDeltaEkxiinf}
{\Delta \avg{\epsilon}_k}=  v^2 \sigma^d  \intdd{q}
\frac{\epsilon_q^0 C_d(q\sigma)}{(2+q^2\xi^2)^2}, % >0,  
\end{align}
is independent of $k$. Incidentally, it is exactly opposite to the
negative average shift $\avg{\mu^{(2)}}$ of
the chemical potential, from Eq.~\eqref{eqMu2}, for fixed average density. At fixed chemical
potential, the dispersion shift \eqref{eqDeltaEkxiinf} would even be twice as big.
%To this order, the dispersion relation of free particles $\epsilon^0_k$ is recovered up to an overall shift,
%$\overline{\epsilon}_k \approx \epsilon^0_k + (gn - \mu^{(2)})$. 
Note that this shift cannot be  na{\"\i}vely accounted for by an
overall shift $gn \to gn+\avg{\mu^ {(2)}}$ in the clean dispersion, Eq.\
\eqref{cleanBgPartcl}. Just like the wrong LDA attempt to explain the sound
velocity, discussed above, such a reasoning misses the
essential scattering dynamics. 

Eq.\ \eqref{eqDeltaEkxiinf} differs also from the chemical potential
shift for noninteracting particles in disorder  \cite{Bruus2004}.   
But it must be kept in mind that the disorder expansion of the
Bogoliubov Hamiltonian in Sec.~\ref{ssecExpansionV} was performed
under the assumption $V \ll g n$. 
For single particles, the interaction energy $gn$ goes to zero, i.e.\ the ratio of $V$ and $g n$ would have to be reversed. 
Therefore, our perturbative theory cannot be expected to apply
universally in this
regime.
  
In any case, the main effect of disorder in the single-particle regime
is to yield the finite scattering rate calculated in
Sec.~\ref{secMeanFreePath}, but it only produces a very small shift in
dispersion.  
In addition, these high-energy excitations are less important for
low-temperature properties of BECs, and will not be considered in the
remainder of this work.  

(iii) Let us turn to the sound-wave regime $k\xi \ll 1$.
In case (i) and Ref.~\cite{Gaul2009a}, this has been achieved by
sending the healing length $\xi$ to zero,  thus yielding the
dispersion as function of $k\sigma$, but only for rather long-range
correlated potentials with 
$\sigma\gg\xi$.  
In order to cover arbitrary correlation ratios $\zeta = \sigma/\xi$, we now change the point of
view and take $k \to 0$. 
This allows in particular to reach the case $\sigma \ll \xi, k^{-1}$
of truly $\delta$-correlated disorder that was inaccessible to
\cite{Gaul2009a}. 
The kernel \eqref{eqKernelDeltaE} simplifies, and we find the relative
shift in the speed of sound 
\begin{align}\label{DeltaCk0}
 \frac{\Delta\avg{c}}{c v^2} = \sigma^d \intdd{q} 
\left\{ 
 \frac{2 q^2\xi^2}{(2+q^2\xi^2)^3} - \frac{2 \cos^2\beta}{(2+q^2\xi^2)^2} 
\right\} 
C_d(q\sigma). 
\end{align}
$\beta = \measuredangle(\k,\q)$ is the angle between the direction
of propagation and $\q$. 
In contrast to the hydrodynamic case (i), 
there are now two competing
contributions with opposite sign (for an interpretation of these
contributions, cf.\ the end of 
Sec.~\ref{lattice.sec} below). 
In case of isotropic correlation, the angular integral maps $\cos^2\beta $
to $1/d$.  Then, only for $d=1$ is the radial integrand strictly
negative, and $\Delta \avg{c}$ is negative as well. For $d>1$, the radial integrand
has no definite sign. 

%----------------------------------
\begin{figure}[btp]
\centerline{\includegraphics[angle=270,width=0.95\linewidth]{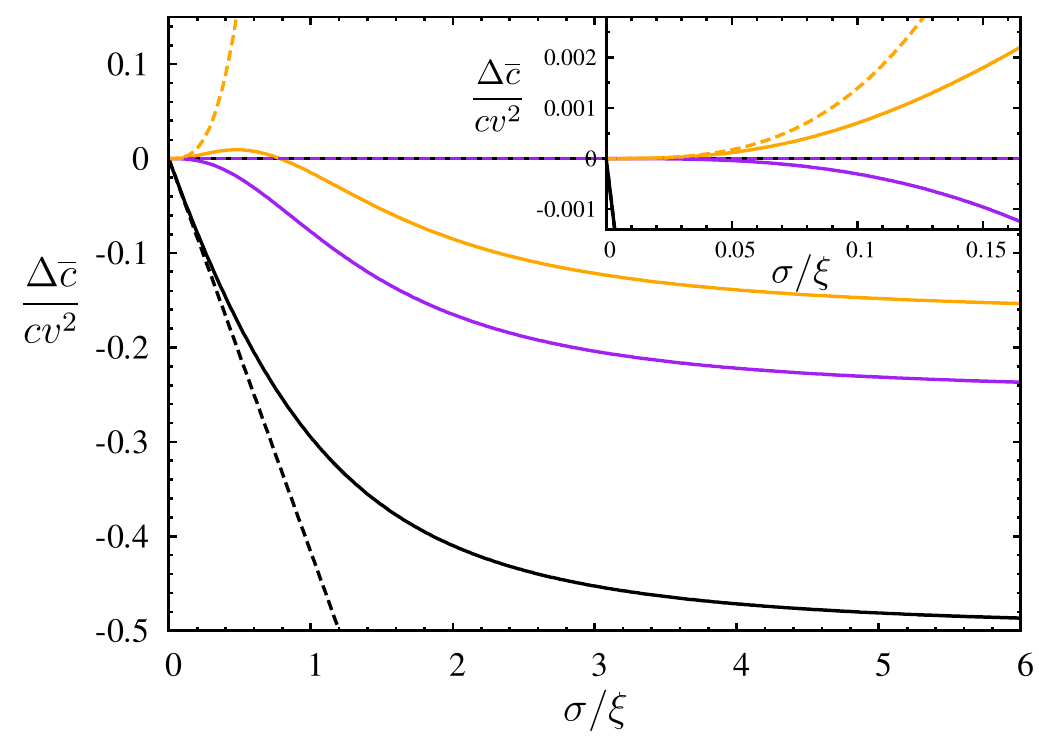}}
\caption{(Color online) Relative correction \eqref{DeltaCk0} of sound
  velocity due to a speckle disorder potential \eqref{eqSpeckleCorr}
  as function of the correlation ratio $\zeta=\sigma/\xi$ in
  $d=1,2,3$. 
The exact formulae are given in Eq.~\eqref{eqLimitK0}.
Dashed: universal limits, collected in \autoref{tabDeltaCLimits}.
Inset: same data around the origin, showing the rapid departure from the leading-order
estimate \cite{Giorgini1994,Lopatin2002,Falco2007}.}
\label{figDeltaC_k0} 
\end{figure}
%------------------------------------
%-----------------------------
\begin{table}
 \begin{tabular}{c|ccc}
$\Delta \avg{c}/c$     & $d=1$                             & $d=2$    & $d=3$\\
\hline
$\sigma \gg \xi$ & $-v^2/2$		             & $-v^2/4$ & $-v^2/6$\\
$\sigma \ll \xi$ & $-\frac{3}{16\sqrt{2}} v_\delta^2$&   $0$    & $+\frac{5}{48\sqrt{2}\pi} v_\delta^2$ 
\end{tabular}
\caption{Limiting corrections of the speed of sound, corresponding to
  the dashed limits of Fig.~\ref{figDeltaC_k0}. For uncorrelated
  disorder, the correction is
  proportional to $v_\delta^2$ of Eq.~\eqref{def.vdelta}. 
}\label{tabDeltaCLimits}
\end{table}
%-----------------------------------------

To survey the possible outcomes, we plot in
Fig.~\ref{figDeltaC_k0}  the correction
\eqref{DeltaCk0} to the speed of sound caused by isotropic speckle
disorder, Eq.~\eqref{eqSpeckleCorr}, as function of the correlation ratio $\zeta=\sigma/\xi$. 
The curves can actually be given in closed form, see 
\eqref{eqLimitK0}, but the details depend of course on the specific
correlation. In contrast, we can extract universal limits for very
small or very large $\zeta$.  

In the long-range correlated limit $\zeta \to \infty$, found on the right edge of the
plot, the correlator $C_d$ acts as a $\delta$-distribution, 
which leads to $\Delta \avg{c}/c = -v^2/(2d)$. This value
coincides with the hydrodynamic limit 
\eqref{limits1.eq}, as it should.   

In the opposite limit  $\zeta \ll1$ of $\delta$-correlated disorder, 
the correlator $C_d(q\sigma)\to C_d(0)$ can be pulled out of the integral,
which contributes
a numerical prefactor to the expected scaling with the disorder
strength $v_\delta^2$ defined in Eq.\ \eqref{def.vdelta}. These
results are plotted as dashed lines in Fig.~\ref{figDeltaC_k0} and collected in \autoref{tabDeltaCLimits}. 
Again, these results cannot be found by LDA, which
would have to assume that the system is homogeneous on relevant length scales ($\sigma \gg \xi, 2\pi/k$),
an assumption that is always violated by the sound-wave limit $2\pi /k \to \infty$.

Our result for $\sigma \ll \xi$ in $d=3$ reproduces the value known
from Refs.~\cite{Giorgini1994,Lopatin2002,Falco2007}. 
Interestingly, this is the only case where the correction to the speed
of sound is positive. Actually, this particular numerical value is of rather
limited use since the Taylor expansion  at
the origin is converging very slowly, and 
already a minor correlation can make a major
difference, as shown by the inset in Fig.~\ref{figDeltaC_k0}. Our
results, Eqs.~\eqref{eqDeltaE} and \eqref{DeltaCk0}, hold for a much
larger range of parameter values and arbitrary dimensions, 
which accomplishes one of the main goals of this work.

%%%%%%%%%%%%%%%%%%%%%%%%%%%%%%%%%%%%%%%%%%%%%%%%%%%%%%%%%%%%%%%%%%%%%%%%%%%
\subsubsection{Numerical mean-field study of the sound velocity}\label{secNumericalStudy} 

We confront the theoretical predictions \eqref{eqDeltaE} and
\eqref{DeltaCk0} with data obtained by a numerical simulation in
$d=1$ on the mean-field level, using the time-dependent Gross-Pitaevskii equation. 
This numerical calculation constitutes an independent check since it 
relies neither on the linearization in the excitations,
nor on perturbation theory in the disorder potential, which are the 
two approximations of our analytical theory. 

The numerical procedure has been briefly described in
Ref.~\cite{Gaul2009a}. 
We generate a 1D speckle disorder potential with correlation
length $\sigma$ by Fourier transformation from the set of random complex 
field amplitudes [see Eq.~\eqref{eqSpeckleFieldCorrelation}] with 
all $k\leq\sigma^{-1}$. 
Then the condensate ground state $\Phi(x)$ solving the GP equation
\eqref{eqStationaryGPE} is computed by imaginary-time
evolution, using the fourth-order Runge-Kutta algorithm, while keeping
the wave function normalized
\cite{Dalfovo1996}. 
Onto this disordered ground state, a plane-wave
Bogoliubov excitation is superposed, with a small, but finite $k$ and amplitude
$\Gamma$. In cold-atom experiments, such an excitation is routinely imprinted using
Bragg spectroscopy
\cite{Stamper-Kurn1999,Vogels2002,Steinhauer2002,Steinhauer2003}. 
The 
Bogoliubov transformation \eqref{eqBgTrafo1} requires the imprints in density
and phase to be  
\begin{align}\label{eqDensityPhaseModulation}
 \dn(x) &= 2   \sqrt{n} a_{k}  \,\Gamma \cos(k x), &
\dph(x) &=  \frac{\Gamma }{\sqrt{n} a_k} \sin(k x) .
\end{align}
In the sound-wave regime where $a_{k} \ll 1$ [cf.\ Eq.~\eqref{eq_ak}],
the phase modulation has a much larger amplitude than
the density modulation, and we choose $ \Gamma = 0.3 \sqrt{n} a_k v $. 
Then, the real-time evolution under the GP equation is computed using again the fourth-order
Runge-Kutta algorithm. The excitation propagates, with a modified speed
of sound, surviving over a long course of time given by the inverse
elastic scattering rate $\gamma_k^{-1}$ [cf.~Eq.~\eqref{els}].
In order to extract the eigenenergy $\epsilon_k$, the deviation
$\delta\Phi(x,t)$ from
the ground state is translated into Bogoliubov excitations by means of
Eq.~\eqref{eqBgTrafo1}. 
Monitoring the phase of $\gamma_{k} \propto e^{-i \epsilon_{k}
  t/\hbar}$ over time, we extract the phase velocity
$v_\text{ph}=\epsilon_k/\hbar k$ by linear regression. 
 Then, the procedure is repeated for different realizations of disorder
with the same correlation length $\sigma$,
leading to a whole distribution of values, from which we compute the
average $\Delta \avg{c}/c$. 
As shown in Ref.~\cite{Gaul2009a}, for weak disorder the distribution is 
clearly single-peaked and allows for meaningful
averages.

%----------------------
\begin{figure}
\includegraphics[width=\linewidth]{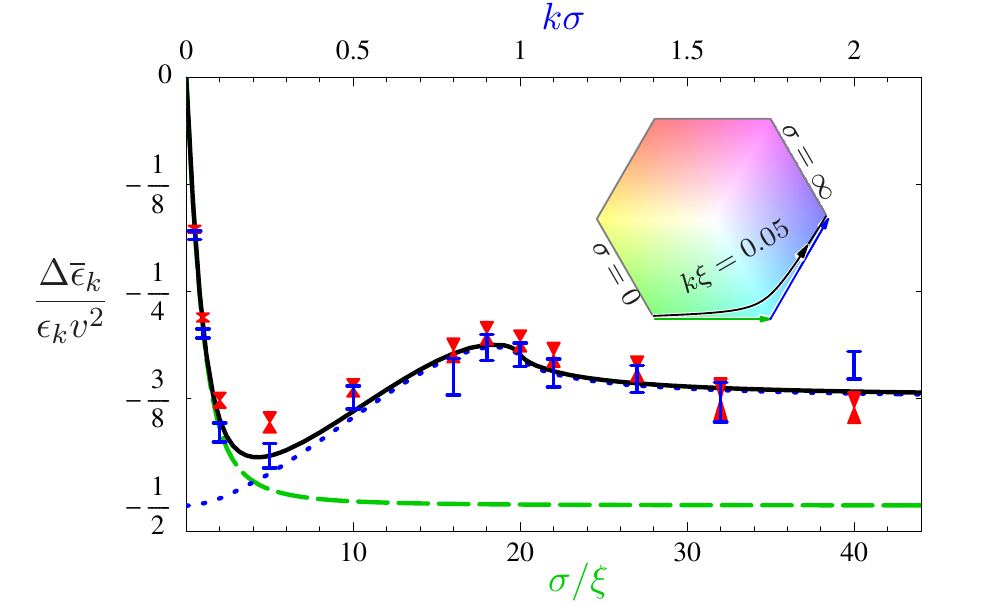}
\caption[Relative correction of the speed of sound (at $k\xi = 0.05$)]%
{(Color online) Relative correction of the Bogoliubov excitation dispersion
  relation due to 1D speckle disorder. The full formula \eqref{eqDeltaE} for
  $k\xi = 0.05$ [solid black] crosses over from the limiting case (iii)
    of low-energy excitations [dashed green, Eq.~\eqref{eqLimitK01D}] to the limiting case (i)
    of the hydrodynamic regime [dotted blue, Eq.~\eqref{eqHydroLimit1D}].  The inset
    shows the corresponding trajectory in parameter space. 
The numerical results (cf.\ Sec~\ref{secNumericalStudy}) for blue- and
red-detuned speckle with $v = +0.03$  [blue lines]
and $v = -0.03$ [red triangles], agree fully with the analytical theory. }
\label{figDeltaC_kxi0.05}
\end{figure}

\autoref{figDeltaC_kxi0.05} shows the numerical data on top of the
theoretical predictions, as function of $\sigma$. Since $k\xi=0.05$ is
fixed, the curve can be read as a function of $k\sigma$ at very small $\xi$
(near the hydrodynamic limit (i) of above) or as a function of $\sigma/\xi$ at very small
$k$ (near the low-energy limit (iii) of above).  The inset of
Fig.~\ref{figDeltaC_kxi0.05} depicts the corresponding trajectory in parameter
space. The full prediction \eqref{eqDeltaE} for
$k\xi = 0.05$ is plotted as a black line. The limiting cases are
available in closed form: Eq.\ \eqref{eqHydroLimit1D} 
for ${\xi = 0}$ as function of $k\sigma$  and
Eq.\ \eqref{eqLimitK01D} for ${k=0}$  as function of $\sigma/\xi$.
Independently of the details, we find of course the relevant universal
limits of Sec.~\ref{secSpeedOfSoundLimits} above. 

The numerical data, 
shown for $v = +0.03 $ (blue straight marks) and $v = -0.03 $ (red
triangular marks), follow the analytical prediction very 
well. Interestingly,  the data points of red and blue detuning, with
opposite sign of $v$, tend to lie on
opposite sides of the curve, indicating beyond-Born effects of
odd order $v^3$, which are expected for a speckle potential with its
asymmetric on-site distribution \eqref{eqSpeckleProbDist}. Attentive
readers will also notice that the
data points are shifted asymmetrically with respect to the curve, which is
an effect of order $v^4$.

%------------------------------
\subsubsection{Disorder-shift of Bogoliubov spectrum} 

\begin{figure}
 \includegraphics[angle=270,width=\linewidth]{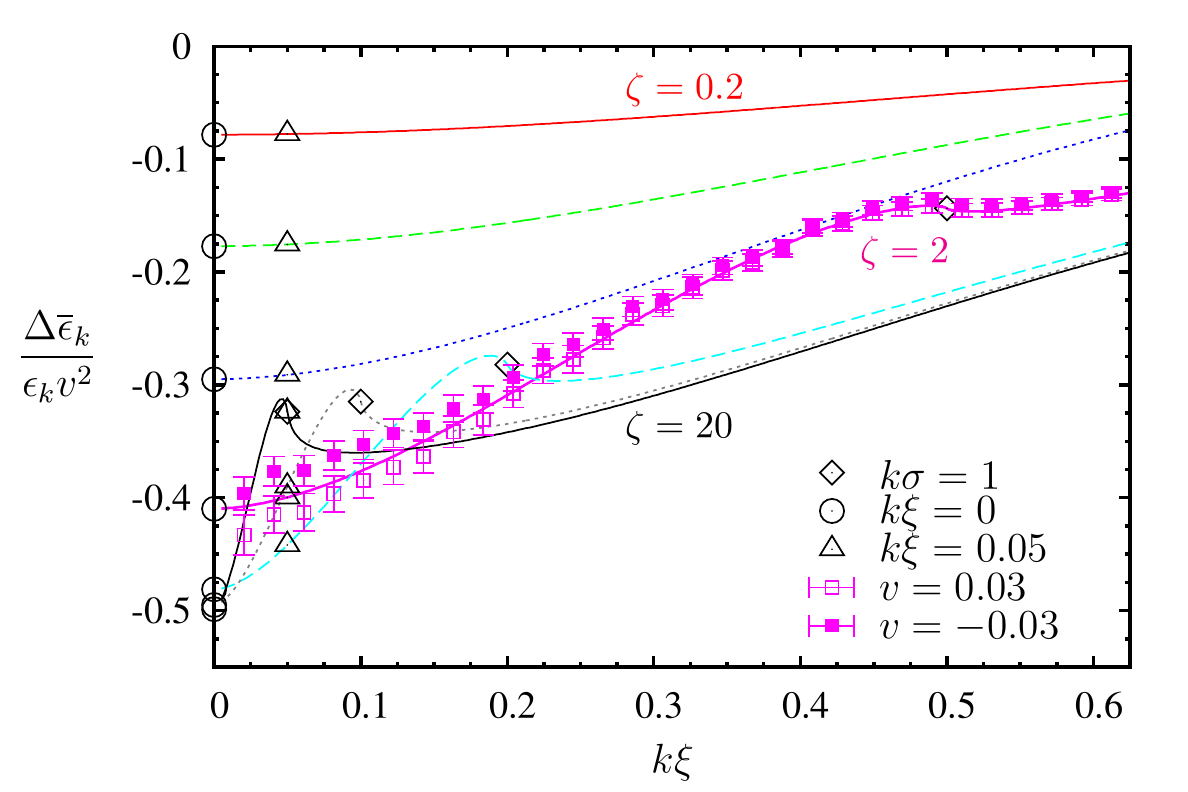}
 \caption{(Color online) Relative correction \eqref{eqDeltaE} of the
   dispersion relation $\Delta \overline{\epsilon}_k$ for different
   correlation ratios $\zeta = \sigma/\xi = 0.2, 0.5, 1, 2, 5, 10, 20$ (top to
   bottom), for 1D speckle disorder, Eq.~\eqref{eqSpeckleCorr1D}. 
 The results of \autoref{figDeltaC_k0} appear at the edge $k\xi=0$
 (circles), whereas the results shown in \autoref{figDeltaC_kxi0.05}
 are found at $k\xi=0.05$ (triangles).  
 Around the points $k\sigma = 1$ (diamonds), the correction behaves non-monotonically.
The points with errorbars close to the curve $\zeta=2$ show data from
the exact diagonalization of the Bogoliubov-de~Gennes eq.~\eqref{eqBg-de-Gn}
(system size $L\approx 300\xi$, correlation $\sigma = 2\xi$) for blue- as well
as red-detuned speckle disorder. 
Each point represents the energy shift of the two modes $\nu = 2j-1,2j$ corresponding to
the degenerate modes $k_j=\pm  2\pi j/L$ of the homogeneous system.
 The data has been averaged over a large number $r$ of realizations
 ($r  L/\sigma \approx 1.9 \times 10^{4}$). 
Errorbars show the estimated error of the mean value. }\label{figDeltaC_zeta} 
\end{figure}

Finally, we explore the correction of the dispersion relation
\eqref{eqDeltaE} as function of $k\xi$. 
\autoref{figDeltaC_zeta} shows a family of curves for different correlation
parameters $\zeta = \sigma/\xi$ in one dimension.
This plot actually contains the information of the previous
Figs.\ \ref{figDeltaC_k0} and \ref{figDeltaC_kxi0.05}, which are taken
at fixed values of $k\xi = 0$ and $k\xi=0.05$, respectively. 

The disorder correction passes through a non-monotonic feature around $k\sigma =1$
(marked by open diamonds), and finally diminishes with increasing $k\xi$.

In dimensions $d>1$, the curves have a slightly different shape.
The curve with $\zeta = 20$, for example, starts with the limit
Eq.~\eqref{limits1.eq} at $k\xi \to 0$ and passes through Eq.~\eqref{limits2.eq} at $k\xi \approx 0.15$.
Thus, while the 1D curve starts at $-1/2$ and passes through $-3/8$, the corresponding curve in $d=3$ will start at $- 1/6$ and pass through an extremum around $- 5/8$, before diminishing in the particle regime.
Also, the sharp speckle features get washed out in higher dimensions.

Complementary to the previous numerical study at constant $k\xi$, we
can verify our predictions also by an exact diagonalization of the
one-dimensional disordered Bogoliubov-de~Gennes equation
\eqref{eqBg-de-Gn}. Thus, we obtain the whole spectrum of the system
characterized by its correlation ratio $\zeta=\sigma / \xi$. For the
case $\zeta=2$, Fig.~\ref{figDeltaC_zeta}  shows excellent agreement
between prediction and numerics. As observed previously, the data for
red- and blue-detuned speckle lie to both sides of the $O(v^2$)
prediction, indicating effects of odd orders.  

The excellent agreement between both numerical approaches and
analytics demonstrates that
our perturbation theory to order $v^2$  gives an impressive account of all relevant
effects over the entire parameter space we set out to cover.

%-----------------
\subsubsection{Weak lattice potentials}
\label{lattice.sec} 

The inhomogeneous Bogoliubov Hamiltonian \eqref{eqInhomBgHamiltonian_Gamma} applies to arbitrary external potentials.
In particular, it covers the important class of weak optical lattice potentials that was studied
in \cite{Liang2008,Taylor2003}.
In a sense, understanding the lattice is a first step to understanding disorder, which can be seen, by virtue of Fourier decomposition, as a superposition of lattices with suitably chosen random amplitudes and phases.

To substantiate this connection, we
briefly show that our formalism reproduces the results of
\cite{Liang2008,Taylor2003} for the
speed of sound in weak lattices. 
In $d$ dimensions, a separable lattice potential $ V(\r) = 
\sum_{j=1}^d V_j \cos(K_j x_j)$ with wave vectors $\K_j=K_j\hat\e_j$ has Fourier components 
\begin{equation}
 V_\q = \frac{1}{2} \sum_{j=1}^d V_j [\delta_{\q\K_j}+\delta_{\q(-\K_j)}].
\end{equation}
The whole formalism developed in Secs.~\ref{secRelevantHamiltonian} and
\ref{secDispersion} applies also to this potential. 
Notably, the equation of motion  \eqref{eqEOMmathcalG}
is still solved by the perturbative expansion 
\eqref{eqPerturbationSeries}, now without the need for averaging over
disorder. The periodic potential scatters Bogoliubov excitations
elastically only at the edges of the Brillouin zone, $k_j = \pm K_j$.
Away from the edges, the dispersion relation is again determined by
the diagonal elements of the Green function $G(\epsilon)_{\k
  \k}$. Expressions like \eqref{eqSigma2} and \eqref{eqDeltaE} are
still valid, where the correlator should now be understood to stand
for
\begin{equation}\label{latticepot.eq}
V^2\sigma^dC_d(q\sigma) = \frac{(2\pi)^d}{4}\sum_{j=1}^d V_j^2 \left[\delta(\q-\K_j)+\delta(\q+\K_j)
\right]. 
\end{equation} 
In the sound-wave limit $k\to 0$, the speed-of-sound correction
\eqref{DeltaCk0} thus reads 
\begin{align}\label{eqDeltaElattice} 
 \frac{\Delta c}{c} = \sum_{j=1}^d \left\lbrace
   \frac{K_j^2\xi^2}{(2+K_j^2\xi^2)^3} - \frac{\cos^2 \beta_j
   }{(2+K_j^2\xi^2)^2}\right\rbrace \frac{V_j^2}{(gn)^2} 
\end{align}
where $\beta_j$ is the angle between the propagation direction and the
lattice direction $\hat\e_j$.
In the case $V_2=V_3=0$, the sound velocity along the $x_1$ direction reproduces the result of Ref.\ \cite[Eq.~(52)]{Taylor2003}.
Also, Eq.~\eqref{eqDeltaElattice} is precisely Eq.~(22) in
Ref.\ \cite{Liang2008} for the potential \eqref{latticepot.eq}. 
This work of Liang et al.~\cite{Liang2008}, while equivalent to ours as
far as the perturbative approach is concerned, has the nice feature
that it permits to attach a thermodynamic meaning to the two
antagonistic contributions appearing in Eq.\ \eqref{eqDeltaElattice}: the first,
positive term stems
from the disorder-shift of the compressibility $\kappa$, while the second, negative one goes back
to the change in the effective mass $m^*$.
Together, these quantities determine the speed of sound $c = 1/\sqrt{\kappa m^*}$.%
\footnote{The compressibility correction in Ref.\ \cite{Liang2008} appears with a wrong sign in Eq.\ (C1), but (C6) agrees with our result.}

%--------------------------------------------------------------
\subsection{Average density of states}\label{secAVDOS_results}

Lastly, we turn to the average density of states (AVDOS), Eq.~\eqref{eqDOS},  at very low
energies or $k \to 0$. As shown
above, the dispersion is linear in this limit, so that the AVDOS  
$\avg{\rho}(\epsilon) = \intdd{k}\delta(\epsilon-\hbar\avg{c}k) =
\rho(\epsilon) (c/\avg{c})^d $ necessarily has the lowest-order
correction 
\begin{equation}\label{eqDOS_k0} 
 \frac{\Delta \avg{\rho}(0)}{\rho(0)} = - d \frac{\Delta \avg{c}}{c}. 
\end{equation}
In other words, a reduced sound velocity entails an enhanced density
of states and vice versa, which is the obvious conclusion one can
already draw from the schematic plot anticipated in Fig.~\ref{figCleanBogoliubov}. 
It is instructive to check that identity \eqref{eqDOS_k0}
also follows from the general equation \eqref{eqAVDOSsound}: 
There, the linear dispersion implies $k\partial_k\epsilon_k=\epsilon_k$, and the fraction approaches 
$\Delta\avg{\epsilon}_k/\epsilon_k$. When
acting on this regular function,
the operator $k \partial_k$ inside
the bracket evaluates to zero at $k=0$, and we arrive at
\eqref{eqDOS_k0}. Thus, the AVDOS shift is entirely determined by the
sound velocity shift \eqref{DeltaCk0}; see also the analytic
solutions \eqref{eqLimitK0}. 

In the hydrodynamic regime $k\xi\ll1$ realized by $\xi\to 0$ at
finite $k\sigma$, again $k\partial_k\epsilon_k=\epsilon_k =\hbar c k$ in
Eq.~\eqref{eqAVDOSsound}, which then reproduces our previous results
\cite{Gaul2009a}. The limiting values for  small or large  energy
compared to the hydrodynamic correlation energy $\hbar c/\sigma$,  corresponding to 
\eqref{eqHydroLimits}, are 
\begin{align}\label{eqHydroDOSLimits}
\frac{\Delta \avg{\rho}(\epsilon)}{\rho(\epsilon)}
 = \frac{v^2}{2} \times 
	\begin{cases} 
		1, 			& \epsilon \ll \hbar c/\sigma, \\ 
		\frac{d}{4}(2+d), 	& \epsilon \gg \hbar c/\sigma. 
	\end{cases} 
\end{align}
In between these limits, the correction $\Delta \avg{\rho}/\rho$  as
function of $\epsilon$ can 
show a surprisingly rich behavior, depending 
on the potential correlations. 
In three dimensions, the correction is smooth and monotonic, but  
in one dimension, speckle correlations are responsible for an
unexpected, non-monotonic behavior with a sharp feature at $k_\epsilon
\sigma = 1$, as discussed in \cite{Gaul2009a}.  

Note  that these  simple perturbative results are indeed expected to hold at low
energy or $k\xi\ll1$, in stark contrast to the case of single particles in disorder,
where the DOS has a (nonperturbative) Lifshitz tail at low  energy 
 \cite{Lifshitz1988,Lugan2007}. 
In the present case, the interparticle repulsion
screens the disorder very effectively at low energy, such that
localization effects are absent (cf.\ the diverging localization
length of Sec.~\ref{ssLocLength})  \cite{Lee1990,Lugan2007}. A transition to the Bose
glass phase \cite{Giamarchi1987,*Giamarchi1988,Fisher1989} 
occurs only for stronger disorder or much weaker interaction,
where the Bogoliubov theory developed here breaks down, and different
approaches are needed \cite{Falco2009a,Pilati2009,Carrasquilla2010}.

%------------------------------------------------------------------------
\section{Conclusions and Outlook}\label{secConclusions}
%--------------------------------------------------------------------------

In conclusion, we have formulated a comprehensive Bogoliubov theory of inhomogeneous 
Bose-Einstein condensates. This analytical theory describes the elementary excitations 
of condensates with s-wave interaction, deformed by  
weak external potentials with  arbitrary spatial correlations  and in
arbitrary spatial dimension. 
Expanding the many-body Hamiltonian
around the deformed ground state, we
have obtained the inhomogeneous Bogoliubov Hamiltonian. 
We have justified our choice of the basis of density and phase
fluctuations that ensures proper orthogonality between the excitations
and the inhomogeneous Bogoliubov vacuum. 
Expressed in terms of Nambu-Bogoliubov spinors, all effects of the
external potential can be collected into a scattering vertex that is
non-perturbative in the external potential. A fully analytical
formulation has been achieved up to second order in weak
potentials, allowing in principle an expansion to even higher orders. 

From this fundamental Hamiltonian, one can derive numerous physically
relevant quantities by means of standard perturbation theory. This
paper has been devoted to a detailed discussion of the single-excitation dispersion
relation. We have calculated the mean-free path and renormalized speed
of sound together with the resulting
average density of states, over the full parameter space of the
disordered Bogoliubov problem, with numerous
analytical results in limiting cases. 
It turns out that the frequently investigated case of $\delta$-correlated disorder in three
dimensions, with its positive
shift in the speed of sound, is far from generic. Over most of the
parameter space, the speed of sound is reduced. We have confirmed 
these predictions in detail by mean-field numerical simulations as
well as exact diagonalization for the experimentally
relevant case of correlated speckle disorder in $d=1$.  

Strictly speaking, the present work is incomplete without proving that
the weak disorder under consideration causes only  a small
condensate depletion.  
Indeed, the Bogoliubov ansatz \eqref{eqBogoliubovPrescription} relies on the macroscopic population of the condensate mode. In
a pure mean-field description, all atoms are in the condensate at
zero temperature. But due to the effect of interaction, even at zero temperature, 
there is a finite fraction of particles not in the condensate,  
which constitutes the so-called \emph{quantum depletion} that can be
calculated within Bogoliubov theory
\cite{Pethick2002,Pitaevskii2003}. The quantum depletion should be
small, thus providing an 
important, self-consistent check of the theory's
validity. 

In the homogeneous setting, the mean-field condensate forms in the
homogeneous mode, i.e., the zero-momentum state. Thus, the quantum depletion 
consists of all particles with non-zero
momentum. The density of these particles can be easily calculated
within Bogoliubov theory \cite{Pethick2002,Pitaevskii2003}, with the
result (we take $d=3$ here) that the fractional quantum depletion  
$\delta n/n=8(n a_{\rm s}^3)^{1/2}/3\sqrt{\pi}$ is proportional to the
root of the gas parameter and thus very small, 
especially so for dilute and weakly interacting cold gases. 

It is not immediately obvious how to generalize the recipe ``count all particles with
non-zero momentum'' to the inhomogeneous case. The vast majority of works dedicated to the inhomogeneous
Bogoliubov problem simply calculates the same quantity, namely the number
of particles with non-zero momentum. But one has no means of knowing whether these
particles belong to the deformed condensate or to the true,
disorder-induced quantum depletion. And really, the supposed ``depletion''
calculated by Huang and Meng, followed by  
\cite{Huang1992,Giorgini1994,Astrakharchik2002,Kobayashi2002},  
involves only the condensate deformation at fixed chemical
potential, which is a mean-field effect as described in Sec.~\ref{secMeanfield}. 
Only few authors seem to have clearly
recognized  that this supposed depletion describes merely the non-uniform
density of the condensate \cite{Lopatin2002}. 

In order to assess the
true condensate depletion, one has to determine the density of
particles not in the condensate at all, irrespective of their
particular momentum. 
With the general Bogoliubov Hamiltonian at hand, we have calculated
this disorder-induced quantum depletion \cite{Gaul2010_letter}. We  find that
it is much smaller than the mean-field condensate deformation. This is no surprise:
To first order, the external potential merely \emph{deforms} the
condensate. The scattering of particles out of the condensate is a
second-order effect, mediated by the interaction between particles and
the condensate. Details of the full calculation, including 
finite-temperature effects, will be discussed in a
forthcoming publication \cite{Muller2010}. 

These results validate and strengthen the Bogoliubov approach, and we expect that
the theory we have developed here should
fare very well in describing the excitations of inhomogeneous BECs. As an
immediate extension of the present work, finite-temperature effects can be captured very
straightforwardly by the Matsubara formalism \cite{Bruus2004},
allowing the calculation of the heat capacity and many other
(thermo-)dynamic response functions. This is left for future work. 

\begin{acknowledgments} 
This work is supported by the National Research Foundation \& Ministry of
Education, Singapore.
Work at Madrid was supported by MEC (Project MOSAICO).
Financial support by Deutsche Forschungsgemeinschaft (DFG) and
Deutscher Akademischer Auslandsdienst (DAAD) is
acknowledged for the time when both authors were 
affiliated with Universit\"at Bayreuth, Germany. 
We are grateful for helpful discussions with, and 
generous hospitality extended by,  
P.~Bouyer,  D.~Delande, B.~Englert, T.~Giamarchi, 
V.~Gurarie, M.~Holthaus, P.~Lugan,
A.~Pelster,  L.~Sanchez-Palencia,  P.~Schlagheck, H.~Stoof, and E.~Zaremba.
\end{acknowledgments}

\appendix

%---------------------------------------------------------------
\section{Optical speckle disorder}\label{secSpeckle}
%--------------------------------------------------------------

\subsection{Statistical properties}
The optical speckle field of a laser defines a disorder potential with very
well controlled statistical properties 
\cite{Goodman1975,Clement2006,Lugan2009,Sanchez-Palencia2010}. 
When coherent laser light is directed on a rough surface or through a
diffusor plate, the elementary waves originating from different points
have random phases and form a random interference pattern in the far
field. By virtue of the central limit theorem, the resulting field
$\mathcal{E}(\r)$ is a complex Gaussian random process. 
For notational simplicity, we consider a scalar field with dipole coupling to a
single atomic transition and neglect polarization issues.  
The electronic atomic ground state is then subject to the light-shift potential induced by the
intensity $I(\r) = |\mathcal{E}(\r)|^2$ \cite{Grimm2000}. 
Magnitude and sign of this potential depend on
the laser detuning from the dipole
transition frequency. For a far-detuned potential, one has 
\begin{align}\label{eqSpeckleVI} 
V(\r) = V \left(\frac{I(\r)}{I_0} - 1\right)  ,
\end{align}
The prefactor $V$ contains the
atomic polarizability, besides all other proportionality factors, and 
we have shifted the potential to zero average,
$\avg{V(\r)}=0$. 
The magnitude of $V$ gives the variance of the potential
fluctuations, $V^2=\avg{V(\r)^2}$, that can be readily adjusted in the
experiment by changing the overall laser intensity. 
Because the intensity is the modulus square of the Gaussian
field $\mathcal{E}$, the potential has a skewed
on-site probability distribution for $w = V(\r)/V$,   
\begin{equation}\label{eqSpeckleProbDist}
P(w) {\rm d} w = \Theta(w+1) e^{-(w + 1)} {\rm d} w. 
\end{equation}
A blue-detuned potential with  $V>0$ consists
of high potential bumps rising over a flat 
baseline.  A red-detuned potential  with $V<0$ rather corresponds to
a random set of deep wells.

The potential correlation between different spatial points is captured
by the  pair correlator
\begin{equation} 
\widehat C_d(r/\sigma) 
= \avg{V(\r)V(0)}/V^2
\end{equation} 
 that decays from the starting value 
$\widehat C_d(0)=1$ to zero on
the scale of the correlation length $\sigma$. In momentum
representation, the normalization implies the useful identity 
\begin{equation}\label{normCd.eq} 
\intdd{(q\sigma)}C_d(q\sigma) = 1.
\end{equation}  

A speckle pattern's correlation is entirely determined by the fact that  
the Fourier components of the field
$\mathcal{E}_\k$ are independent, complex Gaussian random variables with a pair
correlation 
\begin{align}
\avg{\mathcal{E}_\k^* \mathcal{E}_{\k'}}
&=:\gamma(k) \delta_{\k\k'}  \label{eqSpeckleFieldCorrelation}
\end{align}
that defines the degree of coherence $\gamma(k)$. In one dimension,
for a rectangular source, 
the degree of coherence $ \gamma(k) = \gamma 
\Theta(1-k\sigma)$ \cite{Kuhn2007a} has uniform weight inside the
interval of allowed $k$-values. The correlation length $\sigma$
depends on the laser wave length and the imaging optics
and is typically of the order of $1\,\mu$m or smaller \cite{Billy2008}.   
It is not easy to create an isotropic multi-dimensional speckle field
in the laboratory \cite{Robert-de-Saint-Vincent2010}. 
For the purpose of the present paper, we follow Pilati \textit{et al.}\ \cite{Pilati2009} 
and define the isotropic speckle disorder on the grounds of
Eq.~\eqref{eqSpeckleFieldCorrelation} with the same $\gamma(k)$ in all dimensions.
This definition grasps the essential feature of speckle disorder, namely
the finite support of its power spectrum. The extension to a more
realistic, possibly anisotropic correlation function for a given
experimental configuration  is
straightforward within the present formalism.

%-----------------------------------------
\begin{figure}
\includegraphics[height=4.4cm] {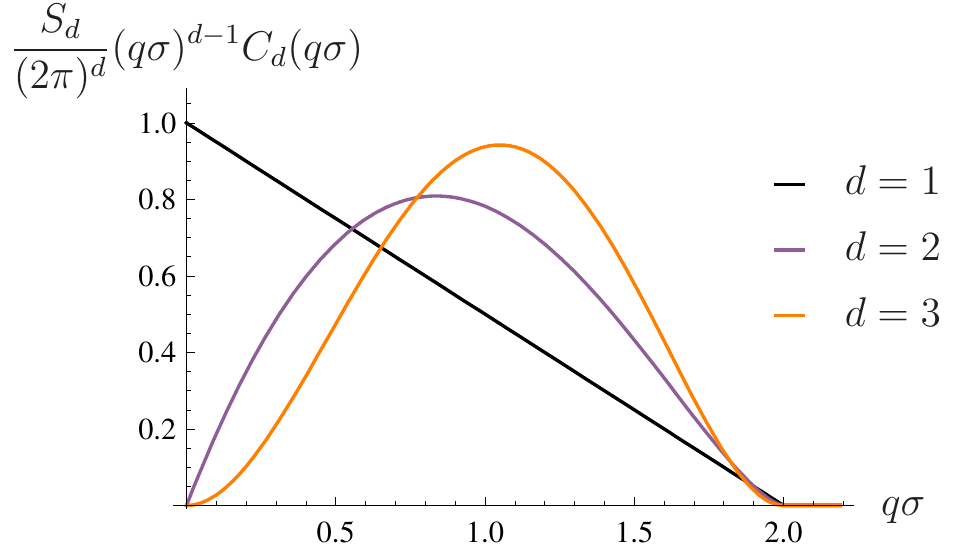}
\caption{Isotropic speckle
  correlation function as function of reduced momentum 
in $d=1,2,3$. In order to show comparable scales,
we plot $[S_d/(2\pi)^{d}] (q\sigma)^{d-1} C_d(q\sigma)$, including the
$d$-dimensional volume integration element.}
\label{figSpeckleCorrelationK}
\end{figure}
%--------------------------------------------

The potential correlator $C_d(k\sigma)$, used in Eq.\ \eqref{eqDisorderCorrelator} and the following,
then obtains as the auto-convolution of the field correlator $\gamma(k)$, i.e.\ of a $d$-dimensional ball of radius $\sigma^{-1}$.
Thus, $C_d(q\sigma)$ is centered at $q=0$ and vanishes like $(2-q\sigma)^{\smash{\frac{d+1}{2}}}$ at $q=2\sigma^{-1}$.
Explicitly, in $d=1,2,3$ one has 
\begin{subequations}\label{eqSpeckleCorr}
\begin{align} 
C_1(q\sigma) 
   &= \frac{\pi}{2} (2 - q\sigma) \, \Theta(2-q\sigma) , \label{eqSpeckleCorr1D}\\
C_2(q\sigma) 
   &= \left[ 8\arccos\left(\textstyle{\frac{q\sigma}{2}}\right) - 2 q\sigma \sqrt{4-q^2\sigma^2}\right] \Theta(2-q\sigma) , \label{eqSpeckleCorr2D}\\
C_3(q\sigma) 
   &= \frac{3\pi^2}{8} (2-q\sigma)^2 (4 + q\sigma)  \,  \Theta(2-q\sigma) .\label{eqSpeckleCorr3D}
\end{align}
\end{subequations}
\autoref{figSpeckleCorrelationK} shows these isotropic correlation functions,
multiplied with the $d$-dimensional integration element
$S_d(q\sigma)^{d-1}/(2\pi)^d$, where $S_d$ is the unit-sphere surface
($S_1=2$, $S_2=2\pi$, $S_3=4\pi$). 
 
%-----------------------------------------------------------------------------------
\subsection{Analytical dispersion 
  corrections}\label{SpeckleDeltaC}

The speckle correlation functions
\eqref{eqSpeckleCorr} yield closed-form expressions for 
the dispersion correction \eqref{eqDeltaE} in 
important limiting cases.

\subsubsection{Hydrodynamic limit \texorpdfstring{$\xi = 0$}{xi = 0}}
In $d=1$, the correlation function \eqref{eqSpeckleCorr1D} is
piecewise linear, and the integral \eqref{eqDeltaE} over the
hydrodynamic kernel \eqref{eqKernelHydro} yields  the dispersion shift
\cite{Gaul2009a,Renner2009}
\begin{align}
\frac{\Delta \avg{\epsilon}_k}{\epsilon_k v^2} &=
-\frac{1}{2} 
- \frac{k\sigma}{8} \ln \left|\frac{1-k\sigma}{1+k\sigma} \right| 
+ \frac{k^2\sigma^2}{8} \ln \left|\frac{1-k^2\sigma^2}{k^2\sigma^2} \right|
 \label{eqHydroLimit1D}.
\end{align}
This result is plotted as the blue dotted curve in \autoref{figDeltaC_kxi0.05}.
It is non-analytic at $k\sigma = 1$, corresponding to the
non-analyticity of the speckle pair correlation function
\eqref{eqSpeckleCorr1D} at the boundary of its support.
Using Eq.\ \eqref{eqHydroLimit1D}, we can also write down the 
AVDOS shift \eqref{eqAVDOSsound} in closed form:  
\begin{equation}\label{eqSpHydro1D}
\frac{\Delta \avg{\rho}(\varepsilon \hbar c /\sigma)}{\rho(
  \varepsilon \hbar c /\sigma)v^2} 
= \frac{1}{2}
+ \frac{\varepsilon}{4}  \ln\left|\frac{1-\varepsilon}{1+\varepsilon}\right|  
	- \frac{3\varepsilon^2}{8} \ln\left|\frac{1-\varepsilon^2}{\varepsilon^2}\right| . 
\end{equation}
It shows a pronounced dip around $\varepsilon \approx 0.7$ and a sharp
logarithmic divergence at $\varepsilon=1$
\cite{Gaul2009a}, resulting from the non-analyticity of \eqref{eqSpHydro1D} at $k\sigma =
1$.

In higher dimensions, the integral  \eqref{eqDeltaE} gets more
complicated, but  we find partial analytical results in two dimensions.
Denote the angular part of the integral \eqref{eqDeltaE} over the hydrodynamic kernel \eqref{eqKernelHydro} by $A_2(q)$. 
We drop the principal value $\PV$ in Eq.\ \eqref{eqKernelHydro} and
re-insert the infinitesimal imaginary shift in the denominator. Then,
we can compute the angular integral analytically as a closed-path
integral in the complex plane $z=e^{i\beta}$, $\beta =
\measuredangle(\k,\q)$ 
\begin{align}\label{eq2DAngularIntegral}
 A_2(q) = - \frac{1}{2} \frac{S_2}{(2\pi)^2} \left[1 - \left(\frac{q}{2k} \right)^2 + k^2 \frac{1- {q^2}/(2k^2)}{q\sqrt{q^2-(2k)^2}} \right].
\end{align}
In general, the last term is too complicated for the remaining radial integral to be evaluated in closed
form. 
For $k\sigma > 1$, however, the integrand of Eq.\ \eqref{eqDeltaE} is
restricted to $q \leq 2/\sigma < 2k$. 
There, the last term in \eqref{eq2DAngularIntegral} is imaginary and does not contribute to $\Re \Sigma$, such that
\begin{equation}\label{eqLambda2Dgtr} 
\frac{\Delta \avg{\epsilon}_k}{\epsilon_k} =
-\frac{v^2}{2}\left(1-\frac{1}{8 k^2\sigma^2}\right), \quad
k\sigma>1. 
\end{equation}
In the AVDOS \eqref{eqAVDOSsound}, this leads to a totally flat plateau, as shown in Fig.~5 of Ref.~\cite{Gaul2009a}.

%%%%%%%%%%%%%%%%%%%%%%%%%%%%%%%%%%%%%%%%%%%%%%%%%%%%%%%%%%%%%%%%%%%%%%%%%%%
\subsubsection{Lowest-energy excitations, \texorpdfstring{$k\to 0$}{k=0}}
Also in the low-energy limit  (iii) of Sec.~\ref{secSpeedOfSoundLimits}, we can find analytical solutions.
The integral \eqref{DeltaCk0} with 
the speckle correlator \eqref{eqSpeckleCorr} evaluates to closed form in all
relevant dimensions:
\newcommand{\z}{\frac{\sigma}{\sqrt{2}\xi}}
\renewcommand{\z}{z}
\begin{subequations}\label{eqLimitK0}
\begin{alignat}{2}
\frac{\Delta \avg{c}}{c v^2}&= -\frac{3}{8} \z
\left[\cot^{-1}\big(\z\big) 
+ \frac{1}{3} \frac{\z}{1+\z^2}\right], & &d=1\label{eqLimitK01D} \\
\frac{\Delta \avg{c}}{c v^2} &= z^3 \frac{2 z \sqrt{1 + z^2} - 1 - 2 z^2}{\sqrt{1 + z^2}} \, , & &d=2	\\
\frac{\Delta \avg{c}}{c v^2} &= z^4\biggl[7 + \frac{5 \cot^{-1}(z)}{2 z}  - 
 (6 + 7 z^2) \log&&\left(\frac{1 + z^2}{z^2}\right)\biggr], \nonumber \\
  & &  &d=3
\end{alignat}
\end{subequations}
with $z = \zeta/\sqrt{2}$. 
All three cases are plotted in \autoref{figDeltaC_k0}; the result \eqref{eqLimitK01D}
also features in \autoref{figDeltaC_kxi0.05}. 

\bibliography{BogoLong}

%merlin.mbs 2010-03-15 4.21a (PWD, AO, DPC)
\begin{thebibliography}{86}%
\makeatletter
\providecommand \@ifxundefined [1]{%
 \@ifx{#1\undefined}
}%
\providecommand \@ifnum [1]{%
 \ifnum #1\expandafter \@firstoftwo
 \else \expandafter \@secondoftwo
 \fi
}%
\providecommand \@ifx [1]{%
 \ifx #1\expandafter \@firstoftwo
 \else \expandafter \@secondoftwo
 \fi
}%
\providecommand \natexlab [1]{#1}%
\providecommand \enquote  [1]{``#1''}%
\providecommand \bibnamefont  [1]{#1}%
\providecommand \bibfnamefont [1]{#1}%
\providecommand \citenamefont [1]{#1}%
\providecommand \href@noop [0]{\@secondoftwo}%
\providecommand \href [0]{\begingroup \@sanitize@url \@href}%
\providecommand \@href[1]{\@@startlink{#1}\@@href}%
\providecommand \@@href[1]{\endgroup#1\@@endlink}%
\providecommand \@sanitize@url [0]{\catcode `\\12\catcode `\$12\catcode
  `\&12\catcode `\#12\catcode `\^12\catcode `\_12\catcode `\%12\relax}%
\providecommand \@@startlink[1]{}%
\providecommand \@@endlink[0]{}%
\providecommand \url  [0]{\begingroup\@sanitize@url \@url }%
\providecommand \@url [1]{\endgroup\@href {#1}{\urlprefix }}%
\providecommand \urlprefix  [0]{URL }%
\providecommand \Eprint [0]{\href }%
\@ifxundefined \urlstyle {%
  \providecommand \doi  [0]{\begingroup \@sanitize@url \@doi}%
  \providecommand \@doi [1]{\endgroup \@@startlink {\doibase
  #1}doi:\discretionary {}{}{}#1\@@endlink }%
}{%
  \providecommand \doi  [0]{doi:\discretionary{}{}{}\begingroup
  \urlstyle{rm}\Url }%
}%
\providecommand \doibase [0]{http://dx.doi.org/}%
\providecommand \Doi [0]{\begingroup \@sanitize@url \@Doi }%
\providecommand \@Doi  [1]{\endgroup\@@startlink{\doibase#1}\@@Doi}%
\providecommand \@@Doi [1]{#1\@@endlink}%
\providecommand \selectlanguage [0]{\@gobble}%
\providecommand \bibinfo  [0]{\@secondoftwo}%
\providecommand \bibfield  [0]{\@secondoftwo}%
\providecommand \translation [1]{[#1]}%
\providecommand \BibitemOpen [0]{}%
\providecommand \bibitemStop [0]{}%
\providecommand \bibitemNoStop [0]{.\EOS\space}%
\providecommand \EOS [0]{\spacefactor3000\relax}%
\providecommand \BibitemShut  [1]{\csname bibitem#1\endcsname}%
%</preamble>
\bibitem [{\citenamefont {Giamarchi}\ and\ \citenamefont
  {Schulz}(1987)}]{Giamarchi1987}%
  \BibitemOpen
  \bibfield  {author} {\bibinfo {author} {\bibfnamefont {T.}~\bibnamefont
  {Giamarchi}}\ and\ \bibinfo {author} {\bibfnamefont {H.~J.}\ \bibnamefont
  {Schulz}},\ }\Doi {10.1209/0295-5075/3/12/007} {\bibfield  {journal}
  {\bibinfo  {journal} {Europhys. Lett.},\ }\textbf {\bibinfo {volume} {3}},\
  \bibinfo {pages} {1287} (\bibinfo {year} {1987})}\BibitemShut {NoStop}%
\bibitem [{\citenamefont {Giamarchi}\ and\ \citenamefont
  {Schulz}(1988)}]{Giamarchi1988}%
  \BibitemOpen
  \bibfield  {author} {\bibinfo {author} {\bibfnamefont {T.}~\bibnamefont
  {Giamarchi}}\ and\ \bibinfo {author} {\bibfnamefont {H.~J.}\ \bibnamefont
  {Schulz}},\ }\Doi {10.1103/PhysRevB.37.325} {\bibfield  {journal} {\bibinfo
  {journal} {Phys. Rev. B},\ }\textbf {\bibinfo {volume} {37}},\ \bibinfo
  {pages} {325} (\bibinfo {year} {1988})}\BibitemShut {NoStop}%
\bibitem [{\citenamefont {Fisher}\ \emph {et~al.}(1989)\citenamefont {Fisher},
  \citenamefont {Weichman}, \citenamefont {Grinstein},\ and\ \citenamefont
  {Fisher}}]{Fisher1989}%
  \BibitemOpen
  \bibfield  {author} {\bibinfo {author} {\bibfnamefont {M.~P.~A.}\
  \bibnamefont {Fisher}}, \bibinfo {author} {\bibfnamefont {P.~B.}\
  \bibnamefont {Weichman}}, \bibinfo {author} {\bibfnamefont {G.}~\bibnamefont
  {Grinstein}}, \ and\ \bibinfo {author} {\bibfnamefont {D.~S.}\ \bibnamefont
  {Fisher}},\ }\Doi {10.1103/PhysRevB.40.546} {\bibfield  {journal} {\bibinfo
  {journal} {Phys. Rev. B},\ }\textbf {\bibinfo {volume} {40}},\ \bibinfo
  {pages} {546} (\bibinfo {year} {1989})}\BibitemShut {NoStop}%
\bibitem [{\citenamefont {Crooker}\ \emph {et~al.}(1983)\citenamefont
  {Crooker}, \citenamefont {Hebral}, \citenamefont {Smith}, \citenamefont
  {Takano},\ and\ \citenamefont {Reppy}}]{Crooker1983}%
  \BibitemOpen
  \bibfield  {author} {\bibinfo {author} {\bibfnamefont {B.~C.}\ \bibnamefont
  {Crooker}}, \bibinfo {author} {\bibfnamefont {B.}~\bibnamefont {Hebral}},
  \bibinfo {author} {\bibfnamefont {E.~N.}\ \bibnamefont {Smith}}, \bibinfo
  {author} {\bibfnamefont {Y.}~\bibnamefont {Takano}}, \ and\ \bibinfo {author}
  {\bibfnamefont {J.~D.}\ \bibnamefont {Reppy}},\ }\Doi
  {10.1103/PhysRevLett.51.666} {\bibfield  {journal} {\bibinfo  {journal}
  {Phys. Rev. Lett.},\ }\textbf {\bibinfo {volume} {51}},\ \bibinfo {pages}
  {666} (\bibinfo {year} {1983})}\BibitemShut {NoStop}%
\bibitem [{\citenamefont {Chan}\ \emph {et~al.}(1988)\citenamefont {Chan},
  \citenamefont {Blum}, \citenamefont {Murphy}, \citenamefont {Wong},\ and\
  \citenamefont {Reppy}}]{Chan1988}%
  \BibitemOpen
  \bibfield  {author} {\bibinfo {author} {\bibfnamefont {M.~H.~W.}\
  \bibnamefont {Chan}}, \bibinfo {author} {\bibfnamefont {K.~I.}\ \bibnamefont
  {Blum}}, \bibinfo {author} {\bibfnamefont {S.~Q.}\ \bibnamefont {Murphy}},
  \bibinfo {author} {\bibfnamefont {G.~K.~S.}\ \bibnamefont {Wong}}, \ and\
  \bibinfo {author} {\bibfnamefont {J.~D.}\ \bibnamefont {Reppy}},\ }\Doi
  {10.1103/PhysRevLett.61.1950} {\bibfield  {journal} {\bibinfo  {journal}
  {Phys. Rev. Lett.},\ }\textbf {\bibinfo {volume} {61}},\ \bibinfo {pages}
  {1950} (\bibinfo {year} {1988})}\BibitemShut {NoStop}%
\bibitem [{\citenamefont {Wong}\ \emph {et~al.}(1990)\citenamefont {Wong},
  \citenamefont {Crowell}, \citenamefont {Cho},\ and\ \citenamefont
  {Reppy}}]{Wong1990}%
  \BibitemOpen
  \bibfield  {author} {\bibinfo {author} {\bibfnamefont {G.~K.~S.}\
  \bibnamefont {Wong}}, \bibinfo {author} {\bibfnamefont {P.~A.}\ \bibnamefont
  {Crowell}}, \bibinfo {author} {\bibfnamefont {H.~A.}\ \bibnamefont {Cho}}, \
  and\ \bibinfo {author} {\bibfnamefont {J.~D.}\ \bibnamefont {Reppy}},\ }\Doi
  {10.1103/PhysRevLett.65.2410} {\bibfield  {journal} {\bibinfo  {journal}
  {Phys. Rev. Lett.},\ }\textbf {\bibinfo {volume} {65}},\ \bibinfo {pages}
  {2410} (\bibinfo {year} {1990})}\BibitemShut {NoStop}%
\bibitem [{\citenamefont {Cl{\'e}ment}\ \emph {et~al.}(2005)\citenamefont
  {Cl{\'e}ment}, \citenamefont {Var{\'o}n}, \citenamefont {Retter},
  \citenamefont {Bouyer}, \citenamefont {Sanchez-Palencia}, \citenamefont
  {Gangardt}, \citenamefont {Shlyapnikov},\ and\ \citenamefont
  {Aspect}}]{Clement2005}%
  \BibitemOpen
  \bibfield  {author} {\bibinfo {author} {\bibfnamefont {D.}~\bibnamefont
  {Cl{\'e}ment}}, \bibinfo {author} {\bibfnamefont {A.~F.}\ \bibnamefont
  {Var{\'o}n}}, \bibinfo {author} {\bibfnamefont {J.~A.}\ \bibnamefont
  {Retter}}, \bibinfo {author} {\bibfnamefont {P.}~\bibnamefont {Bouyer}},
  \bibinfo {author} {\bibfnamefont {L.}~\bibnamefont {Sanchez-Palencia}},
  \bibinfo {author} {\bibfnamefont {D.}~\bibnamefont {Gangardt}}, \bibinfo
  {author} {\bibfnamefont {G.~V.}\ \bibnamefont {Shlyapnikov}}, \ and\ \bibinfo
  {author} {\bibfnamefont {A.}~\bibnamefont {Aspect}},\ }\Doi
  {10.1103/PhysRevLett.95.170409} {\bibfield  {journal} {\bibinfo  {journal}
  {Phys. Rev. Lett.},\ }\textbf {\bibinfo {volume} {95}},\ \bibinfo {pages}
  {170409} (\bibinfo {year} {2005})}\BibitemShut {NoStop}%
\bibitem [{\citenamefont {Schulte}\ \emph {et~al.}(2005)\citenamefont
  {Schulte}, \citenamefont {Denkelforth}, \citenamefont {Kruse}, \citenamefont
  {Ertmer}, \citenamefont {Arlt}, \citenamefont {Sacha}, \citenamefont
  {Zakrzewski},\ and\ \citenamefont {Lewenstein}}]{Schulte2005}%
  \BibitemOpen
  \bibfield  {author} {\bibinfo {author} {\bibfnamefont {T.}~\bibnamefont
  {Schulte}}, \bibinfo {author} {\bibfnamefont {S.}~\bibnamefont
  {Denkelforth}}, \bibinfo {author} {\bibfnamefont {J.}~\bibnamefont {Kruse}},
  \bibinfo {author} {\bibfnamefont {W.}~\bibnamefont {Ertmer}}, \bibinfo
  {author} {\bibfnamefont {J.}~\bibnamefont {Arlt}}, \bibinfo {author}
  {\bibfnamefont {K.}~\bibnamefont {Sacha}}, \bibinfo {author} {\bibfnamefont
  {J.}~\bibnamefont {Zakrzewski}}, \ and\ \bibinfo {author} {\bibfnamefont
  {M.}~\bibnamefont {Lewenstein}},\ }\Doi {10.1103/PhysRevLett.95.170411}
  {\bibfield  {journal} {\bibinfo  {journal} {Phys. Rev. Lett.},\ }\textbf
  {\bibinfo {volume} {95}},\ \bibinfo {pages} {170411} (\bibinfo {year}
  {2005})}\BibitemShut {NoStop}%
\bibitem [{\citenamefont {Lye}\ \emph {et~al.}(2005)\citenamefont {Lye},
  \citenamefont {Fallani}, \citenamefont {Modugno}, \citenamefont {Wiersma},
  \citenamefont {Fort},\ and\ \citenamefont {Inguscio}}]{Lye2005}%
  \BibitemOpen
  \bibfield  {author} {\bibinfo {author} {\bibfnamefont {J.~E.}\ \bibnamefont
  {Lye}}, \bibinfo {author} {\bibfnamefont {L.}~\bibnamefont {Fallani}},
  \bibinfo {author} {\bibfnamefont {M.}~\bibnamefont {Modugno}}, \bibinfo
  {author} {\bibfnamefont {D.~S.}\ \bibnamefont {Wiersma}}, \bibinfo {author}
  {\bibfnamefont {C.}~\bibnamefont {Fort}}, \ and\ \bibinfo {author}
  {\bibfnamefont {M.}~\bibnamefont {Inguscio}},\ }\Doi
  {10.1103/PhysRevLett.95.070401} {\bibfield  {journal} {\bibinfo  {journal}
  {Phys. Rev. Lett.},\ }\textbf {\bibinfo {volume} {95}},\ \bibinfo {pages}
  {070401} (\bibinfo {year} {2005})}\BibitemShut {NoStop}%
\bibitem [{\citenamefont {Chen}\ \emph {et~al.}(2008)\citenamefont {Chen},
  \citenamefont {Hitchcock}, \citenamefont {Dries}, \citenamefont {Junker},
  \citenamefont {Welford},\ and\ \citenamefont {Hulet}}]{Chen2008}%
  \BibitemOpen
  \bibfield  {author} {\bibinfo {author} {\bibfnamefont {Y.~P.}\ \bibnamefont
  {Chen}}, \bibinfo {author} {\bibfnamefont {J.}~\bibnamefont {Hitchcock}},
  \bibinfo {author} {\bibfnamefont {D.}~\bibnamefont {Dries}}, \bibinfo
  {author} {\bibfnamefont {M.}~\bibnamefont {Junker}}, \bibinfo {author}
  {\bibfnamefont {C.}~\bibnamefont {Welford}}, \ and\ \bibinfo {author}
  {\bibfnamefont {R.~G.}\ \bibnamefont {Hulet}},\ }\Doi
  {10.1103/PhysRevA.77.033632} {\bibfield  {journal} {\bibinfo  {journal}
  {Phys. Rev. A},\ }\textbf {\bibinfo {volume} {77}},\ \bibinfo {eid} {033632}
  (\bibinfo {year} {2008})}\BibitemShut {NoStop}%
\bibitem [{\citenamefont {White}\ \emph {et~al.}(2009)\citenamefont {White},
  \citenamefont {Pasienski}, \citenamefont {McKay}, \citenamefont {Zhou},
  \citenamefont {Ceperley},\ and\ \citenamefont {DeMarco}}]{White2009}%
  \BibitemOpen
  \bibfield  {author} {\bibinfo {author} {\bibfnamefont {M.}~\bibnamefont
  {White}}, \bibinfo {author} {\bibfnamefont {M.}~\bibnamefont {Pasienski}},
  \bibinfo {author} {\bibfnamefont {D.}~\bibnamefont {McKay}}, \bibinfo
  {author} {\bibfnamefont {S.~Q.}\ \bibnamefont {Zhou}}, \bibinfo {author}
  {\bibfnamefont {D.}~\bibnamefont {Ceperley}}, \ and\ \bibinfo {author}
  {\bibfnamefont {B.}~\bibnamefont {DeMarco}},\ }\Doi
  {10.1103/PhysRevLett.102.055301} {\bibfield  {journal} {\bibinfo  {journal}
  {Phys. Rev. Lett.},\ }\textbf {\bibinfo {volume} {102}},\ \bibinfo {eid}
  {055301} (\bibinfo {year} {2009})}\BibitemShut {NoStop}%
\bibitem [{\citenamefont {Dries}\ \emph {et~al.}(2010)\citenamefont {Dries},
  \citenamefont {Pollack}, \citenamefont {Hitchcock},\ and\ \citenamefont
  {Hulet}}]{Dries2010}%
  \BibitemOpen
  \bibfield  {author} {\bibinfo {author} {\bibfnamefont {D.}~\bibnamefont
  {Dries}}, \bibinfo {author} {\bibfnamefont {S.~E.}\ \bibnamefont {Pollack}},
  \bibinfo {author} {\bibfnamefont {J.~M.}\ \bibnamefont {Hitchcock}}, \ and\
  \bibinfo {author} {\bibfnamefont {R.~G.}\ \bibnamefont {Hulet}},\ }\Doi
  {10.1103/PhysRevA.82.033603} {\bibfield  {journal} {\bibinfo  {journal}
  {Phys. Rev. A},\ }\textbf {\bibinfo {volume} {82}},\ \bibinfo {pages}
  {033603} (\bibinfo {year} {2010})}\BibitemShut {NoStop}%
\bibitem [{\citenamefont {Bogoliubov}(1947)}]{Bogoliubov1947}%
  \BibitemOpen
  \bibfield  {author} {\bibinfo {author} {\bibfnamefont {N.~N.}\ \bibnamefont
  {Bogoliubov}},\ }\href@noop {} {\bibfield  {journal} {\bibinfo  {journal}
  {Journal of Physics (Moscow)},\ }\textbf {\bibinfo {volume} {11}},\ \bibinfo
  {pages} {23} (\bibinfo {year} {1947})}\BibitemShut {NoStop}%
\bibitem [{\citenamefont {Nozi{\`e}res}\ and\ \citenamefont
  {Pines}(1999)}]{Nozieres1999}%
  \BibitemOpen
  \bibfield  {author} {\bibinfo {author} {\bibfnamefont {P.}~\bibnamefont
  {Nozi{\`e}res}}\ and\ \bibinfo {author} {\bibfnamefont {D.}~\bibnamefont
  {Pines}},\ }\href@noop {} {\emph {\bibinfo {title} {The Theory of Quantum
  Liquids}}}\ (\bibinfo  {publisher} {Perseus Books, Cambridge, MA},\ \bibinfo
  {year} {1999})\BibitemShut {NoStop}%
\bibitem [{\citenamefont {Lee}\ and\ \citenamefont {Gunn}(1990)}]{Lee1990}%
  \BibitemOpen
  \bibfield  {author} {\bibinfo {author} {\bibfnamefont {D.~K.~K.}\
  \bibnamefont {Lee}}\ and\ \bibinfo {author} {\bibfnamefont {J.~M.~F.}\
  \bibnamefont {Gunn}},\ }\Doi {10.1088/0953-8984/2/38/004} {\bibfield
  {journal} {\bibinfo  {journal} {J. Phys.: Condens. Matter},\ }\textbf
  {\bibinfo {volume} {2}},\ \bibinfo {pages} {7753} (\bibinfo {year}
  {1990})}\BibitemShut {NoStop}%
\bibitem [{\citenamefont {Huang}\ and\ \citenamefont {Meng}(1992)}]{Huang1992}%
  \BibitemOpen
  \bibfield  {author} {\bibinfo {author} {\bibfnamefont {K.}~\bibnamefont
  {Huang}}\ and\ \bibinfo {author} {\bibfnamefont {H.-F.}\ \bibnamefont
  {Meng}},\ }\Doi {10.1103/PhysRevLett.69.644} {\bibfield  {journal} {\bibinfo
  {journal} {Phys. Rev. Lett.},\ }\textbf {\bibinfo {volume} {69}},\ \bibinfo
  {pages} {644} (\bibinfo {year} {1992})}\BibitemShut {NoStop}%
\bibitem [{\citenamefont {Giorgini}\ \emph {et~al.}(1994)\citenamefont
  {Giorgini}, \citenamefont {Pitaevskii},\ and\ \citenamefont
  {Stringari}}]{Giorgini1994}%
  \BibitemOpen
  \bibfield  {author} {\bibinfo {author} {\bibfnamefont {S.}~\bibnamefont
  {Giorgini}}, \bibinfo {author} {\bibfnamefont {L.}~\bibnamefont
  {Pitaevskii}}, \ and\ \bibinfo {author} {\bibfnamefont {S.}~\bibnamefont
  {Stringari}},\ }\Doi {10.1103/PhysRevB.49.12938} {\bibfield  {journal}
  {\bibinfo  {journal} {Phys. Rev. B},\ }\textbf {\bibinfo {volume} {49}},\
  \bibinfo {pages} {12938} (\bibinfo {year} {1994})}\BibitemShut {NoStop}%
\bibitem [{\citenamefont {Kobayashi}\ and\ \citenamefont
  {Tsubota}(2002)}]{Kobayashi2002}%
  \BibitemOpen
  \bibfield  {author} {\bibinfo {author} {\bibfnamefont {M.}~\bibnamefont
  {Kobayashi}}\ and\ \bibinfo {author} {\bibfnamefont {M.}~\bibnamefont
  {Tsubota}},\ }\Doi {10.1103/PhysRevB.66.174516} {\bibfield  {journal}
  {\bibinfo  {journal} {Phys. Rev. B},\ }\textbf {\bibinfo {volume} {66}},\
  \bibinfo {pages} {174516} (\bibinfo {year} {2002})}\BibitemShut {NoStop}%
\bibitem [{\citenamefont {Astrakharchik}\ \emph {et~al.}(2002)\citenamefont
  {Astrakharchik}, \citenamefont {Boronat}, \citenamefont {Casulleras},\ and\
  \citenamefont {Giorgini}}]{Astrakharchik2002}%
  \BibitemOpen
  \bibfield  {author} {\bibinfo {author} {\bibfnamefont {G.~E.}\ \bibnamefont
  {Astrakharchik}}, \bibinfo {author} {\bibfnamefont {J.}~\bibnamefont
  {Boronat}}, \bibinfo {author} {\bibfnamefont {J.}~\bibnamefont {Casulleras}},
  \ and\ \bibinfo {author} {\bibfnamefont {S.}~\bibnamefont {Giorgini}},\ }\Doi
  {10.1103/PhysRevA.66.023603} {\bibfield  {journal} {\bibinfo  {journal}
  {Phys. Rev. A},\ }\textbf {\bibinfo {volume} {66}},\ \bibinfo {pages}
  {023603} (\bibinfo {year} {2002})}\BibitemShut {NoStop}%
\bibitem [{\citenamefont {Bilas}\ and\ \citenamefont
  {Pavloff}(2006)}]{Bilas2006}%
  \BibitemOpen
  \bibfield  {author} {\bibinfo {author} {\bibfnamefont {N.}~\bibnamefont
  {Bilas}}\ and\ \bibinfo {author} {\bibfnamefont {N.}~\bibnamefont
  {Pavloff}},\ }\Doi {10.1140/epjd/e2006-00166-3} {\bibfield  {journal}
  {\bibinfo  {journal} {Eur. Phys. J. D},\ }\textbf {\bibinfo {volume} {40}},\
  \bibinfo {pages} {387} (\bibinfo {year} {2006})}\BibitemShut {NoStop}%
\bibitem [{\citenamefont {Lugan}\ \emph
  {et~al.}(2007){\natexlab{a}}\citenamefont {Lugan}, \citenamefont
  {Cl{\'e}ment}, \citenamefont {Bouyer}, \citenamefont {Aspect},\ and\
  \citenamefont {Sanchez-Palencia}}]{Lugan2007a}%
  \BibitemOpen
  \bibfield  {author} {\bibinfo {author} {\bibfnamefont {P.}~\bibnamefont
  {Lugan}}, \bibinfo {author} {\bibfnamefont {D.}~\bibnamefont {Cl{\'e}ment}},
  \bibinfo {author} {\bibfnamefont {P.}~\bibnamefont {Bouyer}}, \bibinfo
  {author} {\bibfnamefont {A.}~\bibnamefont {Aspect}}, \ and\ \bibinfo {author}
  {\bibfnamefont {L.}~\bibnamefont {Sanchez-Palencia}},\ }\Doi
  {10.1103/PhysRevLett.99.180402} {\bibfield  {journal} {\bibinfo  {journal}
  {Phys. Rev. Lett.},\ }\textbf {\bibinfo {volume} {99}},\ \bibinfo {eid}
  {180402} (\bibinfo {year} {2007}{\natexlab{a}})}\BibitemShut {NoStop}%
\bibitem [{\citenamefont {Falco}\ \emph {et~al.}(2007)\citenamefont {Falco},
  \citenamefont {Pelster},\ and\ \citenamefont {Graham}}]{Falco2007}%
  \BibitemOpen
  \bibfield  {author} {\bibinfo {author} {\bibfnamefont {G.~M.}\ \bibnamefont
  {Falco}}, \bibinfo {author} {\bibfnamefont {A.}~\bibnamefont {Pelster}}, \
  and\ \bibinfo {author} {\bibfnamefont {R.}~\bibnamefont {Graham}},\ }\Doi
  {10.1103/PhysRevA.75.063619} {\bibfield  {journal} {\bibinfo  {journal}
  {Phys. Rev. A},\ }\textbf {\bibinfo {volume} {75}},\ \bibinfo {eid} {063619}
  (\bibinfo {year} {2007})}\BibitemShut {NoStop}%
\bibitem [{\citenamefont {Fontanesi}\ \emph {et~al.}(2009)\citenamefont
  {Fontanesi}, \citenamefont {Wouters},\ and\ \citenamefont
  {Savona}}]{Fontanesi2009}%
  \BibitemOpen
  \bibfield  {author} {\bibinfo {author} {\bibfnamefont {L.}~\bibnamefont
  {Fontanesi}}, \bibinfo {author} {\bibfnamefont {M.}~\bibnamefont {Wouters}},
  \ and\ \bibinfo {author} {\bibfnamefont {V.}~\bibnamefont {Savona}},\ }\Doi
  {10.1103/PhysRevLett.103.030403} {\bibfield  {journal} {\bibinfo  {journal}
  {Phys. Rev. Lett.},\ }\textbf {\bibinfo {volume} {103}},\ \bibinfo {pages}
  {030403} (\bibinfo {year} {2009})}\BibitemShut {NoStop}%
\bibitem [{\citenamefont {Hu}\ \emph {et~al.}(2009)\citenamefont {Hu},
  \citenamefont {Liang},\ and\ \citenamefont {Hu}}]{Hu2009}%
  \BibitemOpen
  \bibfield  {author} {\bibinfo {author} {\bibfnamefont {Y.}~\bibnamefont
  {Hu}}, \bibinfo {author} {\bibfnamefont {Z.}~\bibnamefont {Liang}}, \ and\
  \bibinfo {author} {\bibfnamefont {B.}~\bibnamefont {Hu}},\ }\Doi
  {10.1103/PhysRevA.80.043629} {\bibfield  {journal} {\bibinfo  {journal}
  {Phys. Rev. A},\ }\textbf {\bibinfo {volume} {80}},\ \bibinfo {pages}
  {043629} (\bibinfo {year} {2009})}\BibitemShut {NoStop}%
\bibitem [{\citenamefont {Lewenstein}\ \emph {et~al.}(2007)\citenamefont
  {Lewenstein}, \citenamefont {Sanpera}, \citenamefont {Ahufinger},
  \citenamefont {Damski}, \citenamefont {Sen},\ and\ \citenamefont
  {Sen}}]{Lewenstein2007}%
  \BibitemOpen
  \bibfield  {author} {\bibinfo {author} {\bibfnamefont {M.}~\bibnamefont
  {Lewenstein}}, \bibinfo {author} {\bibfnamefont {A.}~\bibnamefont {Sanpera}},
  \bibinfo {author} {\bibfnamefont {V.}~\bibnamefont {Ahufinger}}, \bibinfo
  {author} {\bibfnamefont {B.}~\bibnamefont {Damski}}, \bibinfo {author}
  {\bibfnamefont {A.}~\bibnamefont {Sen}}, \ and\ \bibinfo {author}
  {\bibfnamefont {U.}~\bibnamefont {Sen}},\ }\Doi {10.1080/00018730701223200}
  {\bibfield  {journal} {\bibinfo  {journal} {Adv. Phys.},\ }\textbf {\bibinfo
  {volume} {56}},\ \bibinfo {pages} {243} (\bibinfo {year} {2007})}\BibitemShut
  {NoStop}%
\bibitem [{\citenamefont {Bloch}\ \emph {et~al.}(2008)\citenamefont {Bloch},
  \citenamefont {Dalibard},\ and\ \citenamefont {Zwerger}}]{Bloch2008}%
  \BibitemOpen
  \bibfield  {author} {\bibinfo {author} {\bibfnamefont {I.}~\bibnamefont
  {Bloch}}, \bibinfo {author} {\bibfnamefont {J.}~\bibnamefont {Dalibard}}, \
  and\ \bibinfo {author} {\bibfnamefont {W.}~\bibnamefont {Zwerger}},\ }\Doi
  {10.1103/RevModPhys.80.885} {\bibfield  {journal} {\bibinfo  {journal} {Rev.
  Mod. Phys.},\ }\textbf {\bibinfo {volume} {80}},\ \bibinfo {eid} {885}
  (\bibinfo {year} {2008})}\BibitemShut {NoStop}%
\bibitem [{\citenamefont {Sanchez-Palencia}\ and\ \citenamefont
  {Lewenstein}(2010)}]{Sanchez-Palencia2010}%
  \BibitemOpen
  \bibfield  {author} {\bibinfo {author} {\bibfnamefont {L.}~\bibnamefont
  {Sanchez-Palencia}}\ and\ \bibinfo {author} {\bibfnamefont {M.}~\bibnamefont
  {Lewenstein}},\ }\Doi {10.1038/nphys1507} {\bibfield  {journal} {\bibinfo
  {journal} {Nat. Phys.},\ }\textbf {\bibinfo {volume} {6}},\ \bibinfo {pages}
  {87} (\bibinfo {year} {2010})}\BibitemShut {NoStop}%
\bibitem [{\citenamefont {Modugno}(2010)}]{Modugno2010}%
  \BibitemOpen
  \bibfield  {author} {\bibinfo {author} {\bibfnamefont {G.}~\bibnamefont
  {Modugno}},\ }\Doi {10.1088/0034-4885/73/10/102401} {\bibfield  {journal}
  {\bibinfo  {journal} {Rep. Prog. Phys.},\ }\textbf {\bibinfo {volume} {73}},\
  \bibinfo {pages} {102401} (\bibinfo {year} {2010})}\BibitemShut {NoStop}%
\bibitem [{\citenamefont {{Robert-de-Saint-Vincent}}\ \emph
  {et~al.}(2010)\citenamefont {{Robert-de-Saint-Vincent}}, \citenamefont
  {Brantut}, \citenamefont {Allard}, \citenamefont {Plisson}, \citenamefont
  {Pezz\'e}, \citenamefont {Sanchez-Palencia}, \citenamefont {Aspect},
  \citenamefont {Bourdel},\ and\ \citenamefont
  {Bouyer}}]{Robert-de-Saint-Vincent2010}%
  \BibitemOpen
  \bibfield  {author} {\bibinfo {author} {\bibfnamefont {M.}~\bibnamefont
  {{Robert-de-Saint-Vincent}}}, \bibinfo {author} {\bibfnamefont {J.-P.}\
  \bibnamefont {Brantut}}, \bibinfo {author} {\bibfnamefont {B.}~\bibnamefont
  {Allard}}, \bibinfo {author} {\bibfnamefont {T.}~\bibnamefont {Plisson}},
  \bibinfo {author} {\bibfnamefont {L.}~\bibnamefont {Pezz\'e}}, \bibinfo
  {author} {\bibfnamefont {L.}~\bibnamefont {Sanchez-Palencia}}, \bibinfo
  {author} {\bibfnamefont {A.}~\bibnamefont {Aspect}}, \bibinfo {author}
  {\bibfnamefont {T.}~\bibnamefont {Bourdel}}, \ and\ \bibinfo {author}
  {\bibfnamefont {P.}~\bibnamefont {Bouyer}},\ }\Doi
  {10.1103/PhysRevLett.104.220602} {\bibfield  {journal} {\bibinfo  {journal}
  {Phys. Rev. Lett.},\ }\textbf {\bibinfo {volume} {104}},\ \bibinfo {pages}
  {220602} (\bibinfo {year} {2010})}\BibitemShut {NoStop}%
\bibitem [{\citenamefont {Kleinert}(2009)}]{Kleinert2009}%
  \BibitemOpen
  \bibfield  {author} {\bibinfo {author} {\bibfnamefont {H.}~\bibnamefont
  {Kleinert}},\ }\href@noop {} {\emph {\bibinfo {title} {Path Integrals in
  Quantum Mechanics, Statistics, Polymer Physics, and Financial Markets}}}\
  (\bibinfo  {publisher} {World Scientific, Singapore},\ \bibinfo {year}
  {2009})\BibitemShut {NoStop}%
\bibitem [{\citenamefont {Giamarchi}\ and\ \citenamefont
  {Le~Doussal}(1996)}]{Giamarchi1996}%
  \BibitemOpen
  \bibfield  {author} {\bibinfo {author} {\bibfnamefont {T.}~\bibnamefont
  {Giamarchi}}\ and\ \bibinfo {author} {\bibfnamefont {P.}~\bibnamefont
  {Le~Doussal}},\ }\Doi {10.1103/PhysRevB.53.15206} {\bibfield  {journal}
  {\bibinfo  {journal} {Phys. Rev. B},\ }\textbf {\bibinfo {volume} {53}},\
  \bibinfo {pages} {15206} (\bibinfo {year} {1996})}\BibitemShut {NoStop}%
\bibitem [{\citenamefont {Goldstone}\ \emph {et~al.}(1962)\citenamefont
  {Goldstone}, \citenamefont {Salam},\ and\ \citenamefont
  {Weinberg}}]{Goldstone1962}%
  \BibitemOpen
  \bibfield  {author} {\bibinfo {author} {\bibfnamefont {J.}~\bibnamefont
  {Goldstone}}, \bibinfo {author} {\bibfnamefont {A.}~\bibnamefont {Salam}}, \
  and\ \bibinfo {author} {\bibfnamefont {S.}~\bibnamefont {Weinberg}},\ }\Doi
  {10.1103/PhysRev.127.965} {\bibfield  {journal} {\bibinfo  {journal} {Phys.
  Rev.},\ }\textbf {\bibinfo {volume} {127}},\ \bibinfo {pages} {965} (\bibinfo
  {year} {1962})}\BibitemShut {NoStop}%
\bibitem [{\citenamefont {Dalfovo}\ \emph {et~al.}(1999)\citenamefont
  {Dalfovo}, \citenamefont {Giorgini}, \citenamefont {Pitaevskii},\ and\
  \citenamefont {Stringari}}]{Dalfovo1999}%
  \BibitemOpen
  \bibfield  {author} {\bibinfo {author} {\bibfnamefont {F.}~\bibnamefont
  {Dalfovo}}, \bibinfo {author} {\bibfnamefont {S.}~\bibnamefont {Giorgini}},
  \bibinfo {author} {\bibfnamefont {L.~P.}\ \bibnamefont {Pitaevskii}}, \ and\
  \bibinfo {author} {\bibfnamefont {S.}~\bibnamefont {Stringari}},\ }\Doi
  {10.1103/RevModPhys.71.463} {\bibfield  {journal} {\bibinfo  {journal} {Rev.
  Mod. Phys},\ }\textbf {\bibinfo {volume} {71}},\ \bibinfo {pages} {463}
  (\bibinfo {year} {1999})}\BibitemShut {NoStop}%
\bibitem [{\citenamefont {Pethick}\ and\ \citenamefont
  {Smith}(2002)}]{Pethick2002}%
  \BibitemOpen
  \bibfield  {author} {\bibinfo {author} {\bibfnamefont {C.~J.}\ \bibnamefont
  {Pethick}}\ and\ \bibinfo {author} {\bibfnamefont {H.}~\bibnamefont
  {Smith}},\ }\href@noop {} {\emph {\bibinfo {title} {{B}ose-{E}instein
  condensation in dilute gases}}}\ (\bibinfo  {publisher} {Cambridge Univ.
  Press},\ \bibinfo {year} {2002})\BibitemShut {NoStop}%
\bibitem [{\citenamefont {Pitaevskii}\ and\ \citenamefont
  {Stringari}(2003)}]{Pitaevskii2003}%
  \BibitemOpen
  \bibfield  {author} {\bibinfo {author} {\bibfnamefont {L.}~\bibnamefont
  {Pitaevskii}}\ and\ \bibinfo {author} {\bibfnamefont {S.}~\bibnamefont
  {Stringari}},\ }\href@noop {} {\emph {\bibinfo {title} {{B}ose-{E}instein
  condensation}}}\ (\bibinfo  {publisher} {Clarendon Press, Oxford},\ \bibinfo
  {year} {2003})\BibitemShut {NoStop}%
\bibitem [{\citenamefont {Lopatin}\ and\ \citenamefont
  {Vinokur}(2002)}]{Lopatin2002}%
  \BibitemOpen
  \bibfield  {author} {\bibinfo {author} {\bibfnamefont {A.~V.}\ \bibnamefont
  {Lopatin}}\ and\ \bibinfo {author} {\bibfnamefont {V.~M.}\ \bibnamefont
  {Vinokur}},\ }\Doi {10.1103/PhysRevLett.88.235503} {\bibfield  {journal}
  {\bibinfo  {journal} {Phys. Rev. Lett.},\ }\textbf {\bibinfo {volume} {88}},\
  \bibinfo {pages} {235503} (\bibinfo {year} {2002})}\BibitemShut {NoStop}%
\bibitem [{\citenamefont {Yukalov}\ and\ \citenamefont
  {Graham}(2007)}]{Yukalov2007}%
  \BibitemOpen
  \bibfield  {author} {\bibinfo {author} {\bibfnamefont {V.~I.}\ \bibnamefont
  {Yukalov}}\ and\ \bibinfo {author} {\bibfnamefont {R.}~\bibnamefont
  {Graham}},\ }\Doi {10.1103/PhysRevA.75.023619} {\bibfield  {journal}
  {\bibinfo  {journal} {Phys. Rev. A},\ }\textbf {\bibinfo {volume} {75}},\
  \bibinfo {eid} {023619} (\bibinfo {year} {2007})}\BibitemShut {NoStop}%
\bibitem [{\citenamefont {Yukalov}\ \emph {et~al.}(2007)\citenamefont
  {Yukalov}, \citenamefont {Yukalova}, \citenamefont {Krutitsky},\ and\
  \citenamefont {Graham}}]{Yukalov2007a}%
  \BibitemOpen
  \bibfield  {author} {\bibinfo {author} {\bibfnamefont {V.~I.}\ \bibnamefont
  {Yukalov}}, \bibinfo {author} {\bibfnamefont {E.~P.}\ \bibnamefont
  {Yukalova}}, \bibinfo {author} {\bibfnamefont {K.~V.}\ \bibnamefont
  {Krutitsky}}, \ and\ \bibinfo {author} {\bibfnamefont {R.}~\bibnamefont
  {Graham}},\ }\Doi {10.1103/PhysRevA.76.053623} {\bibfield  {journal}
  {\bibinfo  {journal} {Phys. Rev. A},\ }\textbf {\bibinfo {volume} {76}},\
  \bibinfo {eid} {053623} (\bibinfo {year} {2007})}\BibitemShut {NoStop}%
\bibitem [{\citenamefont {Gaul}\ \emph {et~al.}(2009)\citenamefont {Gaul},
  \citenamefont {Renner},\ and\ \citenamefont {M\"{u}ller}}]{Gaul2009a}%
  \BibitemOpen
  \bibfield  {author} {\bibinfo {author} {\bibfnamefont {C.}~\bibnamefont
  {Gaul}}, \bibinfo {author} {\bibfnamefont {N.}~\bibnamefont {Renner}}, \ and\
  \bibinfo {author} {\bibfnamefont {C.~A.}\ \bibnamefont {M\"{u}ller}},\ }\Doi
  {10.1103/PhysRevA.80.053620} {\bibfield  {journal} {\bibinfo  {journal}
  {Phys. Rev. A},\ }\textbf {\bibinfo {volume} {80}},\ \bibinfo {eid} {053620}
  (\bibinfo {year} {2009})}\BibitemShut {NoStop}%
\bibitem [{\citenamefont {Taylor}\ and\ \citenamefont
  {Zaremba}(2003)}]{Taylor2003}%
  \BibitemOpen
  \bibfield  {author} {\bibinfo {author} {\bibfnamefont {E.}~\bibnamefont
  {Taylor}}\ and\ \bibinfo {author} {\bibfnamefont {E.}~\bibnamefont
  {Zaremba}},\ }\Doi {10.1103/PhysRevA.68.053611} {\bibfield  {journal}
  {\bibinfo  {journal} {Phys. Rev. A},\ }\textbf {\bibinfo {volume} {68}},\
  \bibinfo {pages} {053611} (\bibinfo {year} {2003})}\BibitemShut {NoStop}%
\bibitem [{\citenamefont {Liang}\ \emph {et~al.}(2008)\citenamefont {Liang},
  \citenamefont {Dong}, \citenamefont {Zhang},\ and\ \citenamefont
  {Wu}}]{Liang2008}%
  \BibitemOpen
  \bibfield  {author} {\bibinfo {author} {\bibfnamefont {Z.~X.}\ \bibnamefont
  {Liang}}, \bibinfo {author} {\bibfnamefont {X.}~\bibnamefont {Dong}},
  \bibinfo {author} {\bibfnamefont {Z.~D.}\ \bibnamefont {Zhang}}, \ and\
  \bibinfo {author} {\bibfnamefont {B.}~\bibnamefont {Wu}},\ }\Doi
  {10.1103/PhysRevA.78.023622} {\bibfield  {journal} {\bibinfo  {journal}
  {Phys. Rev. A},\ }\textbf {\bibinfo {volume} {78}},\ \bibinfo {eid} {023622}
  (\bibinfo {year} {2008})}\BibitemShut {NoStop}%
\bibitem [{\citenamefont {Pitaevskii}\ and\ \citenamefont
  {Stringari}(1998)}]{Pitaevskii1998}%
  \BibitemOpen
  \bibfield  {author} {\bibinfo {author} {\bibfnamefont {L.}~\bibnamefont
  {Pitaevskii}}\ and\ \bibinfo {author} {\bibfnamefont {S.}~\bibnamefont
  {Stringari}},\ }\Doi {10.1103/PhysRevLett.81.4541} {\bibfield  {journal}
  {\bibinfo  {journal} {Phys. Rev. Lett.},\ }\textbf {\bibinfo {volume} {81}},\
  \bibinfo {pages} {4541} (\bibinfo {year} {1998})}\BibitemShut {NoStop}%
\bibitem [{\citenamefont {Einstein}()}]{EinsteinBEC}%
  \BibitemOpen
  \bibfield  {author} {\bibinfo {author} {\bibfnamefont {A.}~\bibnamefont
  {Einstein}},\ }\href@noop {} {}\bibinfo {howpublished} {Sitzungsber.\ Preuss.
  Akad. der Wiss. 1924, p.~261; ibid 1925, p.~3}\BibitemShut {NoStop}%
\bibitem [{\citenamefont {Lieb}\ and\ \citenamefont
  {Seiringer}(2002)}]{Lieb2002}%
  \BibitemOpen
  \bibfield  {author} {\bibinfo {author} {\bibfnamefont {E.~H.}\ \bibnamefont
  {Lieb}}\ and\ \bibinfo {author} {\bibfnamefont {R.}~\bibnamefont
  {Seiringer}},\ }\Doi {10.1103/PhysRevLett.88.170409} {\bibfield  {journal}
  {\bibinfo  {journal} {Phys. Rev. Lett.},\ }\textbf {\bibinfo {volume} {88}},\
  \bibinfo {pages} {170409} (\bibinfo {year} {2002})}\BibitemShut {NoStop}%
\bibitem [{\citenamefont {Erd\ifmmode~\mbox{\H{o}}\else \H{o}\fi{}s}\ \emph
  {et~al.}(2007)\citenamefont {Erd\ifmmode~\mbox{\H{o}}\else \H{o}\fi{}s},
  \citenamefont {Schlein},\ and\ \citenamefont {Yau}}]{Erdos2007}%
  \BibitemOpen
  \bibfield  {author} {\bibinfo {author} {\bibfnamefont {L.}~\bibnamefont
  {Erd\ifmmode~\mbox{\H{o}}\else \H{o}\fi{}s}}, \bibinfo {author}
  {\bibfnamefont {B.}~\bibnamefont {Schlein}}, \ and\ \bibinfo {author}
  {\bibfnamefont {H.-T.}\ \bibnamefont {Yau}},\ }\Doi
  {10.1103/PhysRevLett.98.040404} {\bibfield  {journal} {\bibinfo  {journal}
  {Phys. Rev. Lett.},\ }\textbf {\bibinfo {volume} {98}},\ \bibinfo {eid}
  {040404} (\bibinfo {year} {2007})}\BibitemShut {NoStop}%
\bibitem [{\citenamefont {Mora}\ and\ \citenamefont {Castin}(2003)}]{Mora2003}%
  \BibitemOpen
  \bibfield  {author} {\bibinfo {author} {\bibfnamefont {C.}~\bibnamefont
  {Mora}}\ and\ \bibinfo {author} {\bibfnamefont {Y.}~\bibnamefont {Castin}},\
  }\Doi {10.1103/PhysRevA.67.053615} {\bibfield  {journal} {\bibinfo  {journal}
  {Phys. Rev. A},\ }\textbf {\bibinfo {volume} {67}},\ \bibinfo {pages}
  {053615} (\bibinfo {year} {2003})}\BibitemShut {NoStop}%
\bibitem [{\citenamefont {Lee}\ \emph {et~al.}(1957)\citenamefont {Lee},
  \citenamefont {Huang},\ and\ \citenamefont {Yang}}]{Lee1957}%
  \BibitemOpen
  \bibfield  {author} {\bibinfo {author} {\bibfnamefont {T.~D.}\ \bibnamefont
  {Lee}}, \bibinfo {author} {\bibfnamefont {K.}~\bibnamefont {Huang}}, \ and\
  \bibinfo {author} {\bibfnamefont {C.~N.}\ \bibnamefont {Yang}},\ }\Doi
  {10.1103/PhysRev.106.1135} {\bibfield  {journal} {\bibinfo  {journal} {Phys.
  Rev.},\ }\textbf {\bibinfo {volume} {106}},\ \bibinfo {pages} {1135}
  (\bibinfo {year} {1957})}\BibitemShut {NoStop}%
\bibitem [{\citenamefont {Gross}(1963)}]{Gross1963}%
  \BibitemOpen
  \bibfield  {author} {\bibinfo {author} {\bibfnamefont {E.~P.}\ \bibnamefont
  {Gross}},\ }\Doi {10.1063/1.1703944} {\bibfield  {journal} {\bibinfo
  {journal} {J. Math. Phys.},\ }\textbf {\bibinfo {volume} {4}},\ \bibinfo
  {pages} {195} (\bibinfo {year} {1963})}\BibitemShut {NoStop}%
\bibitem [{\citenamefont {Dalfovo}\ and\ \citenamefont
  {Stringari}(1996)}]{Dalfovo1996}%
  \BibitemOpen
  \bibfield  {author} {\bibinfo {author} {\bibfnamefont {F.}~\bibnamefont
  {Dalfovo}}\ and\ \bibinfo {author} {\bibfnamefont {S.}~\bibnamefont
  {Stringari}},\ }\Doi {10.1103/PhysRevA.53.2477} {\bibfield  {journal}
  {\bibinfo  {journal} {Phys. Rev. A},\ }\textbf {\bibinfo {volume} {53}},\
  \bibinfo {pages} {2477} (\bibinfo {year} {1996})}\BibitemShut {NoStop}%
\bibitem [{\citenamefont {Sanchez-Palencia}(2006)}]{Sanchez-Palencia2006}%
  \BibitemOpen
  \bibfield  {author} {\bibinfo {author} {\bibfnamefont {L.}~\bibnamefont
  {Sanchez-Palencia}},\ }\Doi {10.1103/PhysRevA.74.053625} {\bibfield
  {journal} {\bibinfo  {journal} {Phys. Rev. A},\ }\textbf {\bibinfo {volume}
  {74}},\ \bibinfo {eid} {053625} (\bibinfo {year} {2006})}\BibitemShut
  {NoStop}%
\bibitem [{\citenamefont {Wellens}\ and\ \citenamefont
  {Gr\'emaud}(2009)}]{Wellens2009}%
  \BibitemOpen
  \bibfield  {author} {\bibinfo {author} {\bibfnamefont {T.}~\bibnamefont
  {Wellens}}\ and\ \bibinfo {author} {\bibfnamefont {B.}~\bibnamefont
  {Gr\'emaud}},\ }\Doi {10.1103/PhysRevA.80.063827} {\bibfield  {journal}
  {\bibinfo  {journal} {Phys. Rev. A},\ }\textbf {\bibinfo {volume} {80}},\
  \bibinfo {pages} {063827} (\bibinfo {year} {2009})}\BibitemShut {NoStop}%
\bibitem [{\citenamefont {Gaul}(2010)}]{GaulPhD2010}%
  \BibitemOpen
  \bibfield  {author} {\bibinfo {author} {\bibfnamefont {C.}~\bibnamefont
  {Gaul}},\ }\emph {\bibinfo {title} {Bogoliubov Excitations of Inhomogeneous
  Bose-Einstein Condensates}},\ \href
  {http://opus.ub.uni-bayreuth.de/volltexte/2010/678/} {Ph.D. thesis},\
  \bibinfo  {school} {Universit{\"a}t Bayreuth} (\bibinfo {year}
  {2010})\BibitemShut {NoStop}%
\bibitem [{\citenamefont {Katz}\ \emph {et~al.}(2002)\citenamefont {Katz},
  \citenamefont {Steinhauer}, \citenamefont {Ozeri},\ and\ \citenamefont
  {Davidson}}]{Katz2002}%
  \BibitemOpen
  \bibfield  {author} {\bibinfo {author} {\bibfnamefont {N.}~\bibnamefont
  {Katz}}, \bibinfo {author} {\bibfnamefont {J.}~\bibnamefont {Steinhauer}},
  \bibinfo {author} {\bibfnamefont {R.}~\bibnamefont {Ozeri}}, \ and\ \bibinfo
  {author} {\bibfnamefont {N.}~\bibnamefont {Davidson}},\ }\Doi
  {10.1103/PhysRevLett.89.220401} {\bibfield  {journal} {\bibinfo  {journal}
  {Phys. Rev. Lett.},\ }\textbf {\bibinfo {volume} {89}},\ \bibinfo {pages}
  {220401} (\bibinfo {year} {2002})}\BibitemShut {NoStop}%
\bibitem [{\citenamefont {Gaul}\ and\ \citenamefont
  {M{\"u}ller}(2008)}]{Gaul2008}%
  \BibitemOpen
  \bibfield  {author} {\bibinfo {author} {\bibfnamefont {C.}~\bibnamefont
  {Gaul}}\ and\ \bibinfo {author} {\bibfnamefont {C.~A.}\ \bibnamefont
  {M{\"u}ller}},\ }\Doi {10.1209/0295-5075/83/10006} {\bibfield  {journal}
  {\bibinfo  {journal} {Europhys. Lett.},\ }\textbf {\bibinfo {volume} {83}},\
  \bibinfo {pages} {10006} (\bibinfo {year} {2008})}\BibitemShut {NoStop}%
\bibitem [{\citenamefont {Kagan}\ \emph {et~al.}(2003)\citenamefont {Kagan},
  \citenamefont {Kovrizhin},\ and\ \citenamefont {Maksimov}}]{Kagan2003}%
  \BibitemOpen
  \bibfield  {author} {\bibinfo {author} {\bibfnamefont {Y.}~\bibnamefont
  {Kagan}}, \bibinfo {author} {\bibfnamefont {D.~L.}\ \bibnamefont
  {Kovrizhin}}, \ and\ \bibinfo {author} {\bibfnamefont {L.~A.}\ \bibnamefont
  {Maksimov}},\ }\Doi {10.1103/PhysRevLett.90.130402} {\bibfield  {journal}
  {\bibinfo  {journal} {Phys. Rev. Lett.},\ }\textbf {\bibinfo {volume} {90}},\
  \bibinfo {pages} {130402} (\bibinfo {year} {2003})}\BibitemShut {NoStop}%
\bibitem [{\citenamefont {Fetter}(1972)}]{Fetter1972}%
  \BibitemOpen
  \bibfield  {author} {\bibinfo {author} {\bibfnamefont {A.~L.}\ \bibnamefont
  {Fetter}},\ }\Doi {10.1016/0003-4916(72)90330-2} {\bibfield  {journal}
  {\bibinfo  {journal} {Annals of Physics},\ }\textbf {\bibinfo {volume}
  {70}},\ \bibinfo {pages} {67 } (\bibinfo {year} {1972})}\BibitemShut
  {NoStop}%
\bibitem [{\citenamefont {Lewenstein}\ and\ \citenamefont
  {You}(1996)}]{Lewenstein1996}%
  \BibitemOpen
  \bibfield  {author} {\bibinfo {author} {\bibfnamefont {M.}~\bibnamefont
  {Lewenstein}}\ and\ \bibinfo {author} {\bibfnamefont {L.}~\bibnamefont
  {You}},\ }\Doi {10.1103/PhysRevLett.77.3489} {\bibfield  {journal} {\bibinfo
  {journal} {Phys. Rev. Lett.},\ }\textbf {\bibinfo {volume} {77}},\ \bibinfo
  {pages} {3489} (\bibinfo {year} {1996})}\BibitemShut {NoStop}%
\bibitem [{\citenamefont {Bruus}\ and\ \citenamefont
  {Flensberg}(2004)}]{Bruus2004}%
  \BibitemOpen
  \bibfield  {author} {\bibinfo {author} {\bibfnamefont {H.}~\bibnamefont
  {Bruus}}\ and\ \bibinfo {author} {\bibfnamefont {K.}~\bibnamefont
  {Flensberg}},\ }\href {http://books.google.com/books?id=v5vhg1tYLC8C} {\emph
  {\bibinfo {title} {Many-body quantum theory in condensed matter physics}}}\
  (\bibinfo  {publisher} {Oxford Univ. Press},\ \bibinfo {year}
  {2004})\BibitemShut {NoStop}%
\bibitem [{\citenamefont {Mahan}(2000)}]{Mahan2000}%
  \BibitemOpen
  \bibfield  {author} {\bibinfo {author} {\bibfnamefont {G.~D.}\ \bibnamefont
  {Mahan}},\ }\href@noop {} {\emph {\bibinfo {title} {Many-particle physics}}}\
  (\bibinfo  {publisher} {Kluwer, New York},\ \bibinfo {year}
  {2000})\BibitemShut {NoStop}%
\bibitem [{\citenamefont {Akkermans}\ and\ \citenamefont
  {Montambaux}(2007)}]{Akkermans2007}%
  \BibitemOpen
  \bibfield  {author} {\bibinfo {author} {\bibfnamefont {E.}~\bibnamefont
  {Akkermans}}\ and\ \bibinfo {author} {\bibfnamefont {G.}~\bibnamefont
  {Montambaux}},\ }\href@noop {} {\emph {\bibinfo {title} {Mesoscopic physics
  of electrons and photons}}}\ (\bibinfo  {publisher} {Cambridge Univ. Press},\
  \bibinfo {year} {2007})\BibitemShut {NoStop}%
\bibitem [{\citenamefont {Gurarie}\ and\ \citenamefont
  {Chalker}(2003)}]{Gurarie2003}%
  \BibitemOpen
  \bibfield  {author} {\bibinfo {author} {\bibfnamefont {V.}~\bibnamefont
  {Gurarie}}\ and\ \bibinfo {author} {\bibfnamefont {J.~T.}\ \bibnamefont
  {Chalker}},\ }\Doi {10.1103/PhysRevB.68.134207} {\bibfield  {journal}
  {\bibinfo  {journal} {Phys. Rev. B},\ }\textbf {\bibinfo {volume} {68}},\
  \bibinfo {pages} {134207} (\bibinfo {year} {2003})}\BibitemShut {NoStop}%
\bibitem [{\citenamefont {Rammer}(1998)}]{Rammer1998}%
  \BibitemOpen
  \bibfield  {author} {\bibinfo {author} {\bibfnamefont {J.}~\bibnamefont
  {Rammer}},\ }\href@noop {} {\emph {\bibinfo {title} {Quantum Transport
  Theory}}}\ (\bibinfo  {publisher} {Perseus Books, Cambridge, MA},\ \bibinfo
  {year} {1998})\BibitemShut {NoStop}%
\bibitem [{\citenamefont {Vollhardt}\ and\ \citenamefont
  {W{\"o}lfle}(1980)}]{Vollhardt1980}%
  \BibitemOpen
  \bibfield  {author} {\bibinfo {author} {\bibfnamefont {D.}~\bibnamefont
  {Vollhardt}}\ and\ \bibinfo {author} {\bibfnamefont {P.}~\bibnamefont
  {W{\"o}lfle}},\ }\Doi {10.1103/PhysRevB.22.4666} {\bibfield  {journal}
  {\bibinfo  {journal} {Phys. Rev. B},\ }\textbf {\bibinfo {volume} {22}},\
  \bibinfo {pages} {4666} (\bibinfo {year} {1980})}\BibitemShut {NoStop}%
\bibitem [{\citenamefont {Kuhn}\ \emph {et~al.}(2005)\citenamefont {Kuhn},
  \citenamefont {Miniatura}, \citenamefont {Delande}, \citenamefont
  {Sigwarth},\ and\ \citenamefont {M{\"u}ller}}]{Kuhn2005}%
  \BibitemOpen
  \bibfield  {author} {\bibinfo {author} {\bibfnamefont {R.~C.}\ \bibnamefont
  {Kuhn}}, \bibinfo {author} {\bibfnamefont {C.}~\bibnamefont {Miniatura}},
  \bibinfo {author} {\bibfnamefont {D.}~\bibnamefont {Delande}}, \bibinfo
  {author} {\bibfnamefont {O.}~\bibnamefont {Sigwarth}}, \ and\ \bibinfo
  {author} {\bibfnamefont {C.~A.}\ \bibnamefont {M{\"u}ller}},\ }\Doi
  {10.1103/PhysRevLett.95.250403} {\bibfield  {journal} {\bibinfo  {journal}
  {Phys. Rev. Lett.},\ }\textbf {\bibinfo {volume} {95}},\ \bibinfo {eid}
  {250403} (\bibinfo {year} {2005})}\BibitemShut {NoStop}%
\bibitem [{\citenamefont {Kuhn}\ \emph {et~al.}(2007)\citenamefont {Kuhn},
  \citenamefont {Sigwarth}, \citenamefont {Miniatura}, \citenamefont
  {Delande},\ and\ \citenamefont {M{\"u}ller}}]{Kuhn2007a}%
  \BibitemOpen
  \bibfield  {author} {\bibinfo {author} {\bibfnamefont {R.}~\bibnamefont
  {Kuhn}}, \bibinfo {author} {\bibfnamefont {O.}~\bibnamefont {Sigwarth}},
  \bibinfo {author} {\bibfnamefont {C.}~\bibnamefont {Miniatura}}, \bibinfo
  {author} {\bibfnamefont {D.}~\bibnamefont {Delande}}, \ and\ \bibinfo
  {author} {\bibfnamefont {C.~A.}\ \bibnamefont {M{\"u}ller}},\ }\Doi
  {10.1088/1367-2630/9/6/161} {\bibfield  {journal} {\bibinfo  {journal} {New
  J. Phys.},\ }\textbf {\bibinfo {volume} {9}},\ \bibinfo {pages} {161}
  (\bibinfo {year} {2007})}\BibitemShut {NoStop}%
\bibitem [{\citenamefont {Milman}\ and\ \citenamefont
  {Schechtman}(1986)}]{Milman1986}%
  \BibitemOpen
  \bibfield  {author} {\bibinfo {author} {\bibfnamefont {V.}~\bibnamefont
  {Milman}}\ and\ \bibinfo {author} {\bibfnamefont {G.}~\bibnamefont
  {Schechtman}},\ }\href
  {http://books.google.com/books?id=7KJQEwWWhvYC&hl=de&source=gbs_ViewAPI}
  {\emph {\bibinfo {title} {Asymptotic theory of finite-dimensional normed
  spaces}}},\ Lecture Notes in Mathematics 1200\ (\bibinfo  {publisher}
  {Springer-Verlag, Berlin},\ \bibinfo {year} {1986})\BibitemShut {NoStop}%
\bibitem [{\citenamefont {Lugan}\ \emph {et~al.}(2009)\citenamefont {Lugan},
  \citenamefont {Aspect}, \citenamefont {Sanchez-Palencia}, \citenamefont
  {Delande}, \citenamefont {Gr{\'e}maud}, \citenamefont {M{\"u}ller},\ and\
  \citenamefont {Miniatura}}]{Lugan2009}%
  \BibitemOpen
  \bibfield  {author} {\bibinfo {author} {\bibfnamefont {P.}~\bibnamefont
  {Lugan}}, \bibinfo {author} {\bibfnamefont {A.}~\bibnamefont {Aspect}},
  \bibinfo {author} {\bibfnamefont {L.}~\bibnamefont {Sanchez-Palencia}},
  \bibinfo {author} {\bibfnamefont {D.}~\bibnamefont {Delande}}, \bibinfo
  {author} {\bibfnamefont {B.}~\bibnamefont {Gr{\'e}maud}}, \bibinfo {author}
  {\bibfnamefont {C.~A.}\ \bibnamefont {M{\"u}ller}}, \ and\ \bibinfo {author}
  {\bibfnamefont {C.}~\bibnamefont {Miniatura}},\ }\Doi
  {10.1103/PhysRevA.80.023605} {\bibfield  {journal} {\bibinfo  {journal}
  {Phys. Rev. A},\ }\textbf {\bibinfo {volume} {80}},\ \bibinfo {eid} {023605}
  (\bibinfo {year} {2009})}\BibitemShut {NoStop}%
\bibitem [{\citenamefont {Gurevich}\ and\ \citenamefont
  {Kenneth}(2009)}]{Gurevich2009}%
  \BibitemOpen
  \bibfield  {author} {\bibinfo {author} {\bibfnamefont {E.}~\bibnamefont
  {Gurevich}}\ and\ \bibinfo {author} {\bibfnamefont {O.}~\bibnamefont
  {Kenneth}},\ }\Doi {10.1103/PhysRevA.79.063617} {\bibfield  {journal}
  {\bibinfo  {journal} {Phys. Rev. A},\ }\textbf {\bibinfo {volume} {79}},\
  \bibinfo {pages} {063617} (\bibinfo {year} {2009})}\BibitemShut {NoStop}%
\bibitem [{\citenamefont {Thouless}(1973)}]{Thouless1973}%
  \BibitemOpen
  \bibfield  {author} {\bibinfo {author} {\bibfnamefont {D.~J.}\ \bibnamefont
  {Thouless}},\ }\Doi {10.1088/0022-3719/6/3/002} {\bibfield  {journal}
  {\bibinfo  {journal} {J. Phys. C: Solid State Phys.},\ }\textbf {\bibinfo
  {volume} {6}},\ \bibinfo {pages} {L49} (\bibinfo {year} {1973})}\BibitemShut
  {NoStop}%
\bibitem [{\citenamefont {John}\ \emph {et~al.}(1983)\citenamefont {John},
  \citenamefont {Sompolinsky},\ and\ \citenamefont {Stephen}}]{John1983}%
  \BibitemOpen
  \bibfield  {author} {\bibinfo {author} {\bibfnamefont {S.}~\bibnamefont
  {John}}, \bibinfo {author} {\bibfnamefont {H.}~\bibnamefont {Sompolinsky}}, \
  and\ \bibinfo {author} {\bibfnamefont {M.~J.}\ \bibnamefont {Stephen}},\
  }\Doi {10.1103/PhysRevB.27.5592} {\bibfield  {journal} {\bibinfo  {journal}
  {Phys. Rev. B},\ }\textbf {\bibinfo {volume} {27}},\ \bibinfo {pages} {5592}
  (\bibinfo {year} {1983})}\BibitemShut {NoStop}%
\bibitem [{\citenamefont {Stamper-Kurn}\ \emph {et~al.}(1999)\citenamefont
  {Stamper-Kurn}, \citenamefont {Chikkatur}, \citenamefont {G{\"o}rlitz},
  \citenamefont {Inouye}, \citenamefont {Gupta}, \citenamefont {Pritchard},\
  and\ \citenamefont {Ketterle}}]{Stamper-Kurn1999}%
  \BibitemOpen
  \bibfield  {author} {\bibinfo {author} {\bibfnamefont {D.~M.}\ \bibnamefont
  {Stamper-Kurn}}, \bibinfo {author} {\bibfnamefont {A.~P.}\ \bibnamefont
  {Chikkatur}}, \bibinfo {author} {\bibfnamefont {A.}~\bibnamefont
  {G{\"o}rlitz}}, \bibinfo {author} {\bibfnamefont {S.}~\bibnamefont {Inouye}},
  \bibinfo {author} {\bibfnamefont {S.}~\bibnamefont {Gupta}}, \bibinfo
  {author} {\bibfnamefont {D.~E.}\ \bibnamefont {Pritchard}}, \ and\ \bibinfo
  {author} {\bibfnamefont {W.}~\bibnamefont {Ketterle}},\ }\Doi
  {10.1103/PhysRevLett.83.2876} {\bibfield  {journal} {\bibinfo  {journal}
  {Phys. Rev. Lett.},\ }\textbf {\bibinfo {volume} {83}},\ \bibinfo {pages}
  {2876} (\bibinfo {year} {1999})}\BibitemShut {NoStop}%
\bibitem [{\citenamefont {Vogels}\ \emph {et~al.}(2002)\citenamefont {Vogels},
  \citenamefont {Xu}, \citenamefont {Raman}, \citenamefont {Abo-Shaeer},\ and\
  \citenamefont {Ketterle}}]{Vogels2002}%
  \BibitemOpen
  \bibfield  {author} {\bibinfo {author} {\bibfnamefont {J.~M.}\ \bibnamefont
  {Vogels}}, \bibinfo {author} {\bibfnamefont {K.}~\bibnamefont {Xu}}, \bibinfo
  {author} {\bibfnamefont {C.}~\bibnamefont {Raman}}, \bibinfo {author}
  {\bibfnamefont {J.~R.}\ \bibnamefont {Abo-Shaeer}}, \ and\ \bibinfo {author}
  {\bibfnamefont {W.}~\bibnamefont {Ketterle}},\ }\Doi
  {10.1103/PhysRevLett.88.060402} {\bibfield  {journal} {\bibinfo  {journal}
  {Phys. Rev. Lett.},\ }\textbf {\bibinfo {volume} {88}},\ \bibinfo {pages}
  {060402} (\bibinfo {year} {2002})}\BibitemShut {NoStop}%
\bibitem [{\citenamefont {Steinhauer}\ \emph {et~al.}(2002)\citenamefont
  {Steinhauer}, \citenamefont {Ozeri}, \citenamefont {Katz},\ and\
  \citenamefont {Davidson}}]{Steinhauer2002}%
  \BibitemOpen
  \bibfield  {author} {\bibinfo {author} {\bibfnamefont {J.}~\bibnamefont
  {Steinhauer}}, \bibinfo {author} {\bibfnamefont {R.}~\bibnamefont {Ozeri}},
  \bibinfo {author} {\bibfnamefont {N.}~\bibnamefont {Katz}}, \ and\ \bibinfo
  {author} {\bibfnamefont {N.}~\bibnamefont {Davidson}},\ }\Doi
  {10.1103/PhysRevLett.88.120407} {\bibfield  {journal} {\bibinfo  {journal}
  {Phys. Rev. Lett.},\ }\textbf {\bibinfo {volume} {88}},\ \bibinfo {pages}
  {120407} (\bibinfo {year} {2002})}\BibitemShut {NoStop}%
\bibitem [{\citenamefont {Steinhauer}\ \emph {et~al.}(2003)\citenamefont
  {Steinhauer}, \citenamefont {Katz}, \citenamefont {Ozeri}, \citenamefont
  {Davidson}, \citenamefont {Tozzo},\ and\ \citenamefont
  {Dalfovo}}]{Steinhauer2003}%
  \BibitemOpen
  \bibfield  {author} {\bibinfo {author} {\bibfnamefont {J.}~\bibnamefont
  {Steinhauer}}, \bibinfo {author} {\bibfnamefont {N.}~\bibnamefont {Katz}},
  \bibinfo {author} {\bibfnamefont {R.}~\bibnamefont {Ozeri}}, \bibinfo
  {author} {\bibfnamefont {N.}~\bibnamefont {Davidson}}, \bibinfo {author}
  {\bibfnamefont {C.}~\bibnamefont {Tozzo}}, \ and\ \bibinfo {author}
  {\bibfnamefont {F.}~\bibnamefont {Dalfovo}},\ }\Doi
  {10.1103/PhysRevLett.90.060404} {\bibfield  {journal} {\bibinfo  {journal}
  {Phys. Rev. Lett.},\ }\textbf {\bibinfo {volume} {90}},\ \bibinfo {pages}
  {060404} (\bibinfo {year} {2003})}\BibitemShut {NoStop}%
\bibitem [{\citenamefont {Lifshitz}\ \emph {et~al.}(1988)\citenamefont
  {Lifshitz}, \citenamefont {Gredeskul},\ and\ \citenamefont
  {Pastur}}]{Lifshitz1988}%
  \BibitemOpen
  \bibfield  {author} {\bibinfo {author} {\bibfnamefont {I.~M.}\ \bibnamefont
  {Lifshitz}}, \bibinfo {author} {\bibfnamefont {S.~A.}\ \bibnamefont
  {Gredeskul}}, \ and\ \bibinfo {author} {\bibfnamefont {L.~A.}\ \bibnamefont
  {Pastur}},\ }\href@noop {} {\emph {\bibinfo {title} {Introduction to the
  Theory of Disordered Systems}}}\ (\bibinfo  {publisher} {Wiley, New York},\
  \bibinfo {year} {1988})\BibitemShut {NoStop}%
\bibitem [{\citenamefont {Lugan}\ \emph
  {et~al.}(2007){\natexlab{b}}\citenamefont {Lugan}, \citenamefont {Clement},
  \citenamefont {Bouyer}, \citenamefont {Aspect}, \citenamefont {Lewenstein},\
  and\ \citenamefont {Sanchez-Palencia}}]{Lugan2007}%
  \BibitemOpen
  \bibfield  {author} {\bibinfo {author} {\bibfnamefont {P.}~\bibnamefont
  {Lugan}}, \bibinfo {author} {\bibfnamefont {D.}~\bibnamefont {Clement}},
  \bibinfo {author} {\bibfnamefont {P.}~\bibnamefont {Bouyer}}, \bibinfo
  {author} {\bibfnamefont {A.}~\bibnamefont {Aspect}}, \bibinfo {author}
  {\bibfnamefont {M.}~\bibnamefont {Lewenstein}}, \ and\ \bibinfo {author}
  {\bibfnamefont {L.}~\bibnamefont {Sanchez-Palencia}},\ }\Doi
  {10.1103/PhysRevLett.98.170403} {\bibfield  {journal} {\bibinfo  {journal}
  {Phys. Rev. Lett.},\ }\textbf {\bibinfo {volume} {98}},\ \bibinfo {eid}
  {170403} (\bibinfo {year} {2007}{\natexlab{b}})}\BibitemShut {NoStop}%
\bibitem [{\citenamefont {Falco}\ \emph {et~al.}(2009)\citenamefont {Falco},
  \citenamefont {Nattermann},\ and\ \citenamefont {Pokrovsky}}]{Falco2009a}%
  \BibitemOpen
  \bibfield  {author} {\bibinfo {author} {\bibfnamefont {G.~M.}\ \bibnamefont
  {Falco}}, \bibinfo {author} {\bibfnamefont {T.}~\bibnamefont {Nattermann}}, \
  and\ \bibinfo {author} {\bibfnamefont {V.~L.}\ \bibnamefont {Pokrovsky}},\
  }\Doi {10.1103/PhysRevB.80.104515} {\bibfield  {journal} {\bibinfo  {journal}
  {Phys. Rev. B},\ }\textbf {\bibinfo {volume} {80}},\ \bibinfo {eid} {104515}
  (\bibinfo {year} {2009})}\BibitemShut {NoStop}%
\bibitem [{\citenamefont {Pilati}\ \emph {et~al.}(2009)\citenamefont {Pilati},
  \citenamefont {Giorgini},\ and\ \citenamefont {Prokof'ev}}]{Pilati2009}%
  \BibitemOpen
  \bibfield  {author} {\bibinfo {author} {\bibfnamefont {S.}~\bibnamefont
  {Pilati}}, \bibinfo {author} {\bibfnamefont {S.}~\bibnamefont {Giorgini}}, \
  and\ \bibinfo {author} {\bibfnamefont {N.}~\bibnamefont {Prokof'ev}},\ }\Doi
  {10.1103/PhysRevLett.102.150402} {\bibfield  {journal} {\bibinfo  {journal}
  {Phys. Rev. Lett.},\ }\textbf {\bibinfo {volume} {102}},\ \bibinfo {eid}
  {150402} (\bibinfo {year} {2009})}\BibitemShut {NoStop}%
\bibitem [{\citenamefont {Carrasquilla}\ \emph {et~al.}(2010)\citenamefont
  {Carrasquilla}, \citenamefont {Becca}, \citenamefont {Trombettoni},\ and\
  \citenamefont {Fabrizio}}]{Carrasquilla2010}%
  \BibitemOpen
  \bibfield  {author} {\bibinfo {author} {\bibfnamefont {J.}~\bibnamefont
  {Carrasquilla}}, \bibinfo {author} {\bibfnamefont {F.}~\bibnamefont {Becca}},
  \bibinfo {author} {\bibfnamefont {A.}~\bibnamefont {Trombettoni}}, \ and\
  \bibinfo {author} {\bibfnamefont {M.}~\bibnamefont {Fabrizio}},\ }\Doi
  {10.1103/PhysRevB.81.195129} {\bibfield  {journal} {\bibinfo  {journal}
  {Phys. Rev. B},\ }\textbf {\bibinfo {volume} {81}},\ \bibinfo {pages}
  {195129} (\bibinfo {year} {2010})}\BibitemShut {NoStop}%
\bibitem [{\citenamefont {Gaul}\ and\ \citenamefont
  {M{\"u}ller}(2010)}]{Gaul2010_letter}%
  \BibitemOpen
  \bibfield  {author} {\bibinfo {author} {\bibfnamefont {C.}~\bibnamefont
  {Gaul}}\ and\ \bibinfo {author} {\bibfnamefont {C.~A.}\ \bibnamefont
  {M{\"u}ller}},\ }\href@noop {} {}\bibinfo {howpublished}
  {\href{http://arxiv.org/abs/arXiv:1009.5448}{arXiv:1009.5448}} (\bibinfo
  {year} {2010})\BibitemShut {NoStop}%
\bibitem [{\citenamefont {M\"uller}\ and\ \citenamefont
  {Gaul}(2011)}]{Muller2010}%
  \BibitemOpen
  \bibfield  {author} {\bibinfo {author} {\bibfnamefont {C.~A.}\ \bibnamefont
  {M\"uller}}\ and\ \bibinfo {author} {\bibfnamefont {C.}~\bibnamefont
  {Gaul}},\ }\href@noop {} {\bibfield  {journal} {\bibinfo  {journal}
  {unpublished}} (\bibinfo {year} {2011})}\BibitemShut {NoStop}%
\bibitem [{\citenamefont {Goodman}(1975)}]{Goodman1975}%
  \BibitemOpen
  \bibfield  {author} {\bibinfo {author} {\bibfnamefont {J.~W.}\ \bibnamefont
  {Goodman}},\ }\enquote {\bibinfo {title} {\textit{Laser speckle and related
  phenomena}},}\ \ (\bibinfo  {publisher} {Springer-Verlag, Berlin},\ \bibinfo
  {year} {1975})\ Chap.~\bibinfo {chapter} {2}, p.~\bibinfo {pages}
  {9}\BibitemShut {NoStop}%
\bibitem [{\citenamefont {Cl{\'e}ment}\ \emph {et~al.}(2006)\citenamefont
  {Cl{\'e}ment}, \citenamefont {Var{\'o}n}, \citenamefont {Retter},
  \citenamefont {Sanchez-Palencia}, \citenamefont {Aspect},\ and\ \citenamefont
  {Bouyer}}]{Clement2006}%
  \BibitemOpen
  \bibfield  {author} {\bibinfo {author} {\bibfnamefont {D.}~\bibnamefont
  {Cl{\'e}ment}}, \bibinfo {author} {\bibfnamefont {A.~F.}\ \bibnamefont
  {Var{\'o}n}}, \bibinfo {author} {\bibfnamefont {J.~A.}\ \bibnamefont
  {Retter}}, \bibinfo {author} {\bibfnamefont {L.}~\bibnamefont
  {Sanchez-Palencia}}, \bibinfo {author} {\bibfnamefont {A.}~\bibnamefont
  {Aspect}}, \ and\ \bibinfo {author} {\bibfnamefont {P.}~\bibnamefont
  {Bouyer}},\ }\Doi {10.1088/1367-2630/8/8/165} {\bibfield  {journal} {\bibinfo
   {journal} {New J. Phys.},\ }\textbf {\bibinfo {volume} {8}},\ \bibinfo
  {pages} {165} (\bibinfo {year} {2006})}\BibitemShut {NoStop}%
\bibitem [{\citenamefont {Grimm}\ \emph {et~al.}(2000)\citenamefont {Grimm},
  \citenamefont {Weidem{\"u}ller},\ and\ \citenamefont
  {Ovchinnikov}}]{Grimm2000}%
  \BibitemOpen
  \bibfield  {author} {\bibinfo {author} {\bibfnamefont {R.}~\bibnamefont
  {Grimm}}, \bibinfo {author} {\bibfnamefont {M.}~\bibnamefont
  {Weidem{\"u}ller}}, \ and\ \bibinfo {author} {\bibfnamefont {Y.~B.}\
  \bibnamefont {Ovchinnikov}},\ }\href@noop {} {\bibfield  {journal} {\bibinfo
  {journal} {Advances in Atomic, Molecular, and Optical Physics},\ }\textbf
  {\bibinfo {volume} {42}},\ \bibinfo {pages} {95} (\bibinfo {year} {2000})},\
  \Eprint {http://arxiv.org/abs/arXiv:physics/9902072} {arXiv:physics/9902072}
  \BibitemShut {NoStop}%
\bibitem [{\citenamefont {Billy}\ \emph {et~al.}(2008)\citenamefont {Billy},
  \citenamefont {Josse}, \citenamefont {Zuo}, \citenamefont {Bernard},
  \citenamefont {Hambrecht}, \citenamefont {Lugan}, \citenamefont {Clement},
  \citenamefont {Sanchez-Palencia}, \citenamefont {Bouyer},\ and\ \citenamefont
  {Aspect}}]{Billy2008}%
  \BibitemOpen
  \bibfield  {author} {\bibinfo {author} {\bibfnamefont {J.}~\bibnamefont
  {Billy}}, \bibinfo {author} {\bibfnamefont {V.}~\bibnamefont {Josse}},
  \bibinfo {author} {\bibfnamefont {Z.}~\bibnamefont {Zuo}}, \bibinfo {author}
  {\bibfnamefont {A.}~\bibnamefont {Bernard}}, \bibinfo {author} {\bibfnamefont
  {B.}~\bibnamefont {Hambrecht}}, \bibinfo {author} {\bibfnamefont
  {P.}~\bibnamefont {Lugan}}, \bibinfo {author} {\bibfnamefont
  {D.}~\bibnamefont {Clement}}, \bibinfo {author} {\bibfnamefont
  {L.}~\bibnamefont {Sanchez-Palencia}}, \bibinfo {author} {\bibfnamefont
  {P.}~\bibnamefont {Bouyer}}, \ and\ \bibinfo {author} {\bibfnamefont
  {A.}~\bibnamefont {Aspect}},\ }\Doi {10.1038/nature07000} {\bibfield
  {journal} {\bibinfo  {journal} {Nature},\ }\textbf {\bibinfo {volume}
  {453}},\ \bibinfo {pages} {891} (\bibinfo {year} {2008})}\BibitemShut
  {NoStop}%
\bibitem [{\citenamefont {Renner}(2009)}]{Renner2009}%
  \BibitemOpen
  \bibfield  {author} {\bibinfo {author} {\bibfnamefont {N.}~\bibnamefont
  {Renner}},\ }\emph {\bibinfo {title} {Schallgeschwindigkeit von
  {B}ogoliubov-{A}n\-re\-gun\-gen in ungeordneten
  {B}ose-{E}instein-{K}on\-den\-saten}},\ \href
  {http://www.phy.uni-bayreuth.de/~btpj00/Theses/Renner_DA.pdf} {\bibinfo
  {type} {diploma thesis}},\ \bibinfo  {school} {Universit{\"a}t Bayreuth}
  (\bibinfo {year} {2009})\BibitemShut {NoStop}%
\end{thebibliography}%

\end{document}